\documentclass{aa}
\pdfoutput=1 
\usepackage{graphicx}
\usepackage{appendix}
\usepackage{txfonts}
\usepackage{placeins}
\usepackage{subfig}
\usepackage{float}
\usepackage{xcolor}
\usepackage[colorlinks=true,citecolor=blue]{hyperref}%
\begin{document}
\title{Fragments of harmony amid apparent chaos: a closer look at the X-ray quasi-periodic eruptions of the galaxy RX~J1301.9+2747}

   \author{Margherita Giustini\inst{1}
   \and
         Giovanni Miniutti\inst{1}
         \and
         Riccardo  Arcodia\inst{2}
         \and
         Adelle Goodwin\inst{3}
         \and
         Kate~D.~Alexander\inst{4}
         \and
         Joheen~Chakraborty\inst{2}
         \and
         Johannes~Buchner\inst{5}
         \and
         Peter Kosec\inst{6}
         \and
         Richard Saxton\inst{7}
         \and
         Matteo Bonetti\inst{8,9, 10}
         \and
         Alessia~Franchini\inst{11,9,12}
         \and
         Taeho~Ryu\inst{13}
         \and
         Xinwen Shu\inst{14} 
         \and
         Erin Kara\inst{2}
         \and
         Gabriele~Ponti\inst{15,5}
         \and
         Erwan~Quintin\inst{16}
         \and
         Federico~Vincentelli\inst{17, 18}
         \and
         Natalie~Webb\inst{16}
         \and
         Jari Kajava\inst{19}
         \and
         Sebastiano D. von Fellenberg\inst{20}
         }

   \institute{Centro de Astrobiologia (CAB), CSIC-INTA, Camino Bajo del Castillo s/n, Campus ESAC, 28692, Villanueva de la Ca\~nada, Madrid, Spain
   \and
   Kavli Institute for Astrophysics and Space Research, Massachusetts Institute of Technology (MIT), Cambridge, MA, USA
   \and
   International Centre for Radio Astronomy Research (ICRA), Curtin University, GPO Box U1987, Perth WA 6845, Australia
   \and
    Department of Astronomy/Steward Observatory, University of Arizona, 933 N Cherry Avenue, Tucson, AZ 85721-0065, USA
    \and
   Max-Planck-Institut f\"ur extraterrestrische Physik (MPE), Giessnbachstrasse 1, 85748 Garching bei M\"unchen, Germany
   \and
   Center for Astrophysics | Harvard \& Smithsonian, Cambridge, MA, USA
   \and
   Telespazio UK for the European Space Agency (ESA), European Space Astronomy Centre (ESAC), Camino Bajo del Castillo s/n, 28692 Villanueva de la Ca\~nada, Madrid, Spain
   \and
   Dipartimento di Fisica ``G. Occhialini'', Università degli Studi di Milano-Bicocca, Piazza della Scienza 3, I-20126 Milano, Italy
   \and
   INFN, Sezione di Milano-Bicocca, Piazza della Scienza 3, I-20126 Milano, Italy
   \and
   INAF - Osservatorio Astronomico di Brera, via Brera 20, I-20121 Milano, Italy
   \and
   Center for Theoretical Astrophysics and Cosmology, Institute for Computational Science, University of Zurich, Winterthurerstrasse 190, CH-8057 Z\"urich, Switzerland
   \and
   Institut für Astrophysik, Universität Zürich,
   Winterthurerstrasse 190, CH-8057 Zürich, Switzerland
   \and
   Max Planck Institute for Astrophysics (MPA), Karl-Schwarzschild-Strasse 1, D-85748 Garching, Germany
   \and
   Department of Physics, Anhui Normal University, Wuhu, Anhui, 241002, People’s Republic of China
   \and
   INAF, Osservatorio Astronomico di Brera, Via E. Bianchi 46, 23807 Merate (LC), Italy
   \and
   IRAP, Universit\'e de Toulouse, CNRS, UPS, CNES, 9 Avenue du Colonel Roche, BP 44346, 31028 Toulouse Cedex 4, France
   \and
   Instituto de Astrof\'isica de Canarias, E-38205 La Laguna, Tenerife, Spain
   \and
   Departamento de Astrof\'isica, Universidad de La Laguna, E-38206 La Laguna, Tenerife, Spain
   \and
   Serco for the European Space Agency (ESA), European Space Astronomy Centre (ESAC), Camino Bajo del Castillo s/n,  28692 Villanueva de la Ca\~nada, Madrid, Spain
   \and
   Max Planck Institute for Radio Astronomy, Auf dem H\"ugel 69, Bonn 53121, Germany
}

   \date{Received  / Accepted  }

\abstract
{Quasi-periodic eruptions (QPEs) are an extreme X-ray variability phenomenon associated with low-mass ($M_{BH} < 10^7 M_{\odot}$) supermassive black holes (SMBHs). First discovered in the nucleus of the galaxy GSN 069, they have been so far securely detected in five other galaxies, including RX~J1301.9+2747.
When detected, the out-of-QPE emission (quiescence) is consistent with the high-energy tail of thermal emission from an accretion disk.  
In this article we present the X-ray properties of RX~J1301.9+2747, both in quiescence and during QPEs, and complement this information with radio observations.
We analyse X-ray data taken during five \textit{XMM-Newton} observations between 2000 and 2022. The last three observations were taken in coordination with radio observations with the Karl G. Jansky Very Large Array. We also make use of \textit{EXOSAT}, \textit{ROSAT}, and \textit{Chandra} archival observations taken between 1983 and 2009. \textit{XMM-Newton} detected 34 QPEs of which 8 have significantly lower amplitudes than the others. No correlated radio/X-ray variability was observed during QPEs. In terms of timing properties, the QPEs in RX~J1301.9+2747 do not exhibit the striking regularity observed in the discovery source GSN 069. In fact there is no clear repetition pattern between QPEs: the average time separation between their peaks is about four hours, but it can be as short as one, and as long as six hours. The QPE spectral properties of RX~J1301.9+2747 as a function of energy are however very similar to those of GSN 069 and of other QPE sources. During their evolution, X-ray QPEs follow a hysteresis pattern in the temperature-luminosity plane, with a hotter rise than decay. The quiescent emission of RX~J1301.9+2747 is more complex than that of GSN 069, as it requires a soft X-ray excess-like component in addition to the thermal emission from the accretion disk. Its long-term X-ray quiescent flux variations are of low-amplitude and not strictly monotonic, with a general decay over $\sim 22$ years.
We discuss our observational results in terms of some of the ideas and models that have been proposed so far for the physical origin of QPEs.}
    

\titlerunning{A closer look at the X-ray QPEs of RX~J1301.9+2747}
\authorrunning{M. Giustini et al.}

\maketitle
\section{Introduction}\label{sec:intro}
X-ray quasi-periodic eruptions (QPEs) are extreme extragalactic variability phenomenona associated with supermassive black holes (SMBHs).
They are rapid, intense, and recurrent flares of soft X-ray emission. Each QPE typically reaches a soft X-ray peak luminosity on the order of $10^{42-43}$ erg s$^{-1}$, roughly one order of magnitude higher than the quiescent state. 
When detected, the quiescent state emission is most likely associated with thermal disk emission with typical temperature $kT \sim 50-80$ eV, while QPEs have X-ray spectra consistent with blackbody-like thermal emission with typical $kT \sim 100-200$ eV. 
The typical duration of QPEs is between $\lesssim 1$ hour and a few hours, and their time separation about $2.5-20$ hours. 

First discovered in the nucleus of the galaxy GSN 069  \citep{2019Natur.573..381M}, QPEs have been so far identified in other five galaxies, along with two other promising candidates. 
Shortly after the very first QPE discovery, flares of similar duration were identified in \textit{XMM-Newton} archival observations of RX~J1301.9+2747 \citep{2013ApJ...768..167S, 2015MNRAS.446.1312M, 2017ApJ...837....3S} and were confirmed to be QPEs with new observations performed in 2019 \citep{2020A&A...636L...2G}. 
Thanks to blind searches within the all-sky \textit{SRG}/eROSITA survey, X-ray QPEs have been discovered in other four galaxies: 2MASS 02314715-1020112 and 2MASX J02344872-4419325 \citep[eRO-QPE1 and eRO-QPE2;][]{2021Natur.592..704A}, and  2MASS 14005331-2846012 and 2MASS 04453380-1012047 \citep[eRO-QPE3 and eRO-QPE4;][]{2024A&A...684A..64A}. Furthermore, QPE-like soft X-ray bursts have been detected in the nuclei of the galaxies XMMSL1 J024916.6-041244 \citep[J0249;][]{2021ApJ...921L..40C} and 4XMM J123856.3+330957 \citep[Tormund;][]{2023A&A...675A.152Q}.

Both GSN 069 and RX~J1301.9+2747 were known to be active galaxies before the discovery of QPEs, albeit exhibiting only narrow optical emission lines \citep[see respectively][]{2013MNRAS.433.1764M,2000MNRAS.318..309D}.
Also eRO-QPE1, eRO-QPE2, and J0249 were shown to host narrow emission lines in their optical spectra, indicating the presence of an ionizing photon source in excess of pure stellar light, i.e., signatures of nuclear activity \citep{2022A&A...659L...2W}. 
The eRO-QPE3 and eRO-QPE4 optical spectra taken with the Southern African Large Telescope revealed the presence of faint and narrow optical emission lines, although higher-resolution spectra are needed to confirm the exact nature of their nuclear activity \citep{2024A&A...684A..64A}. Despite being basically unobscured in the X-rays, none of the current QPE-emitting galaxies shows signs of broad optical or UV emission lines in their spectra. This rules out the presence of a standard active galactic nucleus (AGN) in the QPE-hosting galaxies. 

The QPE-hosting galaxies have black hole masses estimated through stellar velocity dispersion measurements or through continuum luminosity scaling relations on the lower end of the SMBH distribution, $M_{BH}\sim 10^{5}-5\times 10^6\,M_{\odot}$ \citep{2017ApJ...837....3S,2019Natur.573..381M,2021Natur.592..704A,2022A&A...659L...2W,2024A&A...684A..64A}. Three out of eight QPE sources (GSN 069, J0249, and Tormund) have a clear connection to tidal disruption events \citep[TDEs;][]{2018ApJ...857L..16S, 2021ApJ...920L..25S, 2021ApJ...921L..40C,2023A&A...675A.152Q}. The observation of decaying X-ray quiescent flux in eRO-QPE3 also suggests a connection to a TDE \citep{2024A&A...684A..64A}. 
Optical integral field spectroscopy has revealed the presence of extended emission line regions of ionized gas in 3 out of 5 QPE host galaxies, with properties very similar to those of TDE host galaxies \citep{2024ApJ...970L..23W}.
The accretion flow associated with TDEs is possibly too compact to support the mature broad line region typical of AGN. The association between QPEs and TDEs might thus explain the lack of any broad emission line component in the nuclear optical spectra of QPE hosting galaxies \citep{2022A&A...659L...2W,2024ApJ...970L..23W}.

The signal associated with QPEs has only been observed so far in the soft X-ray band, with a maximum amplitude reached around $E=600-800$ eV, fading away at $E\gtrsim 1.5$ keV. 
During the QPE evolution, the temperature smoothly rises to a peak value, and then returns to its initial level.
The QPE's properties are energy-dependent: the flares observed at higher energies have shorter durations, with high-amplitude peaks occurring at earlier times compared to lower energies \citep{2019Natur.573..381M,2020A&A...636L...2G,2021Natur.592..704A}. Therefore the peak temperature of QPEs is usually reached before their peak luminosity. 
The flares measured in the softest X-ray band begin before those measured in the hardest X-ray band in eRO-QPE1, whose longer duration QPEs allow for better energy-dependent measurements with respect to any other QPE source \citep{2022A&A...662A..49A}.
Both the QPE amplitudes and the time differences between the QPE peaks (hereafer recurrence times) alternate between weak and strong, short and long in GSN 069 and eRO-QPE2 \citep[][but see also \citet{2023A&A...670A..93M,2023A&A...674L...1M,2024arXiv240617020A}]{2019Natur.573..381M,2021Natur.592..704A}. The temporal behavior is more complex in eRO-QPE1, with QPEs of different amplitude repeating with a large scatter in recurrence times \citep{2022A&A...662A..49A,2024ApJ...965...12C}.

There are two main physical scenarios able to provide context to interpret the QPE observations: accretion flow instabilities and orbital phenomena. 
The former can reproduce the recurrence times and luminosities of QPEs only if strong magnetic effects are present \citep[e.g.,][]{2022ApJ...928L..18P,2023MNRAS.524.1269K, 2023ApJ...952...32P,2023A&A...672A..19S}, or in the case of  instabilities leading to accretion disk tearing  \citep{2021ApJ...909...82R}. However, accretion instability models generally predict a flare shape with a slow rise and a fast decay, which is opposite to what is typically observed in QPEs \citep[e.g.,][]{2019Natur.573..381M,2022A&A...662A..49A,2024A&A...684A..64A}.  
Orbital phenomena include both gravitational self-lensing of massive black hole binaries of about equal mass ratio \citep{2021MNRAS.503.1703I}, and purely orbital motions of much smaller mass objects around a more massive black hole (extreme mass ratio inspirals, EMRI).
The EMRI scenario can include single or multiple stars or stellar remnants undergoing Roche Lobe overflow or tidal stripping at each pericenter passage \citep[e.g.,][]{2020MNRAS.493L.120K,2022A&A...661A..55Z,2022ApJ...926..101M,2022ApJ...933..225W, 2022MNRAS.515.4344K, 2022ApJ...941...24K, 2023MNRAS.524.6247L, 2023ApJ...945...86L,2023MNRAS.526L..31K}, as well as impacts between low-mass orbiting companions (stars or black holes) and an accretion flow around the primary SMBH \citep{2021ApJ...917...43S,2021ApJ...921L..32X, 2023ApJ...957...34L, 2023A&A...675A.100F, 2023MNRAS.526...69T,2024arXiv240111190Z,2024arXiv240506429Z}. In any case, no model so far can fully account for all the observational properties of QPEs.

In this work, we focus on RX~J1301.9+2747  (RA=13h 02m 00.138s, dec=+27d 46m 57.855s; redshift $z=0.024$). This is a young post-starburst galaxy belonging to a small group of galaxies in the outskirts of the Coma Cluster \citep{1999AJ....117..140C}.
Its optical spectrum lacks broad optical emission lines. A low black hole mass has been estimated either from the width of the [O III] line \citep[$M_{BH}=8\times 10^5\,M_{\odot}$,][]{2013ApJ...768..167S}, 
from spectral fitting \citep[$M_{BH}=1.5-3\times 10^6\,M_{\odot}$,][]{2017ApJ...837....3S}, or from  velocity dispersion measurements  \citep[$M_{BH}=1.2-4.5\times 10^6\,M_{\odot}$,][]{2022A&A...659L...2W, 2024ApJ...970L..23W}. 
In the X-ray band, RX~J1301.9+2747 was serendipitously discovered  during an \textit{EXOSAT} observation of the Coma Cluster \citep{1985MNRAS.216.1043B}.
It was then observed by \textit{ROSAT} in 1991, by \textit{XMM-Newton} in 2000, and by \textit{Chandra} in 2009, in all cases showing hints of rapid variability in its light curves \citep{2000MNRAS.318..309D, 2013ApJ...768..167S,2017ApJ...837....3S}.
More recently, the rapid X-ray variability has been observed again by \citet{2020A&A...636L...2G}, confirming that RX~J1301.9+2747 is indeed a QPE source.

RX~J1301.9+2747 is the second galaxy where X-ray QPEs have been discovered, and the only one  where  such  phenomenon has been observed – albeit recognised a posteriori – on timescales of decades.
QPEs were already present in 2000 \citep{2013ApJ...768..167S}, and three more QPEs were observed during a 2019 \textit{XMM-Newton} observation. 
While the QPEs observed in RX~J1301.9+2747 have amplitudes and durations comparable to those observed in GSN 069, their recurrence pattern appears to be less regular. Specifically, the time separation between the two QPEs detected in 2000 is about 5 hours (5h; equivalent to about 18 ks), while the time separation between the three QPEs detected in 2019 is about 5.5h (20 ks) and 3h~45m (13.5 ks)  \citep{2020A&A...636L...2G}.

In this article we present X-ray data of RX~J1301.9+2747 taken from 1983 until 2022. These include one \textit{EXOSAT} flux measurement, \textit{ROSAT} and \textit{Chandra} spectra, as well as five \textit{XMM-Newton} observations, of which four are pointed at the source. Each of the last three \textit{XMM-Newton} observations was taken in coordination with 10 hours of exposure with the National Radio Astronomy Observatory (NRAO) Karl G. Jansky Very Large Array (VLA).
We provide a detailed description of the setup for the data reduction and analysis in Sect \ref{sec:data}. We then present the X-ray data analysis in Sects. \ref{sec:lcurves} (light curves) and \ref{Xray} (spectra), and the radio data analysis in Sect.~\ref{sec:radio}. We discuss our results in Sect. \ref{sec:discu} and we conclude in Sect. \ref{sec:conclu}. Details about the analysis procedures used, as well as the complete set of Tables and figures can be found in the Appendices~\ref{appendix:obslog}-\ref{appendix:qpes}. 
Errors are quoted at the $1\sigma$ confidence level throughout the text.
A flat cosmology ($\Lambda=0.73$, $q_0=0$, $H_0=70$ km s$^{-1}$ Mpc$^{-1}$) is assumed for the computation of the intrinsic luminosities and luminosity distances.

\begin{figure*}[ht!]
\centering
\includegraphics[width=18cm]{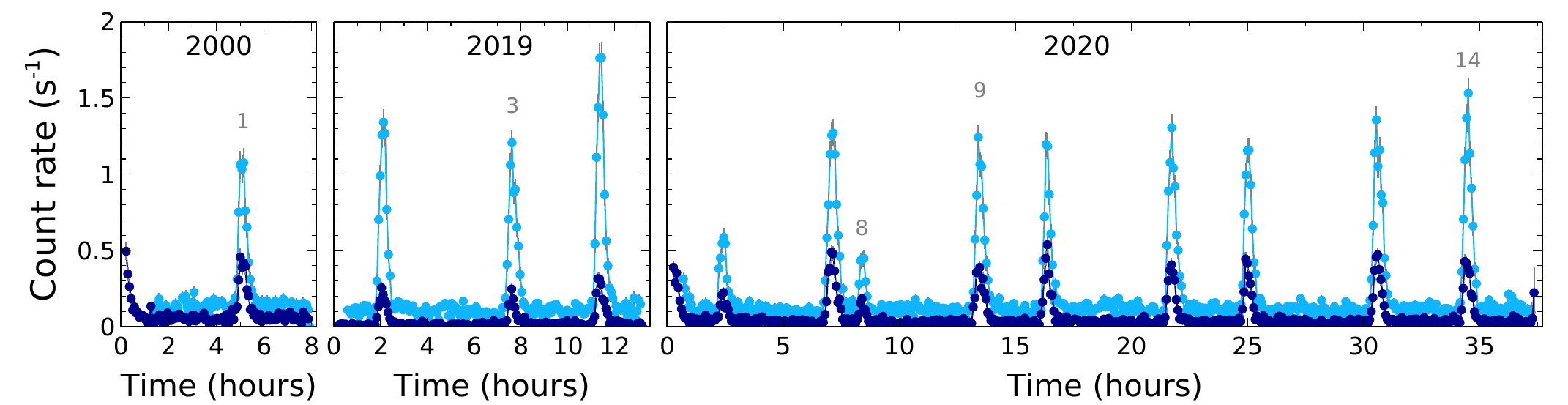}
\includegraphics[width=18cm]{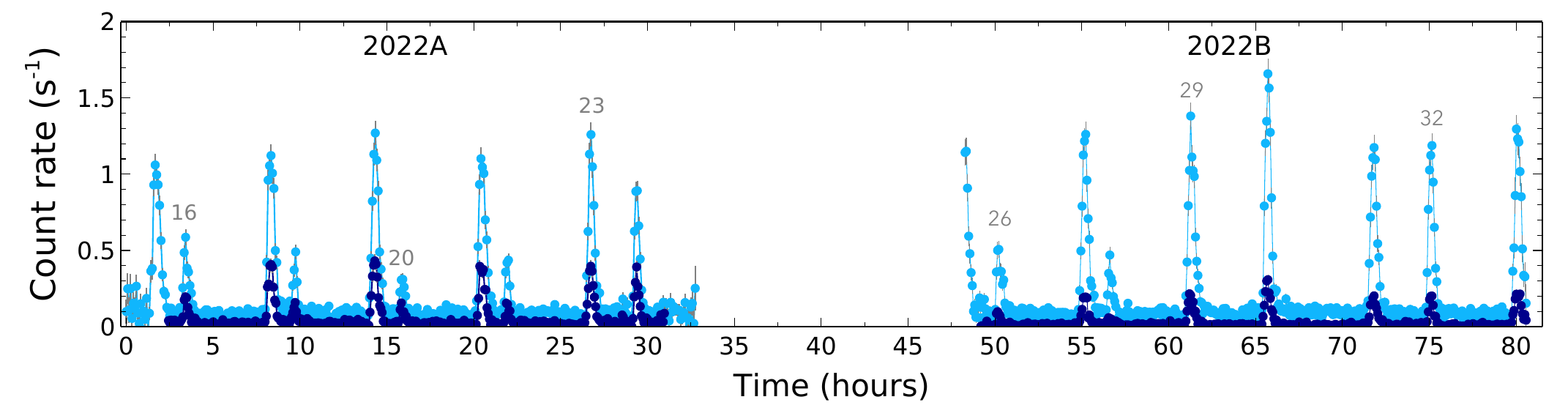}
\caption{Background-corrected \textit{XMM-Newton} light curves of RX~J1301.9+2747, extracted with time bins of 250 s in the $0.2-2$ keV band as observed by the pn in light blue, by the MOS in dark blue. The epoch of observation name used in the text as explained in Table~\ref{table:obslog} is marked in each panel. We number QPEs from number 0 (half-detected by the MOS in 2000) to number 33 (the last detected in 2022B), and mark few of them with grey numbers. 
\label{FIG1}}
\end{figure*}

\section{Data reduction and analysis setup}\label{sec:data}

There are five science observations of RX~J1301.9+2747 in the \textit{XMM-Newton} archive, taken between December 2000 and June 2022, for a total of $\sim 460$ ks of exposure time. Details of the observations are reported in Table~\ref{table:obslog}.
The first two short observations of this dataset have been published by \citet{2013ApJ...768..167S, 2015MNRAS.446.1312M, 2017ApJ...837....3S}, and \citet{2020A&A...636L...2G}, while the latest three observations contain still unpublished data. 
We reduce and analyze all the \textit{XMM-Newton} datasets with calibration files generated in June 2023 and using homogeneous procedures. We also use archival \textit{EXOSAT}, \textit{ROSAT}, and \textit{Chandra} data to investigate the evolution of the properties of the quiescent (non-QPE) X-ray emission of RX~J1301.9+2747.

The \textit{XMM-Newton} data were reduced using the Science Analysis System (\texttt{SAS}) v.18.0.0, following standard threads as recommended by the \textit{XMM-Newton} Science Operation Centre.
Strong background flares were detected at the beginning and at the end of some of the exposures, and were filtered out from the event tables used in the spectral analysis.
The whole event tables were instead used for the analysis of the X-ray light curves, after correcting the observed count rates for the effects of the background flares and other instrumental effects (e.g., vignetting, bad pixels, chip gaps) with the \texttt{SAS} task \texttt{epiclccorr}. The light curves times were also converted from the local satellite frame to barycentric dynamical time, whose spatial origin lies at the solar system barycenter, using the \texttt{SAS} task \texttt{barycen} with the Earth ephemeris table ``DE405''. The effects of the barycentric conversion is a correction of the order of $5-140$~s on photon arrival times, depending on the observation.
The EPIC-MOS1 and MOS2 light curves were merged using the \textsc{Ftool}\footnote{\url{http://heasarc.gsfc.nasa.gov/ftools}} \texttt{lcmath}, while each pair of EPIC-MOS1 and MOS2 spectral products and response files were merged using the \texttt{SAS} task \texttt{epicspeccombine}.
Given the instrumental calibration uncertainty, we retained events with energies greater than 300 eV for the spectral analysis, while for the timing analysis we considered events down to energies of 200 eV. The regions used to extract the source spectra were circles with radii between $20-32''$. These were determined using the \texttt{SAS} task \texttt{eregionanalyse}, maximising the S/N using background regions that were annuli for the EPIC-MOS (hereafter MOS), and circles for the EPIC-pn (hereafter pn) cameras. This choice was based on the strong spatial dependence of the pn instrumental background, for which we choose to extract a background as similar as possible to the source one in detector coordinates. The background regions were always larger than the source ones and all the extraction areas were normalized using the \texttt{SAS} \texttt{backscale} task. 
The optical monitor data were checked for the presence of variability. We found no significant result, with light curves  similar to those of GSN 069 \citep[Extended Data Fig. 2c of][]{2019Natur.573..381M} and of the eRO-QPE sources (Fig. 1 and 2 of \citet{2021Natur.592..704A} and Fig. 4 and 6 of \citet{2024A&A...684A..64A}).

\section{X-ray light curve analysis\label{sec:lcurves}}
 
\begin{figure*}[h!tb]
\centering
\includegraphics[width=17cm]{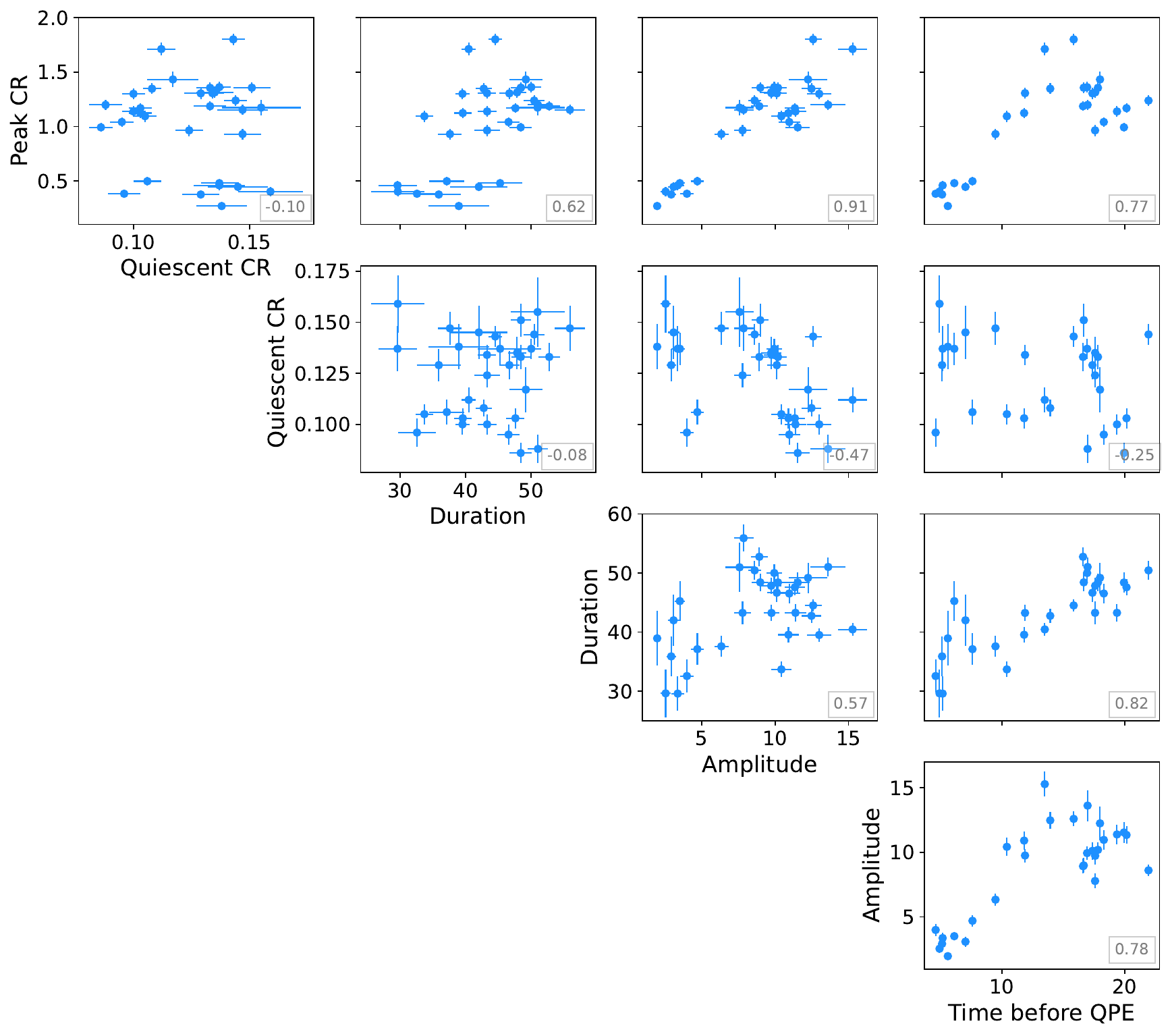}
\caption{Properties of the QPEs of RX~J1301.9+2747, derived from the \textit{XMM-Newton} EPIC light curves: QPE peak count rate (CR), in s$^{-1}$; local quiescence count rate (CR), in s$^{-1}$; QPE duration, in minutes; QPE amplitude; time between QPE number $n$ and QPE number $n-1$, in kiloseconds (ks).
For each quantity we plot the median of the posterior PDF along with its 1 standard deviation error bar. The grey boxes in the lower right corner of each subplot report the correlation coefficients between the parameters. 
\label{FIG2}}
\end{figure*}

Figure~\ref{FIG1} shows the \textit{XMM-Newton} light curves of RX~J1301.9+2747 with a time bin of 250 s, with light blue and dark blue symbols for the pn and MOS data. 
The light curves are extracted in the $0.2-2$ keV band, as the background generally dominates the source signal at higher energies (e.g., see Fig.~\ref{FIGC1}).
A total of 34 QPEs are observed between December 2000 and June 2022. Of these, 3 are only partially detected at the beginning of the exposures\footnote{The pn camera detected 31 and a half QPEs, while the MOS detected 30 QPEs plus two half-QPEs.}. 
There are 25 QPEs with pn count rates of the order of $1-1.5$~s$^{-1}$ (hereafter strong QPEs), and 8 QPEs with about half this count rate (herafter weak QPEs). 
The QPEs are superimposed on a much fainter emission (hereafter quiescence) with a count rate of about $0.04$ s$^{-1}$.

We used a Bayesian statistical framework to compare the pn and MOS light curves to a model comprising a constant, representing the quiescent emission, plus Gaussians representing the QPEs. 
We derive posterior probability distribution functions (hereafter PDFs) and the Bayesian
evidence $\mathcal{Z}$ with the nested sampling Monte Carlo algorithm MLFriends \citep{2016S&C....26..383B,2019PASP..131j8005B} using the \texttt{UltraNest}\footnote{\url{https://johannesbuchner.github.io/UltraNest/}} package \citep{2021JOSS....6.3001B}.
The quiescent level was fitted only around the QPE peak, using both pn and MOS data when available.
From our fit we estimate the most likely constant count rate (quiescent count rate), and the Gaussian centroid (QPE peak arrival time), width $\sigma$, and normalization (QPE peak count rate). Details about our fitting procedure are reported in Appendix~\ref{appendix:lightcurves}, the individual QPE fits are shown in Fig.~\ref{FIGB1}, and the results are reported in Table~\ref{table:QPEprop}.

Figure~\ref{FIG2} shows the median values of the posterior PDFs of the QPE peak count rate, the local quiescence count rate, the QPE duration, the QPE amplitude, and the time preceding each QPE, plotted against each other.
The duration of each QPE was estimated as twice the FWHM of the best-fitting Gaussian, while the amplitude  $A$ as the ratio between the QPE peak count rate and the quiescent one. The small grey boxes in the lower right corner of each subplot report the correlation coefficient between pairs of parameters\footnote{We quote the Pearson product-moment correlation coefficients computed using the \texttt{numpy.corrcoef} routine.}. 
QPEs can be split in two classes depending on their amplitude: weak ($A\sim 5$) and strong ($A\sim 10-15$). They last between $2000-3000$ s, with a tendency of shorter durations for weak QPEs. 
There is a weak positive correlation between the QPE peak count rate (or amplitude) and the QPE duration, with the time preceding the QPE. In particular, whenever the recurrence time exceeds 1h~40m (6~ks), no weak QPEs are observed. 
Remarkably, there is a large scatter in recurrence times between the QPEs of RX~J1301.9+2747.
The most frequent recurrence time is about 5 hours, while the average is 3h 45m (13.5 ks). The shortest recurrence time is 1h~17m (4.6 ks) between QPE7 and QPE8, and the largest is 6h~05m (21.9 ks)  between QPE30 and QPE31.

\begin{figure*}[h!tb]
\centering
\includegraphics[width=17cm]{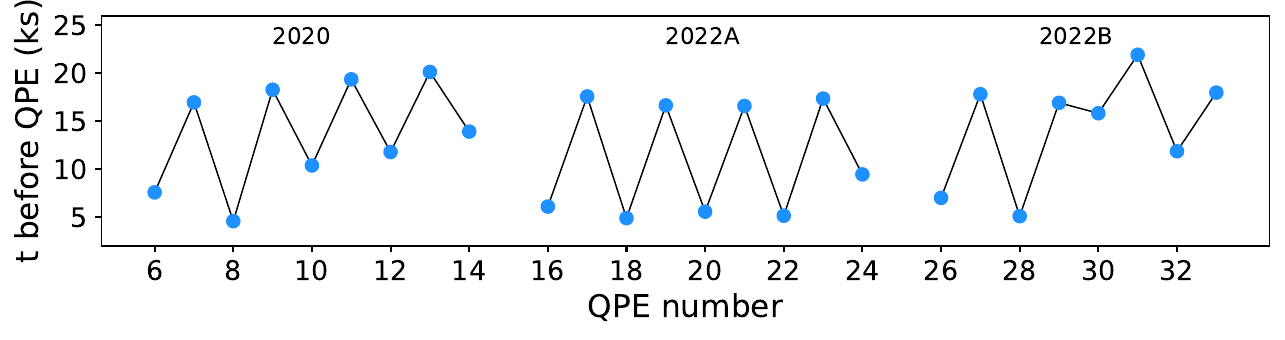}
\caption{Time (in ks) preceding each QPE of the 2020, 2022A, and 2022B \textit{XMM-Newton} observations of RX~J1301.9+2747.
\label{FIG3}}
\end{figure*}

The QPEs of RX~J1301.9+2747 show significantly more irregular timing properties than GSN 069 and eRO-QPE2 \citep{2019Natur.573..381M, 2022A&A...662A..49A}.
However, the QPEs still show an alternate long-short time separation.
This is shown in Fig.~\ref{FIG3}, where we consider
the three long \textit{XMM-Newton} observations taken in 2020 and 2022, and plot for each QPE the time interval preceding it.
The average recurrence time is 14.4 ks in 2020, 11.1 ks in 2022A, and 14.3 ks in 2022B. The light curves of 2020 and 2022B look remarkably similar.
There is no clear repeating pattern, even considering even and odd (or strong and weak) QPEs separately.

Despite the irregular timing properties, the QPEs in RX~J1301.9+2747 are not fully chaotic either: 
at epochs during which weak QPEs are clearly present (roughly the first halves of observations 2020 and 2022B and almost the whole observation 2022A), QPEs clearly come in strong/weak pairs with a short time separation.
On the other hand, when weak QPEs are absent (observation 2019, and roughly the second halves of observations 2020 and 2022B) the pairs are separated by significantly longer intervals. The long/short recurrence time is still present but becomes less dramatic, and the strong/weak amplitude alternation is lost altogether.
We also point out that, if the two different phases are considered separately, the repetition pattern in each phase is reminiscent of more regular QPE activity such as that observed in GSN 069 or eRO-QPE2, with averaged short and long recurrence times of 1h~36m (5.8 ks) and 4h~48m (17.3 ks) in the weak-QPEs phase, and 3h~24m (12.2 ks) and 5h~14m (18.8 ks) during epochs when weak QPEs are absent.

\subsection{Energy-resolved light curves}
We extracted light curves in several observer frame energy bands: $0.2-0.4$, $0.4-0.6$, $0.6-0.8$, $0.8-1.0$, and $1-2$ keV, using time bins of 200 s. The quiescent emission has a very low count rate in energy-resolved bins, hence we compared the QPE light curves to a model comprising a Gaussian line only. We limited the energy-resolved timing analysis to the pn data. 

\begin{figure}[!t]
\centering
\includegraphics[width=7cm]{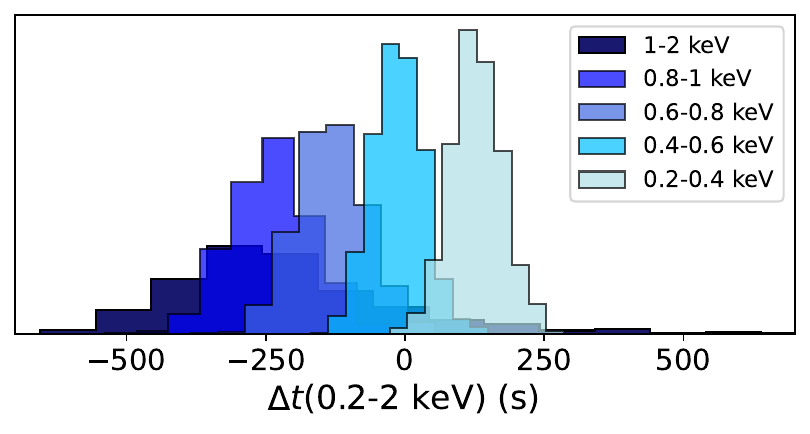}
\includegraphics[width=7cm]{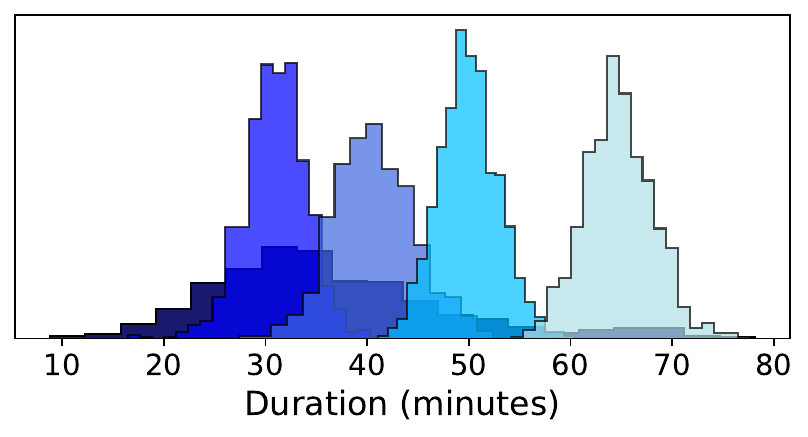}
\includegraphics[width=7cm]{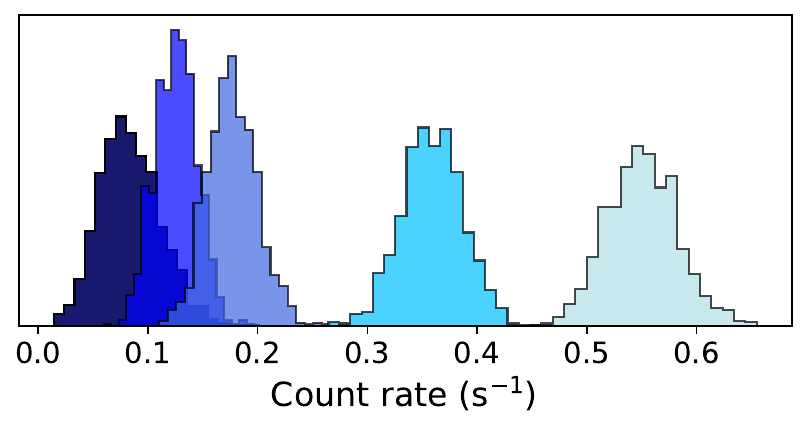}
\caption{Properties of QPEs as a function of energy plotted as histograms of their posterior PDFs. From upper to lower panels: difference between the QPE arrival times measured in different energy bands and the QPE arrival times measured in the $0.2-2$ keV band, QPE duration, and count rate at the QPE peak. The energies considered are $0.2-0.4$, $0.4-0.6$, $0.6-0.8$, $0.8-1$, and $1-2$ keV from lighter to darker histograms.}
\label{FIG4}
\end{figure}

Figure~\ref{FIG4} shows the posterior PDF  of the QPE properties measured in the five energy bands,  with increasing darkness for increasing energy. From upper to lower panels we plot the time delays of energy-resolved QPE peak times with respect to those measured in the full $0.2-2$ keV band, the QPE duration, and the QPE peak count rate. 
As found already by \citet{2020A&A...636L...2G} for the 2000 and 2019 datasets, and as observed in all QPE sources and QPE candidates so far, the QPEs in RX~J1301.9+2747 last less and peak earlier at higher energies. 
The QPE peak times in the $0.4-0.6$ keV band are comparable to those measured in the full band; those measured at lower energies arrive later by about two minutes, while those measured at higher energies arrive earlier, by up to about 10 minutes.
The duration of QPEs measured between $0.2-0.4$ keV is twice the one measured at $E > 0.8$ keV. 
There doesn't appear to be any significant difference between strong and weak QPEs in terms of their properties as a function of energy, although 
the hardest energy band considered ($1-2$ keV) could not always be modeled in weak QPE data because of the very low observed count-rate. 
In general, as the energy increases, the QPE count rate and duration decrease, while the peak time occurs earlier. The decrease in peak count rate with energy is simply a consequence of the spectral shape of QPEs \citep[a blackbody-like spectrum with $kT\sim 100$ eV at peak, see][and Sect.~\ref{Xray}]{2019Natur.573..381M}.
The QPE amplitude increases with energy because the spectral decline of the quiescence is steeper than that of QPEs, as the quiescent spectrum has the signal disappearing alread at $\sim 1 $~keV, while QPEs can be measured at peak up to $\sim 2 $~keV.
This behavior is similar to what has been observed for other QPE sources, thus likely representing a defining property of QPE emission.

\subsection{Hardness Ratio}

\begin{figure}[!t]
\centering
\includegraphics[width=6.5cm]{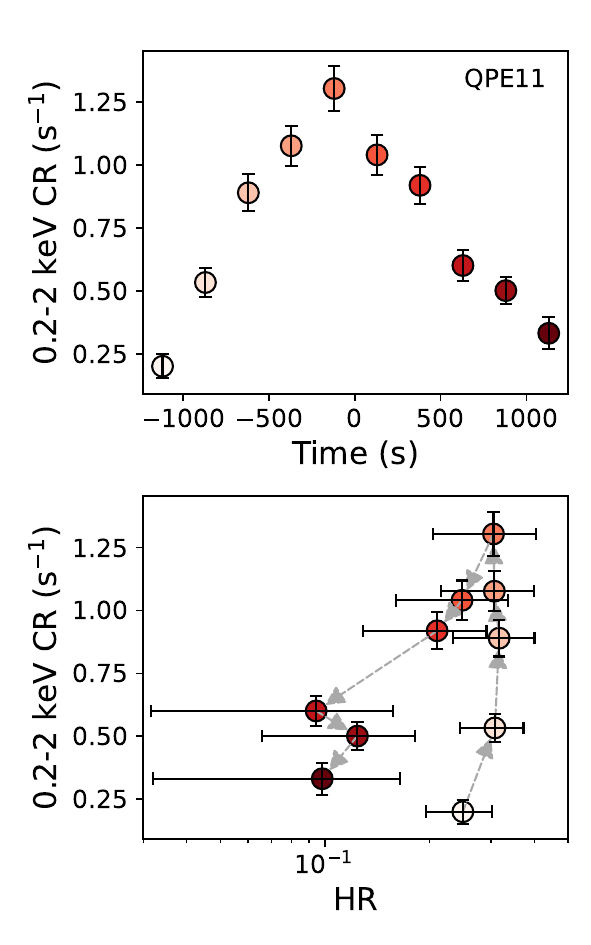}
\caption{Top panel: representative $0.2-2$ keV background-subtracted QPE light curve, centered at the peak and binned to 250 s. Lower panel: hardness ratio (HR) between the $0.6-2$ keV and the $0.2-2$ keV count rates as a function of the total count rate, where each point is color-coded following the time evolution of the QPE in the panel above. The HR shows a counter-clockwise evolution, as tracked by the dashed grey arrows. The complete set of HR plots can be found in Fig.~\ref{FIGB2}.
\label{FIG5}}
\end{figure}

In Figure~\ref{FIGB2} we plot in each panel the individual QPE $0.2-2$ keV light curve and the corresponding hardness ratio (HR) computed between $0.6-2$ keV and $0.2-2$ keV, following \citet{2022A&A...662A..49A}. One example is shown in Fig.~\ref{FIG5} for QPE11.
The QPE evolution in RX~J1301.9+2747 is chromatic, as is in the other QPE sources: at a given count rate, the spectral hardness is greater during the rising phase of QPEs than during their decaying phase.
The QPE peak temperature is not reached at the QPE peak count rate, but during the QPE rise.
While for most of the QPEs the hysteresis cycle is clear, a sub-sample of them show a more complex behaviour, with the HR appearing to further increase during or right after the QPE peak (e.g., QPEs number 3, 13, 15, 19, 20, 21, and 29). The weak QPEs number 8, 18, 26 and 28 also show a rather complex evolution with e.g. a decay as hard as the rise, or a hardening decay.
These complexities might be explained by the presence of sub-structures within the QPE, e.g., overlapping bursts that are not resolved by the current instruments.

\section{X-ray spectral analysis\label{Xray}}

The goals of the X-ray spectral analysis are twofold: (i) understanding the X-ray spectral shape and temporal evolution of the quiescent emission (i.e., QPE excluded) of RX~J1301.9+2747, and compare it with that of known accreting SMBH systems such as AGN and TDEs; and (ii) deriving the X-ray spectral and variability  properties of all the QPEs detected in RX~J1301.9+2747, in order to look for potential differences (e.g. between weak and strong QPEs).
We thus analyse separately the quiescent (Sect.~\ref{sec:quiescence} and Appendix~\ref{appendix:quiescence}) and the QPE spectra (Sect.~\ref{sec:qpes} and Appendix~\ref{appendix:qpes}) of RX~J1301.9+2747 taken by the EPIC cameras onboard \textit{XMM-Newton} between 2000 and 2022. 
We add to our spectral analysis also archival \textit{ROSAT} and \textit{Chandra} data taken respectively in 1991 and 2009 (Sect.~\ref{sec:history}). 

We used the software \texttt{HEASoft} v.6.27.2 \citep{2014ascl.soft08004N} with \texttt{Xspec} v.12.11.01 \citep{1996ASPC..101...17A} and the \texttt{pyXspec} python interface \citep{2021ascl.soft01014G}. 
For the spectral analysis we use the Bayesian X-ray Analysis software \texttt{BXA} v.4.1.1 \citep{2014A&A...564A.125B} that links the Bayesian statistical framework implemented in \texttt{UltraNest} with \texttt{Xspec}.
All the spectral fits assumed the photoionization cross-sections and interstellar medium abundances provided by \citet{2000ApJ...542..914W}, and included a minimum Galactic hydrogen-equivalent column density $N_H=8.8\times 10^{19}$ cm$^{-2}$ along the line of sight in the direction of RX~J1301.9+2747 \citep{2016A&A...594A.116H}, modeled with \texttt{TBabs}. In all the models for the source emission we include the \texttt{zashift} component for the cosmological redshift $z=0.024$.  

\subsection{The quiescent spectra\label{sec:quiescence}} 

\begin{figure}[h!tb]
\centering
\includegraphics[width=8cm]{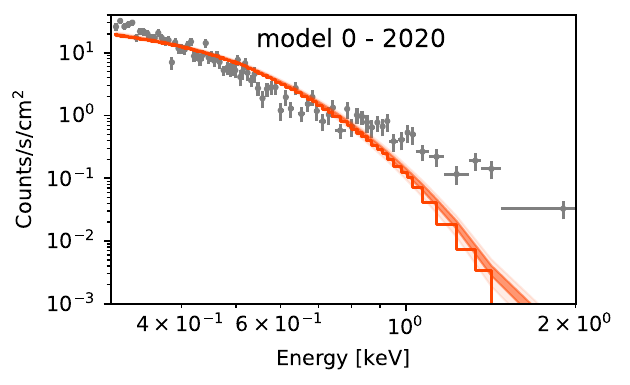 }
\includegraphics[width=8cm]{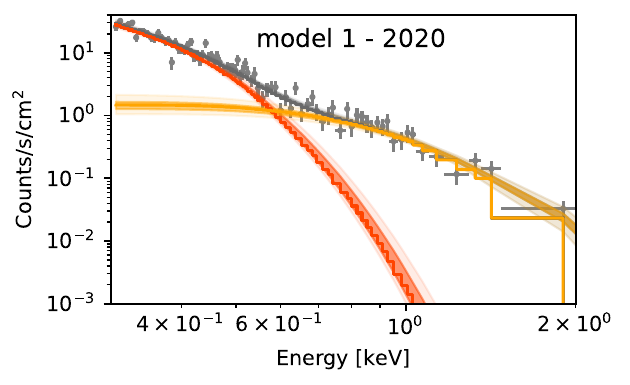 }
\includegraphics[width=8cm]{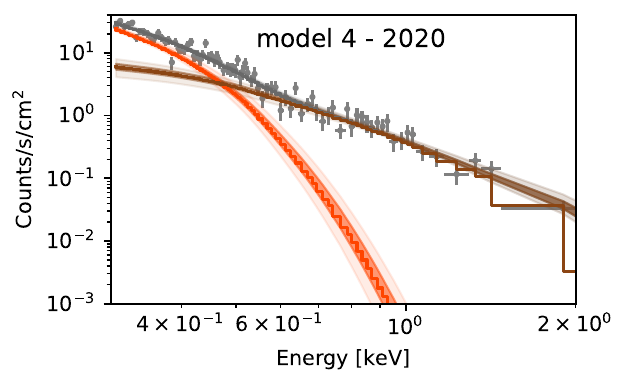 }
\caption{Quiescent pn spectrum of RX~J1301.9+2747 during the 2020 epoch, visually rebinned to 3$\sigma$ significance and overplotted to the posterior PDF of model 0 (\texttt{[diskbb]}, top panel), model 1 (\texttt{[diskbb + bbody]}, middle panel), and model 4 (\texttt{[diskbb + compTT]}, lower panel). The \texttt{[diskbb]} component is plotted in red, the \texttt{[bbody]} in orange, the \texttt{[compTT]} in brown, and the total model in grey. The complete set of spectra for all epochs compared to models 0 and 1 are shown in Fig.~\ref{FIGC2}. Spectra are shown up to $2$~keV, as at higher energies the signal is dominated by the background, however the data were always fitted up to $10$~keV.}
\label{FIG6}
\end{figure}

\begin{figure*}[htb]
\centering
\includegraphics[width=18cm]{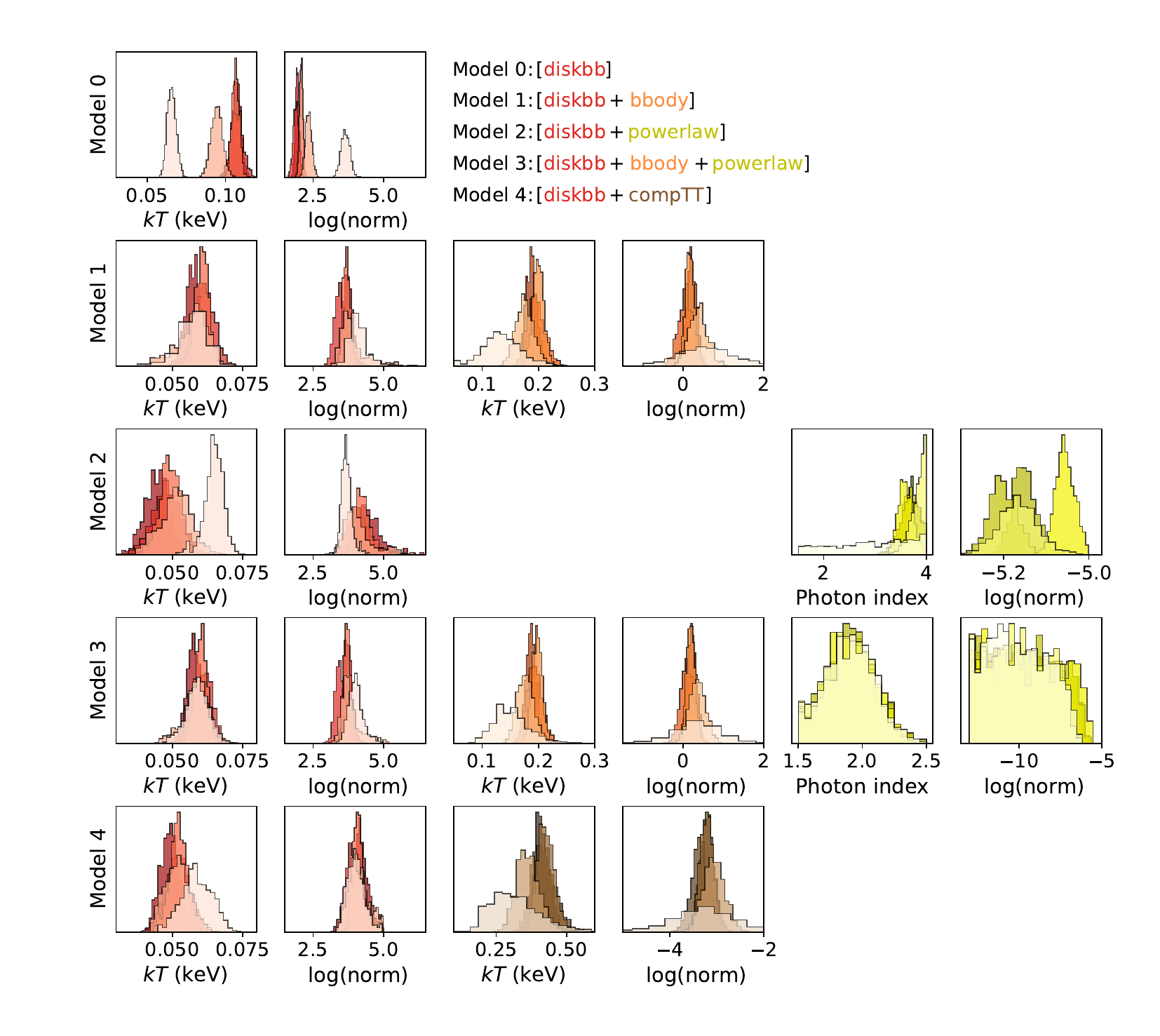}
\caption{Synoptic view of the quiescent spectral properties of RX~J1301.9+2747 in five epochs of \textit{XMM-Newton} observation: 2000, 2019, 2020, 2022A, and 2022B (Table~\ref{table:obslog}), plotted with color tones from lighter to darker. The histograms report the posterior PDF of the parameter of interest considering five different models, 0 to 4 from the first to the fifth row. We plot in red the parameters of the \texttt{diskbb} component, in orange the parameters of the \texttt{bbody} component, in yellow the parameters of the \texttt{powerlaw} component, and in brown the parameters of the \texttt{compTT} component.
\label{FIG8}}
\end{figure*} 

\begin{figure}[ht!]
\centering
\includegraphics[width=9cm]{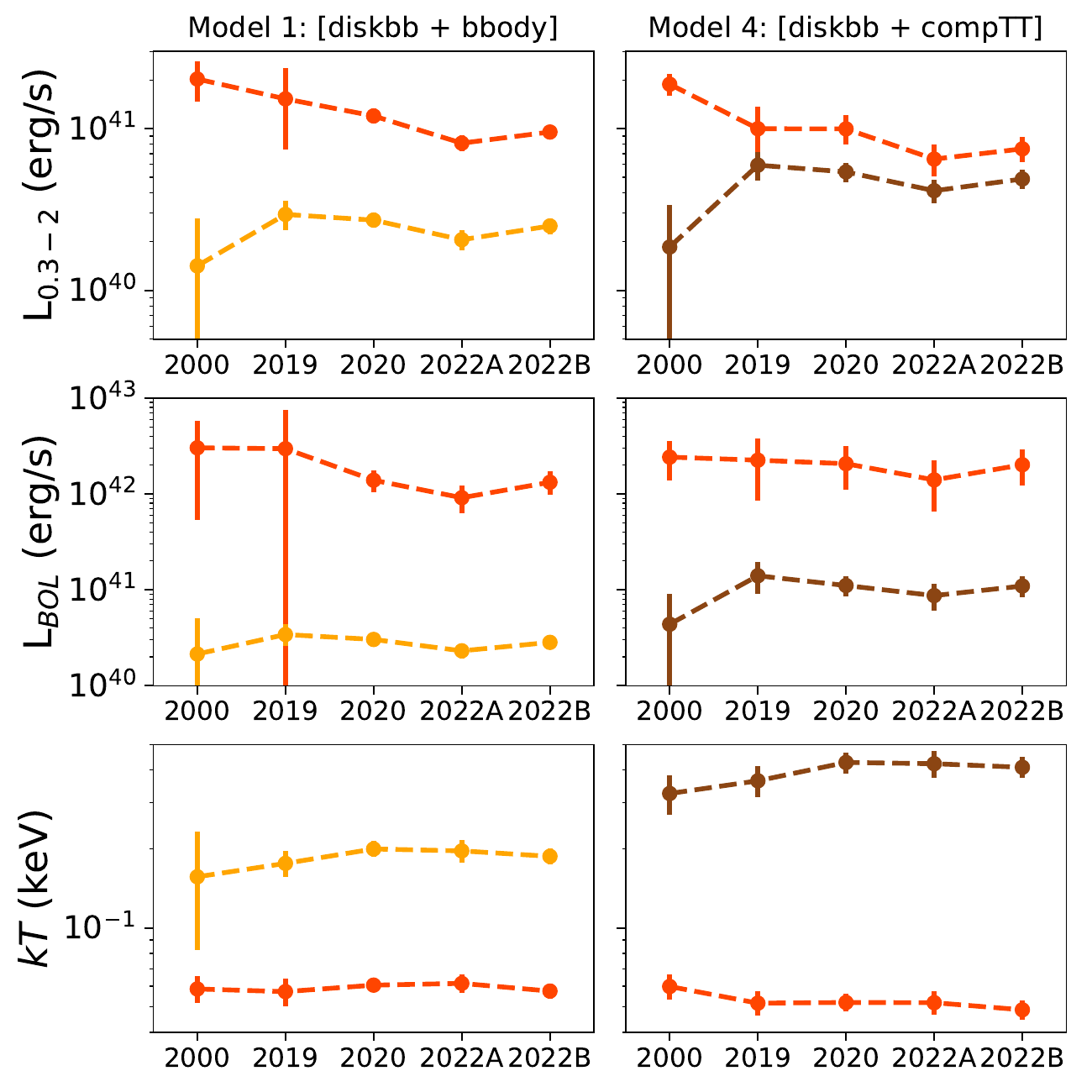}
\caption{Evolution of the $0.3-2$ keV luminosity  (first row), bolometric luminosity (second row), and temperature (third row) of the quiescent emission of RX~J1301.9+2747 in each epoch of \textit{XMM-Newton} observation obtained using model 1 \texttt{[diskbb + bbody]} (left column) and model 4 \texttt{[diskbb + compTT]} (right column). Red symbols represent \texttt{diskbb}, orange ones \texttt{bbody}, and the brown ones \texttt{compTT}.
Points represent the medians of the posterior PDFs, error bars one standard deviation. 
\label{FIG7}}
\end{figure}

The quiescent (source + background)  and background spectra of RX~J1301.9+2747 observed by \textit{XMM-Newton} between 2000 and 2022 are shown in  Fig.~\ref{FIGC1}, rebinned for visual purposes. 
The source and background spectra were obtained by excising time intervals containing QPEs, and, for each observation, we consider the average quiescent spectrum as the intra-observation variability of the quiescent count rate is minimal.
The overall spectral variability between epochs is small.
As in other QPE sources, the X-ray spectra of RX~J1301.9+2747 are super-soft, with most of the source emission at $E < 1$ keV. 

First, we compared the quiescent spectra to the two thermal models \texttt{[bbodyrad]} and \texttt{[diskbb]}. 
We found that at all epochs the \texttt{[diskbb]} model is preferred over the \texttt{[bbodyrad]} one, given the difference in logarithm of the Bayesian evidence $\mathcal{Z}$ always greater than 2 (the most likely model has the highest Bayesian evidence; details about our fitting procedure are reported in Appendix~\ref{appendix:quiescence}).
The posterior PDFs of the model \texttt{[diskbb]} (hereafter model 0) folded with the instrumental response are plotted in the left column of Fig.~\ref{FIGC2} compared to the observed data for the five epochs of  \textit{XMM-Newton} observation, and in the top panel of Fig. \ref{FIG6} for the 2020 epoch.
It is evident that \texttt{[diskbb]} alone is unable to account for the data above $\sim 700$~eV as excess emission is present, especially 2019 onwards.

We then fit the five epochs with a series of multi-component models. These are model 1: \texttt{[diskbb + bbody]}; model 2: \texttt{[diskbb + powerlaw]}; model 3: \texttt{[diskbb + bbody + powerlaw]}; and model 4: \texttt{[diskbb + compTT]}, where we use the Comptonization model by \citet{1994ApJ...434..570T}. 
Results of the fit are reported in Table~\ref{table:quiescence} and Fig.~\ref{FIG8}, where we plot the histograms of the posterior PDF of various parameters for each epoch of observation, for each model, using darker tones for later epochs.
Models 1, 2, 3, and 4 are always strongly preferred to model 0 except for epoch 2000, when the Bayesian evidences $\mathcal{Z}$ are comparable.
Given the likelihoods and the number of degrees of freedom \citep[AIC,][]{1974ITAC...19..716A}, model 3 is never the most probable data representation.
When using model 2, the photon index $\Gamma>3$ is very steep. Thus the power law component of model 2 mostly accounts for the photons detected in the soft X-ray band. 
Indeed, for epoch 2000 when the excess of photons at $E > 700$~eV is the weakest, the posteriors of the \texttt{bbody} component of model 1 and those of the \texttt{powerlaw} component of model 2 are only barely constrained (Fig.~\ref{FIG8}).
Model 1 is the most probable representation of the data at early epochs, while model 4 is the most probable one at late epochs; the two models are equiprobable during 2020.
A visual example of the comparison between models 0, 1, and 4 is shown in Fig.~\ref{FIG6} for the 2020 epoch. We also plot a comparison of models 0 and 1 in Fig.~\ref{FIGC2} for all the epochs, where it is evident that the addition of the \texttt{bbody} component to \texttt{diskbb} (i.e., going from model 0 to model 1)  reproduces the data very well, even if the \texttt{bbody} spectral parameters are largely unconstrained during 2000 (as expected, given the low $\Delta\log\mathcal{Z}$ found between  model 0 and model 1 for this epoch).

On average, the quiescent spectra of RX~J1301.9+2747 are best reproduced by the thermal emission of an accretion disk with inner temperature $kT\sim 50-60$ eV plus a harder spectral component, that can be modeled with \texttt{bbody} (as in model 1) or \texttt{compTT} (as in model 4). 
The median values of the posterior PDFs for the parameters of models 1 and  4 are plotted  in Fig.~\ref{FIG7}, with the \texttt{diskbb} component in orange, the \texttt{bbody} component in yellow, and the  \texttt{compTT} component in brown. From upper to lower panels, we plot the $0.3-2$ keV luminosity $L_{0.3-2\, \rm{keV}}$, the bolometric luminosity\footnote{The bolometric luminosity was estimated using a dummy spectral response 
matrix in \texttt{Xspec} extended between $10^{-5}-10$ keV and computing the total luminosity of the model with the absorption along the line of sight set to negligible values.} $L_{BOL}$, and the temperature $kT$. 
The main difference between model 1 and model 4 is the spectral energy distribution shape of the harder spectral component, that is broader for \texttt{compTT} than for \texttt{bbody}. As a consequence, the inferred inner temperature of the \texttt{diskbb} component is lower, and the estimated total bolometric luminosities are slightly higher, when using model 4 than when using model 1.

The disk inner temperature is consistent with being constant,  $\sim 60$ eV,  between epochs when using model 1. When using model 4 it is slightly decreasing, from $58\pm{5}$ eV of 2000 to $49\pm{3}$ eV of 2022B. 
The hard spectral component is barely constrained during 2000, and appears to slightly increase its temperature from 2019 to later epochs. This is independent of the underlying model adopted.
The luminosity of the harder spectral component estimated using \texttt{compTT} is a factor of $(2-2.4)\times$ higher in the $0.3-2$ keV band compared to the one estimated using \texttt{bbody}, while its bolometric luminosity is a factor $(4-6)\times$ higher, depending on the assumed plasma optical depth (the larger $\tau$, the smaller the difference compared to \texttt{bbody}).
This is however always much lower than the luminosity carried by the disk.
The observed disk luminosity shows an overall decay of about 50\% between 2000 and 2022, from $L_{0.3-2\, \rm{keV}}^{disk}\sim 2 \times 10^{41}$ erg s$^{-1}$ to $L_{0.3-2\, \rm{keV}}^{disk}\lesssim 10^{41}$ erg s$^{-1}$.
The $0.3-2$ keV disk luminosity  is about $2-4$ times higher than the $0.3-2$ keV luminosity of the harder component in all epochs except for 2000, when the luminosity of the former is $> 10\times$ the one of the latter (this component not being constrained by the data). The bolometric luminosity of the disk is instead a factor of $\sim 50\times$ higher than the bolometric luminosity of the harder component when using \texttt{bbody}, and a factor of $\sim 15\times$ higher when using \texttt{compTT}. 
When modeled with \texttt{bbody}, the hard spectral component has an average temperature of $kT\sim 180$ eV. 
This temperature is typical of the soft X-ray excess observed in the spectra of most radiatively efficient AGN \citep{2004MNRAS.349L...7G}.
When modeled with \texttt{compTT}, the Comptonizing plasma temperature $kT$ is degenerate with its optical depth $\tau$. By assuming $\tau=10$, the hard spectral component has an average temperature of $kT\sim 400$ eV.  These values are comparable to those found for the warm corona used to model the soft X-ray excess in local AGN \citep[e.g.,][]{2011A&A...534A..39M,2018A&A...611A..59P}.
We will therefore refer to this hard spectral component as the ``soft X-ray excess-like'' component.

The total $0.3-2$ keV quiescent luminosity of RX~J1301.9+2747 is maximum during 2000 with $L_{0.3-2\, \rm{keV}}=2.0\pm{0.5} \times 10^{41}$ erg s$^{-1}$ and is minimum during 2022A with  $L_{0.3-2\, \rm{keV}}=1.01\pm{0.07} \times 10^{41}$ erg s$^{-1}$. There are variations of quiescent luminosity between 2022A and 2022B, when $L_{0.3-2\, \rm{keV}}=1.20\pm{0.07}\times 10^{41}$ erg s$^{-1}$.
The estimated $L_{BOL}$ of each component is about one order of magnitude larger than the $0.3-2$ keV luminosity at all epochs. The total quiescent luminosity due to the disk plus the soft X-ray excess-like component
is $L_{BOL}\sim 2-3\times 10^{42}$ erg s$^{-1}$, depending on the epoch of observation.

We estimated the black hole mass and mass accretion rate by fitting the quiescent emission to the \texttt{optxagnf} and \texttt{tdediscspec} models, that assume respectively a geometrically thin accretion disk plus a warm and hot corona, and a non-stationary TDE accretion flow 
\citep[see respectively][]{2012MNRAS.420.1848D,2023MNRAS.519.5828M}. By taking into account the uncertainties in the black hole spin, we found a large range of admitted values for the black hole mass $M_{BH}=[2\times 10^5-3\times 10^6]\,M_{\odot}$ and for the Eddington ratio $\dot{m}=[0.03-0.15]$. We note how the black hole mass estimate from galactic bulge stars velocity dispersion is at the higher end of the estimated range from X-ray spectral fitting \citep{2022A&A...659L...2W,2024ApJ...970L..23W}. On the other hand, by taking the $M_{BH}$ optical estimate and the estimated $L_{bol}$ from X-ray spectral fitting and assuming an accretion efficiency of 10\%, one would obtain an $\dot{m}<0.02$, in contrast with the thermal X-ray spectrum observed that is typical of highly-accreting BHs. We also point out that the X-ray luminosity of RX J1301.9+2747 is more than one order of magnitude lower than in GSN 069 despite very similar disk temperatures \citep{2019Natur.573..381M}. While the velocity dispersion is very similar in the two galaxies \citep{2024ApJ...970L..23W}, assuming a similar mass accretion rate would lead to a significantly lighter BH in RX J1301.9+2747 than in GSN 069.

\subsubsection{On the hard X-ray power law\label{sec:powerlaw}}

The hard X-ray power law emission is a defining characteristic of accreting BHs. 
In their X-ray spectra, actively accreting SMBHs (AGN) usually show the non-thermal hard X-ray power law, a soft X-ray excess of emission compared to the power law, and, depending on the central BH mass, the accretion disk thermal emission \citep[e.g.,][]{2012MNRAS.420.1848D}. 
The origin of the hard X-ray power law emission is commonly assumed to be due to thermal Comptonization of the accretion disk seed photons in a hot ($kT\sim 100-200$ keV) optically thin plasma \citep[the hot corona,][]{1991ApJ...380L..51H}. The origin of the soft X-ray excess \citep{1985MNRAS.217..105A} is more debated and thought to be caused by either ionized reflection off the inner accretion disk \citep{2006MNRAS.365.1067C} or by Comptonization in a warm ($kT\sim 100-500$ eV), optically thick plasma \citep[the warm corona,][]{1998MNRAS.301..179M}.

The hard X-ray power law appears to be very weak or absent in QPE sources, that show instead a quiescent emission dominated by soft X-ray thermal-like spectra, most likely associated to the high-energy tail of the disk emission \citep{2019Natur.573..381M,2020A&A...636L...2G,2021Natur.592..704A,2024A&A...684A..64A, 2024ApJ...965...12C}.
It is possible that a standard AGN corona is still present in the quiescent spectra of RX~J1301.9+2747, but very weak. To test this possibility and place limits on the presence of a hard X-ray power law emission, we compared the quiescent spectra to model 3: \texttt{[diskbb + bbody + powerlaw]}.
We assumed informed Gaussian priors for the \texttt{diskbb} and \texttt{bbody} temperatures typical of low-mass AGN, and as found for the parameters of model 1 (respectively $\sim 60$ eV and $\sim 180$ eV), and for the \texttt{powerlaw} photon index with a value typical of unabsorbed AGN ($\Gamma\sim 1.9$). 
The priors for the other parameters were uninformed (see details in Appendix~\ref{appendix:quiescence}). 

Results for the fit to model 3 are reported in Table~\ref{table:quiescence}, and are visually represented in the fourth row of Fig.~\ref{FIG8}.
The posterior PDF for $\Gamma$ is equal to the prior adopted, meaning that the fit is not sensitive to this parameter. The normalization is also unconstrained and consistent with the lowest values probed in our parameter investigation at all epochs.
Thus the power law component of model 3 is unconstrained by the data. 
We conclude that a typical hard X-ray power law is not present or is very weak in the quiescent spectra of RX~J1301.9+2747. An upper limit on its $2-10$ keV luminosity can be placed by considering the highest upper limit in the power law normalization of epoch 2020,
$L_{2-10\, \rm{keV}} < 4\,\times\, 10^{38}$ erg s$^{-1}$.
This luminosity value is extremely small and corresponds to a bolometric correction of $\kappa_{2-10\, \rm{keV}}> 2700$. Similar conclusions were reported for the case of GSN 069, that has $\kappa_{2-10\, \rm{keV}}> 4000$ during the highest X-ray luminosity \textit{XMM-Newton} observation \citep{2019Natur.573..381M}. 

\subsubsection{On the soft X-ray excess}

The quiescent spectra of RX J1301.9+2747 present a soft X-ray excess-like component emerging between 2000 and 2019. 
Since no hard power law emission is ever detected, here the excess is defined with respect to the softer X-ray disk component, rather than to the $2-10$~keV emission as in AGN.
Its characteristics are nonetheless typical of the soft X-ray excess observed in AGN: when modeled with a phenomenological blackbody, its temperature is typical of the soft X-ray excess observed in AGN \citep[e.g.,][]{2004MNRAS.349L...7G}; when modeled with a Comptonization model, it has properties commonly inferred for the warm coronae used to reproduce the AGN soft X-ray excess \citep[e.g.,][]{2020A&A...634A..85P}.
Its temperature appears to slightly increase from 2019 to 2022, from $\sim 175$ eV to $\sim 190$ eV when modeled with \texttt{bbody}, and from $\sim 360$ eV to $\sim 415$ eV when modeled with \texttt{compTT}.
The soft X-ray excess-like component emits about 25\% of the $0.3-2$ keV disk luminosity in epochs from 2019 to 2022.
It was much dimmer in 2000, having 10\% of the disk X-ray luminosity at most.
In terms of bolometric luminosity, it ranges from $1.6-3.3\times 10^{40}$ erg s$^{-1}$ when modeled with \texttt{bbody}, to $4.5-13\times 10^{40}$ erg s$^{-1}$ when modeled with \texttt{compTT}. 

The super-soft X-ray spectrum and the lack of a standard AGN-like hard X-ray continuum of RX J1301.9+2747 and all other QPE sources are highly reminiscent of the properties of most thermal X-ray TDEs.
In both cases, the most likely interpretation for the super-soft component is that it represents the high-energy tail of the emission from  a radiatively efficiently accretion flow around a relatively low mass SMBH. As mentioned in Sect.~\ref{sec:intro}, a connection between QPE sources and TDEs is emerging, based not only on the X-ray properties, but also on those of their host galaxies and black hole masses \citep{2024ApJ...970L..23W}.
A fraction of thermal X-ray TDEs develop harder X-ray emission components over time. For example, AT2019azh and AT2019ehz exhibit a soft excess component with modest luminosity (compared to disk emission) as well as peculiar X-ray variability properties in analogy with RX J1301.9+2747 \citep{2021MNRAS.500.1673H,2024ApJ...966..160G}, and AT2020ocn developed, on timescales of hundreds of days, a high luminosity warm corona component  that later appeared to make a transition to a hot corona one \citep{2024ApJ...970...89C}.
Whether RX J1301.9+2747 is experiencing a similar evolution and will form a hot corona component in the future remains to be seen.
Fast evolution of soft excess and, subsequently,  power law continuum components have also been observed in the changing-look AGN/TDE 1ES 1927+654/ASASSN-18el when the Comptonized components reappeared after destruction \citep{2022ApJ...934...35M}. Some of the soft excess spectral properties, in addition to the exceptional X-ray variability of 1ES 1927+654, are indeed reminiscent of RX J1301.9+2747 as well.

The physical connection between the soft X-ray excess (and, to some extent, the harder power law continuum) observed in AGN, TDEs, and QPE sources is yet to be understood. Future dense monitoring observations of fastly evolving accreting SMBH such as TDEs and QPE sources are likely key to make significant progress in the field.

\subsubsection{Historical evolution of the RX~J1301.9+2747 X-ray quiescent emission\label{sec:history}}

\begin{figure*}[!htb]
\centering
\includegraphics[width=18cm]{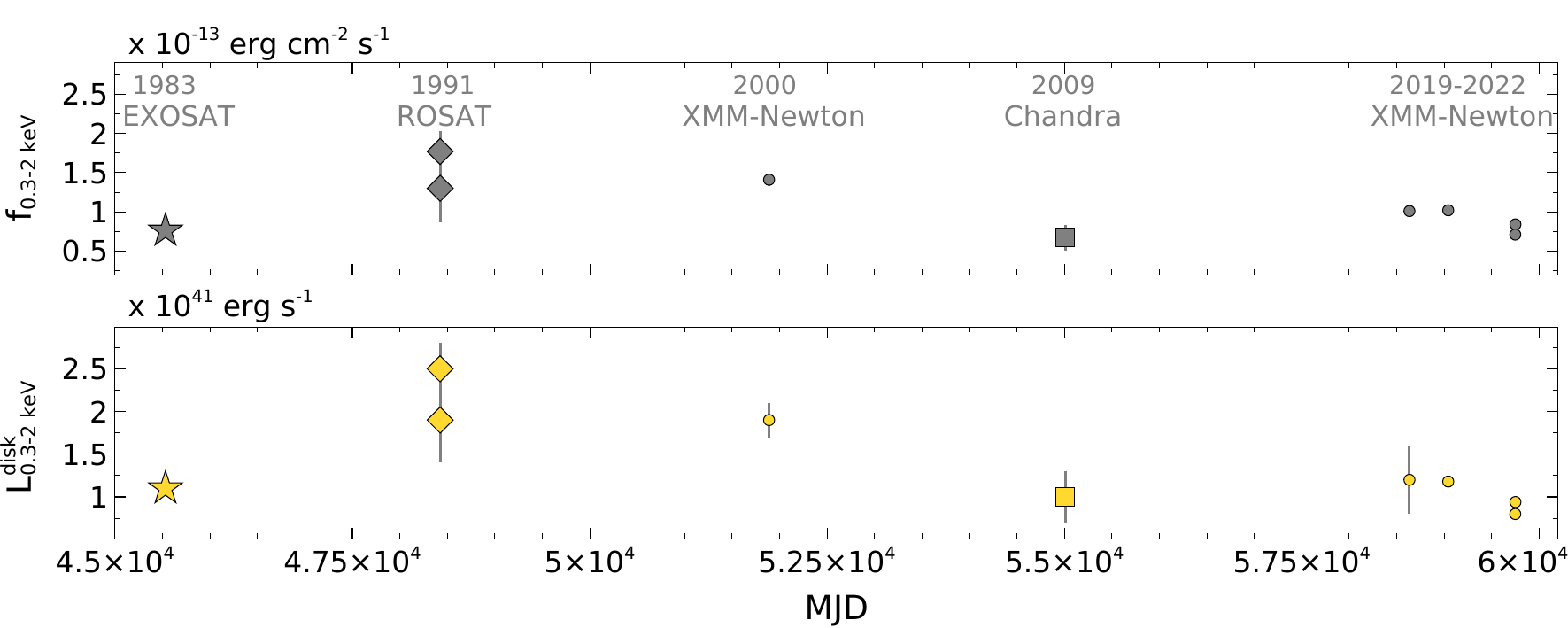}
\caption{Top panel: $0.3-2$ keV quiescent flux observed from RX~J1301.9+2747 by various telescopes in different epochs, marked by their Modified Julian Day. For all the observations where this was possible (all except the EXOSAT one) the QPE emission was removed, leaving the quiescence only. Lower panel: $0.3-2$ keV disk luminosity, corrected for absorption.
For most of the \textit{XMM-Newton} measurements the error bars are smaller than the symbol size.
 \label{FIG9}}
\end{figure*}

RX~J1301.9+2747 has been observed by the \textit{EXOSAT} Low-Energy Imaging Telescopes (LE) in July 1983, by the \textit{ROSAT}  Position Sensitive Proportional Counters (PSPC) in June 1991, and by the Advanced CCD Imaging Spectrometer (ACIS) onboard \textit{Chandra} in June 2009 (Table~\ref{tab:otherobs}).

The \textit{EXOSAT} observation allowed \citet{1985MNRAS.216.1043B} to estimate a $0.02-2.5$ keV flux of $\sim 1.7\times 10^{-13}$ erg cm$^{-2}$ s$^{-1}$ assuming a powerlaw emission with $\Gamma=3.5$ affected by Galactic absorption. We reproduced these values with the \texttt{Xspec} model $\texttt{[tbabs*powerlaw]}$ and a dummy response file extended to low energies, and estimate a $0.3-2$ keV flux of $\sim 7.6\times 10^{-14}$ erg cm$^{-2}$ s$^{-1}$. The luminosity corrected for Galactic absorption is $L_{0.3-2\, \rm{keV}}\sim 1.1\times 10^{41}$ erg s$^{-1}$. 

The \textit{ROSAT} PSPC spectra were retrieved from the White-Giommi-Angelini (WGA) Catalog \citep{1994IAUC.6100....1W}, which contains processed high-level products such as calibrated source and background spectra and light curves. Among the three \textit{ROSAT} observations of RX~J1301.9+2747 performed between 16-19 June 1991, the WGA Catalog contains the spectra extracted during the first two observations. These do not show significant variability, thus we consider them representative of the quiescence. We fit separately the $0.1-1$ keV spectra of the two epochs of observation using \texttt{BXA}. We found that model 0  \texttt{[zashift*diskbb]} gives a reasonable representation of the data, that do not require further complexities. We computed the observed flux in the $0.3-2$ keV band and found $f_{0.3-2\, \rm{keV}}=1.7\pm{0.3}\times 10^{-13}$ erg cm$^{-2}$ s$^{-1}$ for the observation of June 16, and  $f_{0.3-2\, \rm{keV}}=1.3\pm{0.4}\times 10^{-13}$ erg cm$^{-2}$ s$^{-1}$ for the observation of June 17. These correspond to unabsorbed luminosities $L_{0.3-2\, \rm{keV}}= 2.5\pm{0.3}\times 10^{41}$ erg s$^{-1}$ and $1.9\pm{0.5}\times 10^{41}$ erg s$^{-1}$. 

The \textit{Chandra} ACIS spectrum of RX~J1301.9+2747 taken in June 2009 has an exposure of $\sim 5$ ks, divided into a quiescent state lasting $\sim 4$ ks and a flaring state lasting $\sim 0.5$ ks \citep{2013ApJ...768..167S}. Due to the very low count statistics we only considered the $0.4-1$ keV quiescent spectrum. We fitted it to model 0, obtaining a $0.3-2$ keV flux estimate $6.7\pm{1.6}\times 10^{-14}$ erg cm$^{-2}$ s$^{-1}$. The corresponding luminosity corrected for Galactic absorption is $L_{0.3-2\, \rm{keV}}= 1.0\pm{0.3}\times 10^{41}$ erg s$^{-1}$. 

These historical X-ray flux values for the quiescence of RX~J1301.9+2747 are plotted  in the top panel of Fig.~\ref{FIG9} together with the measurements obtained from the \textit{XMM-Newton} observations. The lower panel of the same figure reports the $0.3-2$ keV disk luminosity.
The long-term evolution of the quiescent emission of RX~J1301.9+2747 is complex, with variations that can happen on timescales as short as two days, and that can even be non-monotonic. In particular, the observations 2022A and 2022B are separated by only about 15 hours and show a significant difference in disk luminosity, from $L_{0.3-2\, \rm{keV}}^{\rm{disk}}= 8.0\pm{0.5}\times 10^{40}$ erg s$^{-1}$ to $L_{0.3-2\, \rm{keV}}^{\rm{disk}}= 9.4\pm{0.4}\times 10^{40}$ erg s$^{-1}$.

By assuming a constant $0.3-2$ keV luminosity of $10^{41}$ erg s$^{-1}$, a total of  $10^{50}$ erg would have been emitted by  RX~J1301.9+2747 in quiescence during the 40 years elapsed between the \textit{EXOSAT} observation of the Coma Cluster and the latest \textit{XMM-Newton} pointing at the source in 2022.

\subsection{The QPE spectra\label{sec:qpes}}
\begin{figure}[h!tb]
\centering
\includegraphics[width=7cm]{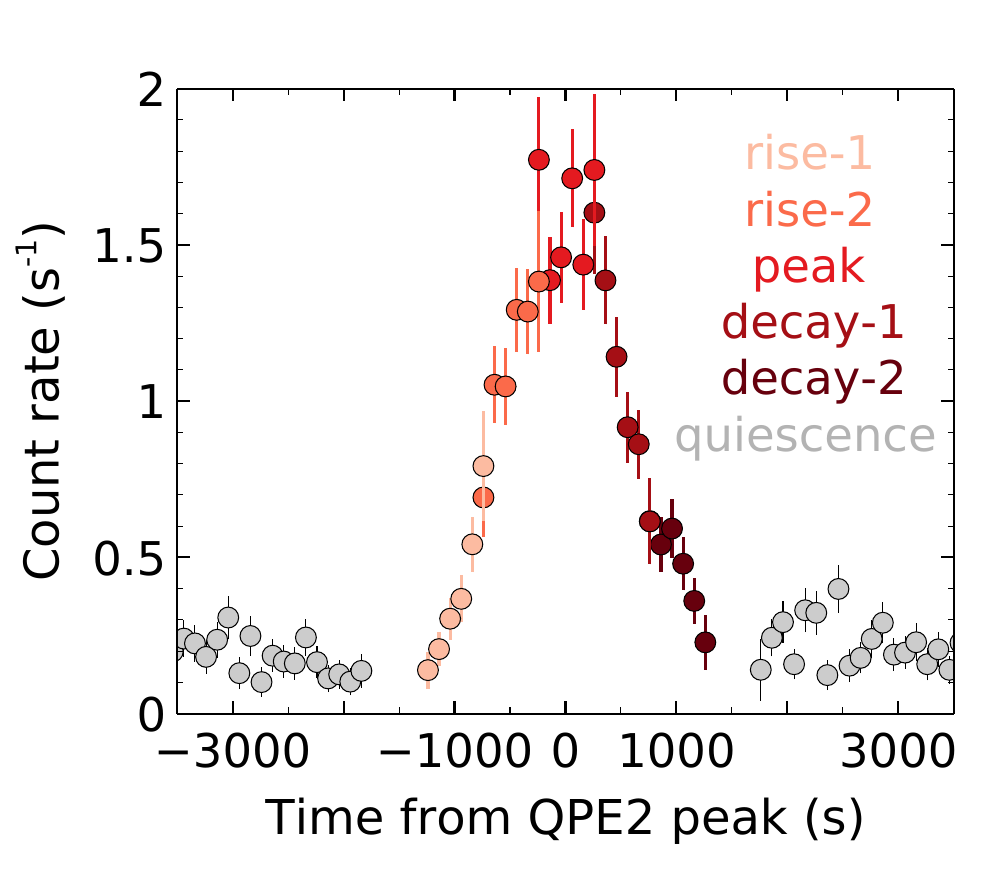}
\caption{Example of the QPE spectral decomposition adopted: the pn data of QPE2 are plotted binned to 100 s. Rise-1 in pink, rise-2 in orange, peak in red, decay-1 in dark red, decay-2 in brown, and quiescence in grey.
 \label{FIG10}}
\end{figure}

We analysed individual QPEs one by one, dividing them in five spectral slices as done by \citet{2019Natur.573..381M} and \citet{2023A&A...670A..93M} for GSN 069 and by \citet{2024A&A...684A..64A} for eRO-QPE3 and eRO-QPE4: two rises (rise-1 and rise-2), one peak, and two decays (decay-1 and decay-2). An example of the adopted spectral decomposition is illustrated in Fig.~\ref{FIG10} for QPE2. Details about the QPE spectral fitting procedure are reported in Appendix~\ref{appendix:qpes}.
In brief, after comparing the \texttt{compTT}, \texttt{bbodyrad} and \texttt{bremss} components, the model \texttt{[TBabs*(zashift*bbodyrad)]} was compared to each QPE spectral slice, using both pn and MOS data when available. 
The QPEs were assumed to be an additive component with respect to the underlying quiescent emission: the quiescent spectrum of the appropriate epoch was used as background spectrum for the QPE one \footnote{We also performed a fit where no such assumption has been made, applying the quiescent spectrum best-fitting model 4 \texttt{[diskbb + compTT]} to each QPE spectral slice with a standard background subtraction. We found that the QPE spectra are dominated by the \texttt{diskbb} component at any stage of their evolution, with a disk inner temperature evolution consistent with the one found when using the \texttt{bbodyrad} model and subtracting the quiescent spectrum. It is thus not possible to disentangle the two scenarios with the available data.}. 
As the neutral absorption modeled with \texttt{TBabs} was found to be consistent with the Galactic value for all spectral slices, we fixed its value to the $8.8\times 10^{19}$ cm$^{-2}$ estimated by the \citet{2016A&A...594A.116H}. The \texttt{zashift} redshift component was fixed to $0.024$. 

\begin{figure}[h!bt]
\centering
\includegraphics[width=7cm]{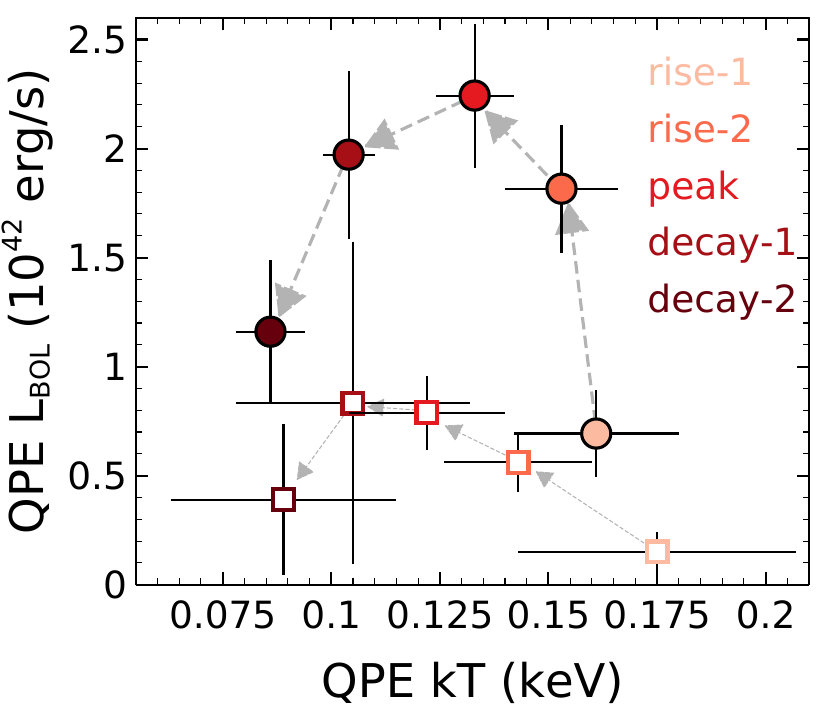}
\includegraphics[width=7cm]{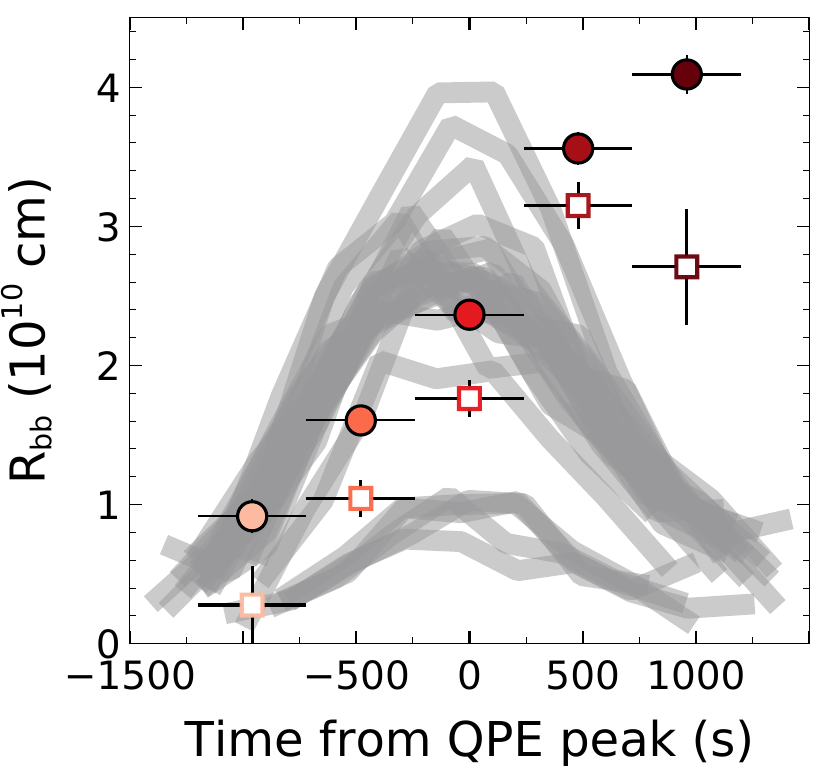}
\caption{Top panel: average of the posterior PDFs of the temperature $kT$ and bolometric luminosity $L_{\rm{BOL}}$ of a redshifted blackbody model (\texttt{[zashift*bbodyrad]} in Xspec) compared to the five spectral slices of each QPE of RX~J1301.9+2747. Weak QPEs are plotted with open squares, strong QPEs with filled circles. The color-code during the QPE evolution is the same of Fig.~\ref{FIG10}.
Lower panel: blackbody radius $R_{bb}$ for each spectral slice, superimposed to the profiles of the 27 QPEs (grey thick lines) with spectral analysis available.
 \label{FIG11}}
\end{figure}

The \texttt{bbodyrad} component allows to recover a temperature $kT$ and a physical size for the blackbody-emitting region, if the distance to the source is known. We adopted a luminosity distance of 100 Mpc for RX~J1301.9+2747.  
The bolometric luminosity $L_{BOL}$ was estimated using a dummy spectral response 
matrix extended between $0.01-10$ keV applied to the \texttt{[TBabs*bbodyrad]} model, setting the absorption to negligible values. The bolometric luminosity values are conservative, as the  \texttt{bremss} and \texttt{compTT} models would provide a QPE bolometric luminosity estimate compared to the \texttt{bbodyrad} model larger by a factor of about 4 and 2.
The median values of the posterior PDFs of $kT$ and $L_{BOL}$ estimated at the peak of each QPE are reported in Table~\ref{table:QPEprop}.
The medians of the QPE temperature and bolometric luminosity posterior PDFs during each spectral slice are plotted in the top panel of Fig.~\ref{FIG11} for the weak (open squares) and strong (filled circles) QPEs, color-coded as in Fig.~\ref{FIG10}. The lower panel of Fig.~\ref{FIG11} reports instead the corresponding blackbody radius as a function of the spectral slice, with the profiles of the QPEs centered around their peak plotted in the background as a reference.
Both the $kT-L_{BOL}$ and $R_{bb}$-time relations are very similar to those observed in GSN 069 and eRO-QPE1 \citep{2023A&A...670A..93M,2024ApJ...965...12C}. However,  the decay in temperature during the QPE evolution in RX~J1301.9+2747 is monotonic, while in GSN 069 and eRO-QPE1 rise-2 is generally hotter than rise-1. 
If interpreted as blackbody emission from a spherical surface, the $R_{bb}$ evolution is consistent with an expanding emitting surface with initial/final radii of the order of $1-4\times 10^{10}$ cm. This is comparable to $\sim 0.14-0.6\,R_{\odot}$, or to about $0.07-0.27(10^6M_{\odot}/M_{BH})\,R_g$ where the gravitational radius $R_g=GM_{BH}/c^2$. 

Assuming a Gaussian parent distribution, we computed the stacked posterior probability density for the temperature of the accretion disk, the soft X-ray excess-like component (excluding 2000), the strong QPE peak, and the weak QPE peak. We found respectively $kT=59\pm{2}$~ eV, $191\pm{18}$~eV, $132\pm{2}$~eV, and $119\pm{12}$~eV. 
As the peak temperatures of both weak and strong QPEs are always much lower than the temperature of the soft excess-like component measured in quiescence, it appears impossible that QPEs are associated with transient enhancements of the soft X-ray excess emission.

\begin{figure*}[htb]
\centering
\includegraphics[width=17.cm]{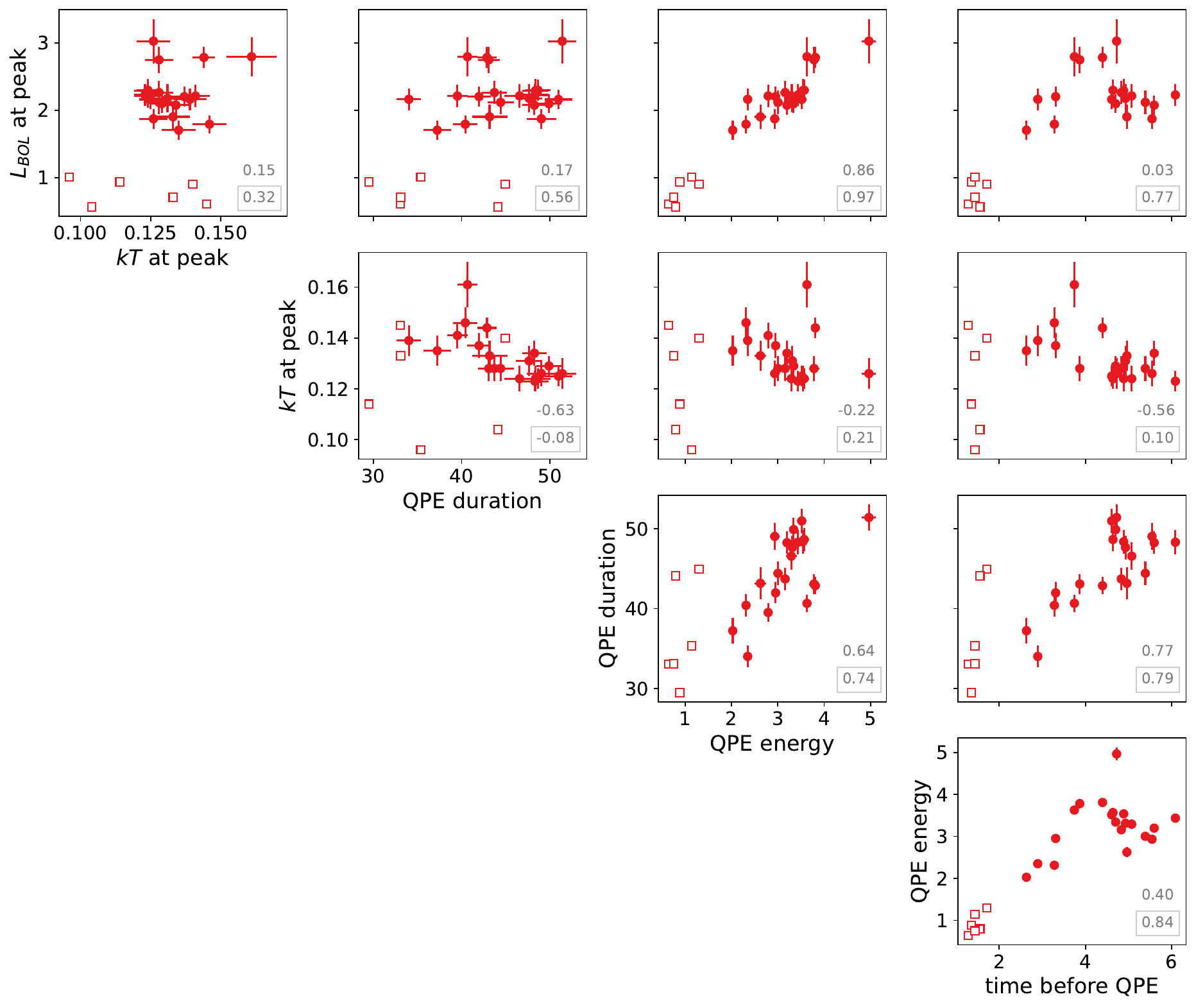}
\caption{Corner plot for the medians of the posterior PDFs of various  QPE parameters of RX~J1301.9+2747. The quantities plotted for each QPE are: the bolometric luminosity at the QPE peak ($10^{42}$ erg s$^{-1}$); the temperature at the QPE peak ($keV$); the QPE duration (m); the QPE total emitted energy ($10^{45}$ erg); and the time preceding each QPE (hr). Strong QPEs are plotted with filled circles, weak QPEs with open squares. Error bars for weak QPEs have been omitted for clarity. A couple of correlation coefficients between the parameters is reported in the lower right corner of each plot: strong QPEs only on top, and both strong and weak QPEs in the grey boxes below.
\label{FIG12}}
\end{figure*} 

The general properties of the QPEs of RX~J1301.9+2747 are plotted in Fig.~\ref{FIG12}. Here we plot against each other the median values of the posterior PDFs of the QPE bolometric luminosity and temperature at the peak, the QPE duration, the total QPE energy, and the time preceding each QPE. Strong QPEs are plotted with filled circles, weak QPEs with open squares. The total QPE energy was estimated as the integral of the best-fitting Gaussian emission line as $E_{QPE}=\sqrt{2\pi}\, L_{BOL}\times \sigma$, where $L_{BOL}$ is estimated at the QPE peak and $\sigma$ is the best-fitting Gaussian standard deviation.
The $L_{BOL}$ and $kT$ emitted at the QPE peak are weakly correlated, while a stronger correlation is present between the $L_{BOL}$ emitted by each QPE and the time preceding the QPE in question. Also the QPE duration correlates with the time preceding each QPE, thus providing the strong observed correlation between $E_{QPE}$ and the time before each QPE.
The temperature at the QPE peak appears to be independent of the QPE duration, energy emitted, or time preceding the QPE. In general, more powerful QPEs (i.e., with a higher amplitude, Fig.~\ref{FIG2}, or with a larger emitted energy, Fig.~\ref{FIG12}) last longer than less powerful QPEs.

\begin{figure}[h!tb]
\centering
\includegraphics[width=7.cm]{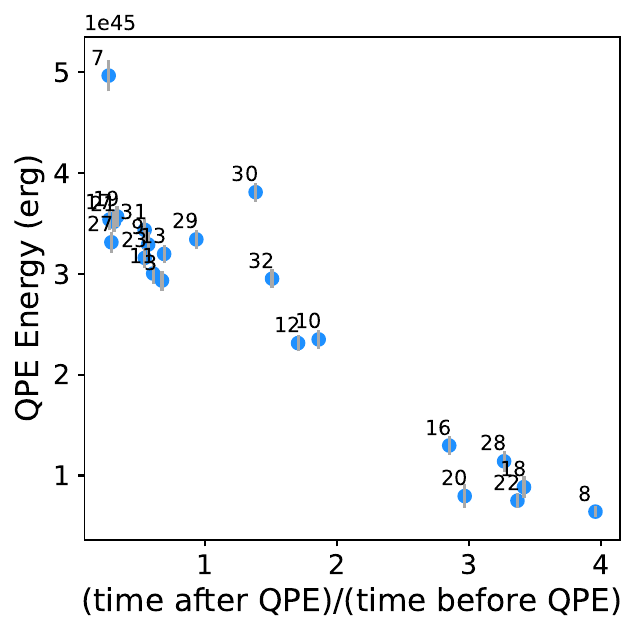}
\caption{For each QPE number $n$ of RX~J1301.9+2747 with pn spectral analysis available, the total emitted energy is plotted against the ratio between the time interval between the peaks of QPE number $n$ and  QPE number $n-1$ and the time interval between the peaks of QPE number $n+1$ and  QPE number $n$. 
 \label{FIG13}}
\end{figure}

In Fig.~\ref{FIG13} the QPE total energy radiated is plotted against the ratio between the time following and the time preceding each QPE.
The time intervals that precede the most energetic QPEs happen to be longer than those that follow them. The weakest QPEs are instead preceded by much shorter time intervals compared to those that follow them.
This is the opposite of what is seen in GSN 069 and eRO-QPE2 \citep{2023A&A...674L...1M, 2024arXiv240617020A} where longer time intervals systematically follow stronger QPEs, although the difference between strong and weak QPEs in those two sources is much less extreme than in RX J1301.9+2747.
 
\subsubsection{Comparison with GSN 069\label{sec:cfrGSN}}

\begin{figure*}[h!tb]
\centering
\includegraphics[width=17cm]{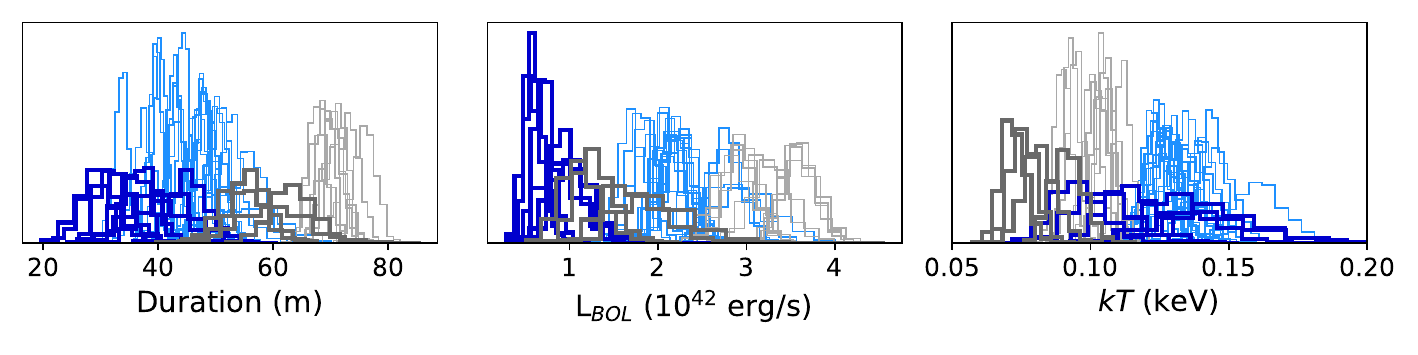}
\caption{Posterior PDFs of the QPE duration (left), computed as twice the FWHM of the best-fitting Gaussian to the $0.2-2$ keV band; the QPE bolometric luminosity at the peak (center) and QPE temperature at the peak (right), computed comparing the data to a redshifted blackbody model. Data of RX~J1301.9+2747 is plotted in blue, data of GSN 069 in grey; for both sources weak QPEs are plotted with thicker and darker lines.
 \label{FIG14}}
\end{figure*}

We perform a quantitative comparison of the QPE properties of RX~J1301.9+2747 with those of GSN 069. This galaxy has QPEs qualitatively similar to those of RX~J1301.9+2747 in terms of duration, amplitude, and time separations, and is the best-monitored QPE source by \textit{XMM-Newton} so far. For the analysis of GSN 069 we used the pn data from December 2018 until August 2022 of all the observations where QPEs are present, for a total of five \textit{XMM-Newton} observations and 18 QPEs. Of these QPEs, 14 show no strong background flares and their data were used for the spectral analysis as well: four are weak QPEs and ten are strong QPEs.
For fair comparison, we consider only the pn data also for RX~J1301.9+2747, i.e., we exclude QPE0 and QPE5 that are detected only by the MOS, and we discard for the spectral analysis also QPEs number 4, 7, and 16 that are contaminated by background flares in the pn data. We therefore consider 32 QPEs for the timing analysis\footnote{The timing properties of the half-QPE number 25 detected by the pn at the beginning of the 2022B exposure can be well-constrained.} of RX~J1301.9+2747, and 24 for the spectral analysis results comparison. 

The distribution of the QPE duration, computed as twice the FWHM of the best-fitting Gaussian in the $0.2-2$ keV band, is plotted in the left panel of Fig.~\ref{FIG14}. 
Here we use blue histograms for the QPEs of RX~J1301.9+2747, and grey histograms for the QPEs of GSN 069; in both cases strong QPEs are plotted with thin solid lines, while weak QPEs with thicker and darker lines. The distribution of duration of QPEs in RX~J1301.9+2747 peaks around 45 minutes, while the one of GSN 069's around 70 minutes. For both sources, the weak QPEs duration appears to be lower than the strong ones.

\begin{figure}[h!tb]
\centering
\includegraphics[width=8.8cm]{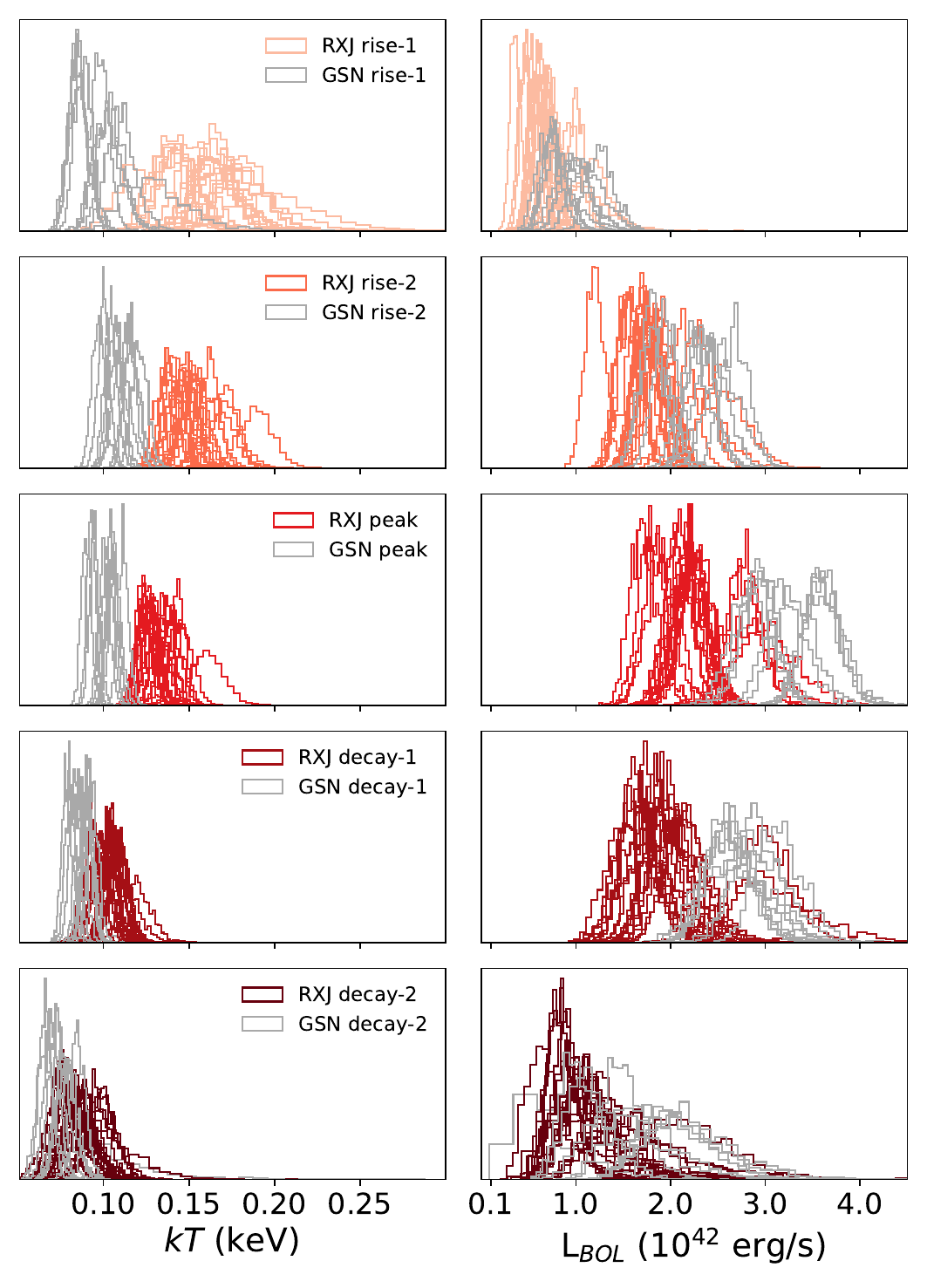}
\caption{Histograms of the posterior PDFs of the temperature (left column) and of the bolometric luminosity (right column) of a blackbody model compared to the each strong QPE spectral slice. From upper to lower panels: rise-1 in pink, rise-2 in orange, peak in red, decay-1 in dark red, and decay-2 in brown. The grey histograms report the  GSN 069 strong QPEs data.  
\label{FIG15}}
\end{figure}

Assuming that the QPEs are additive components on top of a constant quiescent emission, we compared the QPE spectral data to a redshifted blackbody model (\texttt{[zashift*bbodyrad]}) for both RX~J1301.9+2747 and GSN 069. 
The posterior PDFs of the bolometric luminosity and the temperature at the QPE peak assuming such a model are reported in the central and right panels of Fig.~\ref{FIG14}.
The QPEs of RX~J1301.9+2747 are systematically hotter and less luminous than those of GSN 069. As is the case for the QPE duration, also the $L_{BOL}$ and $kT$ of weak QPEs are systematically lower than those of strong QPEs, for both RX~J1301.9+2747 and GSN 069.
A more detailed analysis of the spectral properties of the QPEs during their spectral evolution is presented in Fig.~\ref{FIG15}, where we plot the posterior PDFs of strong QPEs of GSN 069 in grey, and of RX~J1301.9+2747 with tones of red of increasing darkness going from rise-1 to decay-2.
The difference in temperature between the strong QPEs of RX~J1301.9+2747 and those of GSN 069 decreases going from the beginning of the QPE (rise-1) to the end of the QPE (decay-2), when the temperatures of the strong QPEs of the two sources converge to the same value of $kT\sim 80$ eV.  The average temperature during the QPE peak is $\sim 130$ eV for RX~J1301.9+2747 and $\sim 100$ eV for GSN 069, while the average temperature during the QPE rise-1 is respectively $\sim 160$ eV and $\sim 90$ eV.   

\subsubsection{Weak and strong QPEs\label{weakstrong}}

\begin{figure}[h!tb]
\centering
\includegraphics[width=8cm]{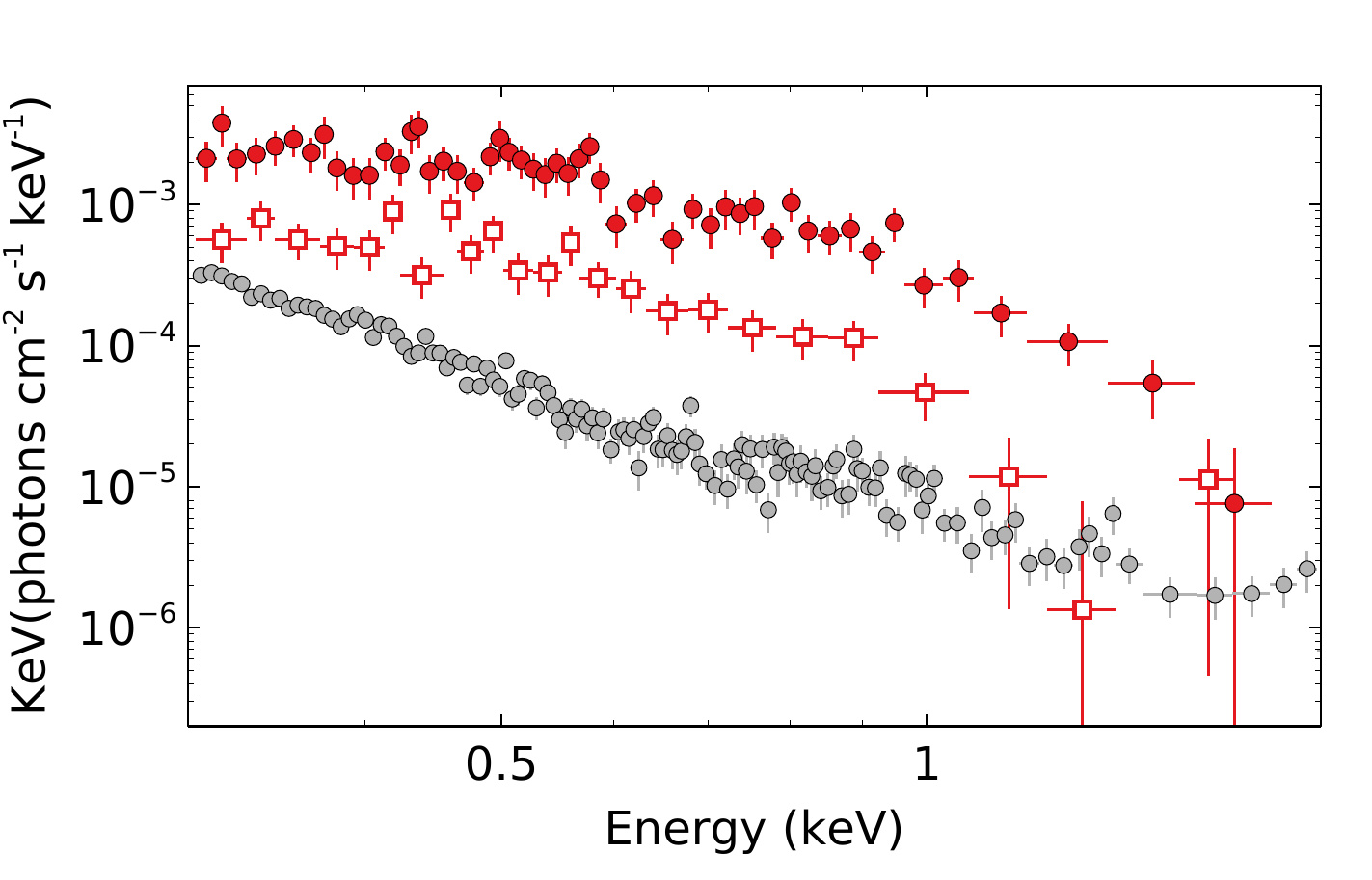}
\caption{Aaverage $0.3-2$ keV $EF(E)$ quiescent spectrum of RX~J1301.9+2747 (grey) plotted together with the average weak QPE (empty squares) and strong QPE (filled circles) spectra extracted during the peak. The QPE spectra are background-subtracted with the average quiescent spectrum used as background.
\label{FIG16}}
\end{figure}

The average QPE-only spectra (that is, quiescence-subtracted spectra) for the weak and strong QPEs are shown in Fig.~\ref{FIG16} compared to the average quiescent spectrum. Here the spectra are plotted unfolded against a power law model with $\Gamma=2$. There appear not to be dramatic differences in the spectral properties of strong and weak QPEs.
The main difference between strong and weak QPEs are the energy released and the timing properties.
The peak $0.3-2$ keV (bolometric) luminosity of strong QPEs is $1.7\pm{0.3}\times10^{42}$ erg s$^{-1}$ ($2.2\pm{0.3}\times10^{42}$ erg s$^{-1}$). 
The typical luminosity of strong QPEs is a factor of $\sim 2.5\times$ higher than the luminosity of weak QPEs.
The duration of weak QPEs is at the lower end of the distribution (left panel of Fig.~\ref{FIG14}), thus making the total energy output of weak QPEs about $1/3$ of that of strong QPEs.
The temperature at peak of strong QPEs is on average $kT=130\pm{10}$~eV. 
Assuming a Gaussian parent distribution, the median temperature at the peak of weak QPEs is $kT = 120\pm{20}$ eV, comparable to the strong QPEs one. 
However, looking at the right panel of Fig.~\ref{FIG14}, where the full PDFs for the QPE temperature are shown, it can be seen that the weak QPEs have systematically lower $kT$ than the strong ones, as observed in GSN 069 \citep{2023A&A...674L...1M}.
The difference in timing properties of weak and strong QPEs is evident in Fig.~\ref{FIG1} and Fig.~\ref{FIG13}: the recurrence time preceding weak QPEs is much shorter than the one following them. 
This is the opposite of what is observed in GSN 069 \citep{2023A&A...674L...1M}, although the difference in QPE amplitude in this source are not as dramatic as in RX J1301.9+2747.
\section{Radio observations\label{sec:radio}}

The last three \textit{XMM-Newton} observations of RX~J1301.9+2747 were partially covered by simultaneous Karl G. Jansky VLA observations, in order to check for the presence of correlated radio and X-ray variability. We observed the coordinates of RX~J1301.9+2747 at 6\,GHz (C-band) on 2020 July 11 and 2020 July 12 (Program ID SJ6456) and 2022 June 17 and 2022 June 19 (Program ID SL0464) where the VLA was in the extended configurations B and A respectively. In each observation, flux and bandpass calibration was performed with 3C286 and phase calibration was performed with J1310+3220 (2020) or J1327+2210 (2022). While the 2020 radio observation had to be split in two segments due to instrumental limitations\footnote{The VLA is not able to perfectly track sources when they pass directly overhead, so in 2020 we followed the standard observatory recommendation to not observe at elevations $> 80^{\circ}$. In 2022 we decided that getting a continuous light curve was more important than getting a perfect pointing.}, the 2022 light curves were continuous and allowed us to check for the presence of radio variability both before and after strong X-ray QPEs. This resulted in the two VLA observations in 2020 being 5\,hr long each, and the two observations in 2022 being 10\,hr long each. 

All data were reduced in the Common Astronomy Software Package \citep[CASA, v5.6.3]{CASA_paper} using standard procedures, including the VLA pipeline. Images of the target field were initially made for each of the 4 observations using the CASA task \texttt{tclean}. An unresolved point source at the location of RX~J1301.9+2747 was detected in each observation. We extracted the flux density using the CASA task \texttt{imfit} and by fitting an elliptical Gaussian the size of the synthesised beam. Next, 1-minute interval images of the target field were created using the CASA task \texttt{tclean} in order to generate a lightcurve for each of the observations.  Again, the flux density of the target was extracted using the CASA task \texttt{imfit} and by fitting an elliptical Gaussian the size of the synthesised beam. Additionally, we extracted the flux density of a nearby source in the field (after applying a primary beam correction), NVSS J130146+274629, in order to serve as a check source. In general, the check source appears to show some variability of the order of $\sim20\%$, uncorrelated with any variability observed from RX~J1301.9+2747.
Both of the 2022 observations suffered from severe gain compression due to radio-frequency interference (RFI), particularly in the first $\approx3$\,hr of the observations, resulting in significantly lower flux density measured for the target and check source in the field and correlated variability observed for both sources in this time. We therefore excluded these data from the analysis.  

\begin{figure*}[h!tb]
\centering
\includegraphics[width=18cm]{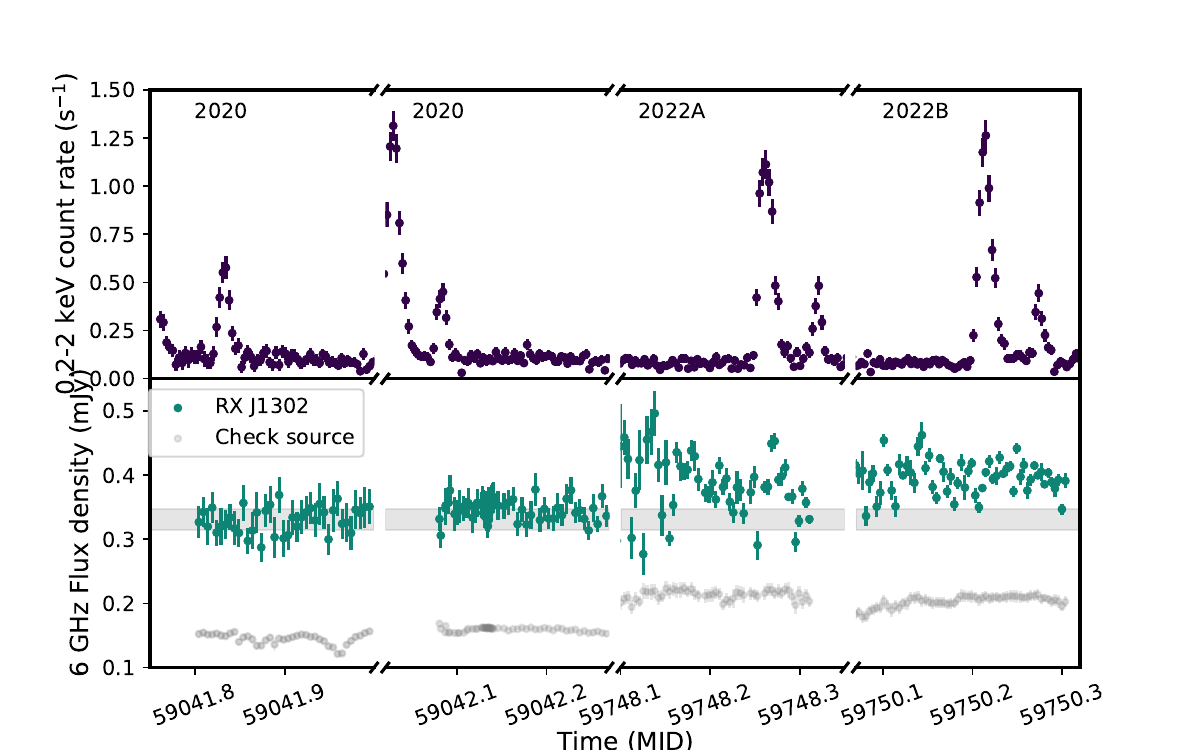}
\caption{Simultaneous X-ray and radio light curves for the 2020, 2022A, and 2022B epochs: the top panel reports the $0.2-2$ keV light curve, and the lower panel the radio light curve at 6 GHz as measured by the VLA. In the lower panel, RX~J1301.9+2747 is shown in green and a different radio source in the field is shown in grey to serve as a check source. Each portion of the broken axes represents 0.25\,d. Note that the flux density of the check source has been scaled down by a factor of 15. The grey shaded region shows the mean value for the 2020 observation plus or minus 5$\%$, the absolute accuracy of the VLA flux density calibration.  
\label{FIG17}}
\end{figure*}

The 6\,GHz flux densities measured for RX~J1301.9+2747 for each of the 4 observations are reported in Table \ref{tab:radio_fluxes}. The VLA light curves of RX~J1301.9+2747 for each of the 3 epochs are shown in Fig.~\ref{FIG17}. The top panel shows the $0.2-2$ keV light curve, while the lower panel shows the 6\,GHz radio flux density in $\approx$1-minute intervals. No evident radio variability is observed during the X-ray QPEs, except for the QPE in the 2022A observation (QPE17), where a small increase in radio flux density is seen. The flux density increases by $\approx$0.15\,mJy during the QPE. The overall variance of the flux density during the observation is 0.09\,mJy, so the small flare is detected at $<2\sigma$ significance. This increase is smaller than the variability seen earlier in the observation (and during   X-ray quiescence) and we deduce it is unlikely to be associated with the X-ray QPE, especially given there is no evidence of radio variability associated to the X-ray QPEs seen in 2020 and 2022B. 

As a cross-check, we also used the {\tt dftphotom} task within the {\tt pwkit} package to fit the calibrated VLA visibilities in the uv plane directly \citep{pwkit}. We then binned the data into $\sim1$ minute chunks to enable a direct comparison with the radio light curves produced by the image-based analysis.  We found consistent results for all four observations, including during the time intervals most affected by RFI in the 2022 data. We therefore conclude that there is no statistically significant evidence for correlated radio/X-ray flaring activity in RX~J1301.9+2747.

\subsection{The radio variability of RX~J1301.9+2747\label{radioresults}}

Whilst there does not appear to be significant radio variability associated with the X-ray QPEs in the radio data presented in this work, there is variability on the order of 5--10$\%$ at 6\,GHz observed between epochs, and within the 2022 observations over the 10\,hr tracks. Additionally, \citet{2022ApJ...935..115Y} observed RX~J1301.9+2747 over 3 days in 2015 at 9\,GHz and over 5 days in 2019 at 14\,GHz, finding a variability on the order of 26$\%$ at 9\,GHz and 12$\%$ at 14\,GHz on timescales as short as a few hours. Here we assess whether the observed radio variability is consistent with interstellar scintillation (ISS) of a compact source, or whether it is due to intrinsic variability of the radio source. 

Using the NE2001 electron density model \citep{Cordes2002}, the transition frequency between strong and weak scintillation regimes occurs at $\nu_0 = 6.4$\,GHz and the angular size limit of the first Fresnel zone at the transition is $\theta_{F0}=4.6$\,microarcsecond at the Galactic coordinates of RX~J1301.9+2747. 
Adopting the \citet{Walker1998} formalism for ISS as appropriate for compact extragalactic sources, RX~J1301.9+2747 will therefore be in the weak scattering regime at 9 and 14\,GHz and could be in the strong scattering regime at 6\,GHz, but we note that it is very close to the transition frequency. 

The modulation expected due to ISS at the observing frequencies of 6, 9, and 14\,GHz is heavily dependent on the source size. If the source is completely unresolved to scintillation (a source size $\lesssim$0.001\,pc at the distance of RX~J1301.9+2747), we would expect to see variation at 6\,GHz of up to 69$\%$ on a timescale of 3\,hr, of 62$\%$ at 9\,GHz on a timescale of 1.7\,hr, and 33$\%$ at 14\,GHz on a timescale of 1.35\,hr. Clearly, the radio emission from RX~J1301.9+2747 was not observed to vary with such high modulation amplitudes at any of the observing frequencies. However, the modulation due to ISS reduces as the source becomes resolved to ISS, by a factor $(\theta_r/\theta_s)^{7/6}$, where $\theta_r$ = $\theta_{F0} (\nu/\nu_0)^{11/5}$ and $\theta_s$ is the true angular size of the source. Assuming a maximum modulation of 10$\%$ at 6\,GHz was observed, this variability could be explained by ISS for a source size 0.008\,pc and the timescale for this modulation would be 15\,hr. A modulation of 26$\%$ at 9\,GHz requires a source size of 0.0011\,pc with a ISS timescale of 3.5\,hr and a modulation of 12$\%$ at 14\,GHz requires a source size of 0.0005\,pc with a ISS timescale of 3.1\,hr. These modulation timescales are consistent with the observed radio data. 

Therefore if the observed radio variability at 6, 9, and 14\,GHz were caused by ISS, the radio-emitting region of RX~J1301.9+2747 must be between 0.0005--0.008\,pc, and smaller at higher frequencies. For AGN jets, the emitting region is commonly observed to be larger at lower frequencies due to the effect of core shift \citep[e.g.,][]{1998A&A...330...79L}. Therefore, if the radio emission from RX~J1301.9+2747 is dominated by a sub-pc jet, the emitting region would be expected to be smallest at the highest observing frequencies and larger at the lowest observing frequencies, as inferred from the scintillation analysis. Indeed, VLBA imaging of RX~J1301.9+2747 at 1.6\,GHz did not resolve any source structure, constraining the source size to $<0.7$\,pc \citep{2022ApJ...935..115Y}. We therefore conclude that the radio variability we detect in this analysis at 6\,GHz, as well as the 9 and 14\,GHz variability detected by \cite{2022ApJ...935..115Y} is not inconsistent with ISS if the radio-emitting region is $<0.008$\,pc. As pointed out by \cite{2022ApJ...935..115Y}, if the radio variability timescales observed are associated to a light-travel time, the inferred radio-emitting region could be as small as 0.0008\,pc. This is consistent with the variability being primarily due to ISS.

\section{Discussion\label{sec:discu}}

The two competing physical scenarios proposed so far to interpret QPEs are accretion flow instabilities and orbital phenomena.
The former can have various origins: thermal, dynamical (Sect.~\ref{discu:instabilities}), or magnetic (Sect.~\ref{sec:mag}).
Orbital phenomena can include both gravitational self-lensing of massive black hole binaries with order unity mass ratio (Sect.~\ref{sec:discu:gravlensing})
and interactions between a massive black hole with  mass $M_{BH}$ (and/or an accretion flow around it) and one or more secondary objects with much smaller mass $m\ll M_{BH}$. When the secondary object is of stellar-mass size these systems are called extreme mass ratio inspirals (EMRIs; Sect.~\ref{discu:emri}).
We discuss the results on RX~J1301.9+2737 in comparison to other QPE sources (Sect.~\ref{discu:comparison}) and to other repeating nuclear transients (Sect.~\ref{discu:rnt}).

\subsection{Accretion flow instabilities\label{discu:instabilities}}

Classical radiation pressure instabilities \citep[RPI,][]{1974ApJ...187L...1L,2002ApJ...576..908J} of a thin accretion disk around a SMBH cannot explain the generally short timescales and high luminosities of QPEs \citep{2021Natur.592..704A}.
Magnetic or dynamical effects can shrink the disk region subject to instabilities and thus also the predicted timescales.
A magnetic wind \citep{2022ApJ...928L..18P,2023ApJ...952...32P} or a strong disk magnetization \citep{2023MNRAS.524.1269K} can shorten the timescale of instability cycles enough to reproduce the timescales and spectra of QPE sources fairly well. 
However, the energy-dependence of the QPE flare evolution (shown in the top panel of Fig.~\ref{FIG4} for the case of RX~J1301.9+2747, and observed in all the QPE sources so far)  appears not to be reproduced by this class of models \citep[e.g.,][]{2022ApJ...928L..18P}.
Dynamical instabilities of the inner accretion flow have also been proposed to explain QPEs \citep{2021ApJ...909...82R}.
In principle, accretion flow instabilities might explain the higher energy released by  QPEs happening after longer recurrence times observed in RX~J1301.9+2747 (Fig.~\ref{FIG14}).
Stronger QPEs would have had more time to accumulate energy to be released in the flare compared to weak QPEs, leading to a larger energy budget that can be radiated away.
However, the behaviour in RX~J1301.9+2747 is opposite to that observed in GSN 069 and eRO-QPE2, where longer recurrence times lead to weaker QPEs \citep{2023A&A...670A..93M, 2024arXiv240617020A},
thus weakening the overall argument.
Furthermore, the flare shape predicted by instability models does not well describe the observed QPE shape. While QPEs generally show a faster rise and a slower decay, the opposite is predicted by RPI, independent of the existence of magnetic fields, and by dynamical instabilities models.

\subsubsection{Magnetic reconnection \label{sec:mag}}
Magnetic instabilities \citep[e.g., relativistic reconnection,][]{2014ApJ...783L..21S} can occur in the innermost regions around SMBHs, and might be responsible for heating up the plasma generating the hard X-ray coronae \citep[e.g.,][]{2017ApJ...850..141B,2021MNRAS.507.5625S,2022arXiv220302856S, 2023ApJ...945L..34M}.
A model for the production of episodic jets driven by magnetic reconnection around SMBHs, in analogy with the coronal mass ejections production in the Sun, was developed by \citet{2009MNRAS.395.2183Y} and expanded by \citet{2017MNRAS.468.2552L,2023MNRAS.520.1271L}.
These models predict a flare shape similar to that observed in QPEs \citep[e.g., Fig. 10 of][]{2017MNRAS.468.2552L}, resulting from synchrotron emission by thermal and power-law electron distributions in an expanding hot spot close to the black hole, with analogies with our findings for QPEs (see Sect.~\ref{sec:qpes} and Fig.~\ref{FIG11}).
In analogy with solar flares, the reconnection events are not completely random but are governed by self-organized criticality \citep{1987PhRvL..59..381B,2016SSRv..198...47A}. The  resulting distribution of energy release and time of arrival of flares follows power law distributions \citep{1991ApJ...380L..89L} and therefore could produce QPEs of different amplitudes and different recurrence times. 

RX~J1301.9+2747 is a known radio source, serendipitously detected at 1.4 GHz in the VLA map of the Coma Cluster \citep{2009AJ....137.4450M}.
An extensive set of VLA observations performed between July 2015 and January 2019 has been presented by \citet{2022ApJ...935..115Y}, and showed significant variability on timescales as short as days, 
implying a compact size of the emission region $< 10^{-3}$ pc.
This scenario is compatible with our 2020-2022 VLA observations, that provided limits on the size of the radio source of $< 0.008$ pc, given the observed variability attributed to ISS.
These results, together with VLBA imaging observations that did not resolve the source scales of $<0.7$ pc and a steep radio spectrum indicating optically thin synchrotron emission \citep{2022ApJ...935..115Y}, rule out various origins for the radio emission in RX~J1301.9+2747 such as star formation or pc-scales outflows, leaving open the scenario of episodic compact jet ejections. 

If X-ray QPEs were the results of episodic compact jet ejections, one might expect to observe correlated radio variability due to, for example, broadband synchrotron emission or synchrotron self-Compton.
However, our simultaneous \textit{XMM-Newton}/VLA observations revealed no correlated variability between the X-ray and radio bands (Fig.~\ref{FIG17}).
This suggests two physical mechanisms responsible for the short-term X-ray and radio variability of RX~J1301.9+2747. 

The radio emission of RX~J1301.9+2747 could be nonetheless connected to its quiescent X-ray emission, and in particular to the weak soft X-ray excess-like spectral component that emerges since 2019.
This can be modeled with either a thermal component with $kT\sim 190$ eV, or a very steep power law with $\Gamma \gtrsim 3.5$. 
One possibility is that this weak spectral component is  
due to magnetic reconnection in a plasma with very low magnetization.
The plasma magnetization around the central BH of RX~J1301.9+2747 should be very low, due to the dominant contribution in energy density of disk particles. 
We might thus be able to observe a few magnetic coronal loops forming. 
These would produce discrete and rapid radio variability, and the expected X-ray properties due to bulk Comptonization in the plasmoid chains would be consistent with those observed in RX~J1301.9+2747  during quiescence, i.e., there would be a very steep X-ray power law-like emission \citep{2014ApJ...783L..21S}.

As for the X-ray QPEs of RX~J1301.9+2747, in principle their different amplitudes and recurrence times could be explained by magnetic reconnection. However, the regularity of appearance of weak QPEs always shortly after a strong QPE, as well as the alternate long-short separation time, are more difficult to explain in such a scenario.   
It is even more difficult to explain the regularity of QPE sources such as GSN 069 and eRO-QPE2 \citep{2019Natur.573..381M,2021Natur.592..704A,2022A&A...662A..49A}.
In fact magnetic reconnection events are expected to be, albeit non completely random, very much irregularly spaced in time.
A magnetic reconnection model has been invoked also to explain the flares observed from our Galactic center \citep[e.g.,][]{2017MNRAS.468.2552L,2022MNRAS.511.3536S,2022ApJ...924L..32R}. However, the observed spectral properties of RX~J1301.9+2747, the general self-similarity, the timescales, and the overall asymmetry of its QPE shape is in (stark) contrast with what has been observed in Sgr A* \citep{2017MNRAS.468.2447P,2023A&A...669L..17V}, casting doubts on the interpretation of QPEs within the same physical context.

\subsection{Gravitational self-lensing of massive black hole binary mini-disks\label{sec:discu:gravlensing}}
 
To the first order, models invoking massive black hole binaries (MBHB) with self-lensing mini-accretion disks predict a  symmetric flare shape and a wavelength-independent (achromatic) signal, which is different than the one observed in QPEs \citep{2021MNRAS.503.1703I}.
Therefore this class of models with binaries with equal (or almost) mass ratio has received less attention so far than the 
high mass ratio (EMRI) scenarios. 
However, detailed simulations by \citet{2023arXiv231019766M} show that the expected signal can be strongly energy-dependent, with similar properties to those observed in the QPE light curves and spectra. Many deviations from a simple symmetric flare shape can occur. For example, 
\citet{2022PhRvD.105j3010D, 2022PhRvL.128s1101D} predict a dip near the middle of the flare. 
The time separations between the QPEs of RX~J1301.9+2747  are however likely too short to be explained by orbital motion of a MBHB, as given the black hole mass involved \citep[$\sim 0.4-4\times 10^6\,M_{\odot}$,][]{2022A&A...659L...2W, 2024ApJ...970L..23W} the merger of the two BHs should have already happened \citep[e.g., Fig. 3 of][]{2021MNRAS.503.1703I}. Only an orbiter with significantly smaller mass than the central massive black hole could be invoked, as its time-to-merger would be much longer.
In any case the requirements of a high orbital eccentricity and of having both orbiting BHs's mini-disks to be lensing each other are necessary to explain the alternate long-short time separations between QPEs and the large difference between such time separations observed in RX~J1301.9+2747.

\subsection{Extreme mass ratio inspirals\label{discu:emri}}

EMRI scenarios are promising in explaining the general properties of the QPE-emitting sources \citep{2021Natur.592..704A}.
Two main classes of EMRI scenarios have been proposed for QPEs: those invoking interactions between the primary SMBH and the secondary object at the pericenter, and those invoking impacts of the secondary on the accretion flow. The first class of scenarios has been explored in a variety of configurations, including both eccentric and circular orbits of single or multiple stars or stellar remnants, experiencing Roche Lobe overflow or tidal stripping, or shocking with the stellar debris \citep[e.g.,][]{2020MNRAS.493L.120K,2022A&A...661A..55Z,2022ApJ...926..101M,2022ApJ...933..225W, 2022MNRAS.515.4344K, 2022ApJ...941...24K, 2023MNRAS.524.6247L, 2023ApJ...945...86L, 2023MNRAS.526L..31K}.
The second class of scenarios has been investigated as impacts of the secondary lower-mass EMRI component on a radiatively inefficient, torus-like \citep{2021ApJ...917...43S} or on a disk-like accretion flow \citep{2021ApJ...921L..32X, 2023ApJ...957...34L, 2023A&A...675A.100F, 2023MNRAS.526...69T, 2024arXiv240111190Z, 2024arXiv240506429Z}.

In EMRI scenarios with mass transfer at the pericenter, QPEs happen once per orbit.
QPEs are then powered either by accretion of the stripped mass on the primary BH \citep{2020MNRAS.493L.120K,2022A&A...661A..55Z,2022ApJ...933..225W,2023ApJ...945...86L}, or by shocks between different streams of matter \citep{2022ApJ...941...24K}, or between the streams and the accretion disk around the primary \citep{2023MNRAS.524.6247L}.
To the first order, the time between QPEs corresponds to the EMRI orbital period. In practice, the passages at the pericenter will occur slightly later or earlier at each orbit, due to the effects of apsidal precession of the EMRI orbit and light travel time.
The QPE duration and intensity depend on the orbital eccentricity and on the relative position of the Roche lobe. The variety of stellar progenitors and orbital parameters \citep[e.g.,][]{2009MNRAS.392..332L,2013ApJ...767...25G} might explain the varieties of QPE phenomenology, with longer duration QPEs due to more gentle pericenter passages (i.e., more circular orbits), more luminous QPEs due to a more intense mass transfer. 
An alternation of long-short recurrence times in these scenarios can in principle be explained by thermodynamic readjustments of the stellar structure after each mass transfer \citep[e.g.,][]{2022MNRAS.515.4344K}. This seems applicable to QPE sources with more regular timing properties such as GSN 069 and eRO-QPE2, where differences between the short and long recurrence times are of the order of few \% \citep{2022A&A...662A..49A,2023A&A...670A..93M}. 
It is however extremely difficult to apply this class of scenarios to the cases of RX~J1301.9+2747 or eRO-QPE1, where a large scatter in consecutive recurrence times is observed, as shown in this work and in \citet{2024ApJ...965...12C}. 


In the case of impacts on the accretion flow, QPEs happen (at least) twice per orbit. The time separation between impacts depend on the EMRI orbital semi-major axis and eccentricity $e$.
Unless the orbit is perfectly circular ($e=0$), there is a natural expectation of alternate long-short time separation between QPEs. In general, orbits with a lower eccentricity  would cause impacts (and thus QPE flares) to be more equally spaced in time, while a higher eccentricity would cause a larger difference in time preceding and following any QPE. 
The presence of large differences in time separation between consecutive QPEs leads to a relatively high-$e$ scenario for the EMRI of RX~J1301.9+2747, eRO-QPE1, eRO-QPE4, and to a more circular EMRI for eRO-QPE2 and GSN 069 \citep{2023A&A...675A.100F, 2024arXiv240111190Z,2024arXiv240506429Z}.
In these scenarios the flare emission is likely due to an expanding bubble at the impact location, consistent with the lower panel of Fig.~\ref{FIG11} in the case of RX~J1301.9+2747, with Fig. 18 of \citet{2023A&A...670A..93M} in the case of GSN 069, and with Fig. 2 and 3 of \citet{2024ApJ...965...12C} in the case of eRO-QPE1. The light curve is characteristic of a mini-supernova \citep{2023ApJ...957...34L} with a fast rise and a slower decay, as observed in QPEs. 

\citet{2023A&A...675A.100F} were the first to qualitatively reproduce the specific timing properties of RX~J1301.9+2747 in a EMRI scenario, focusing on the 2022A light curve that is characterised by the presence of weak QPEs.
In their scenario  QPEs are due to impacts between the secondary EMRI object and the precessing accretion disk surrounding the primary EMRI object. 
The two EMRI objects are assumed to be BHs, with the secondary having a $M_{BH}=100 M_{\odot}$. 
The orbit of the secondary BH is subject to apsidal and nodal precession and this complicates the expected light curve, that can contain up to three impacts (QPEs) per orbit. 
By assuming a prograde and eccentric ($e=0.4$) 
orbit of the secondary BH with a semi-major axis of $50\,R_g$ around a SMBH with $M_{BH}=2\times 10^6 M_{\odot}$ (giving an orbital period of  about 6 hours, or 22 ks), the complex timing properties of RX~J1301.9+2747 can be qualitatively reproduced.
However, the QPE temperature expected in this scenario is $kT\sim 180-450$ eV, a factor of about $2-3$ above what we observe, and the relative QPE amplitudes are not always exactly matched by the model.

\citet{2024arXiv240506429Z} were the second to apply the impacts scenario to the specific case of RX~J1301.9+2747. They assumed a generic stellar-mass object impacting on a TDE-disk and were able to reproduce the timing properties of the 2019 and 2020 light curves, but only when ignoring the presence of weak QPEs. In this case the orbit is still eccentric but with a best-fitting $e=0.25$ and an orbital period of about 9 hours (32 ks). Also in this case the secondary EMRI object reaches distances very close to the central SMBH at the pericenter, and a tight constrain on its mass $ < 0.5\,M_{\odot}$ can be placed in the case of it being a star, in order not to reach its tidal radius and thus being destroyed. 

The two different solutions found by \citet{2023A&A...675A.100F}
and \citet{2024arXiv240506429Z} reflect the difference in light curve behavior observed in 2022A compared to the one observed in 2020 as well as the fact that weak QPEs were ignored by \citet{2024arXiv240506429Z}, thus selecting more regularly spaced QPEs and therefore deriving a lower EMRI orbital eccentricity. The 2022B light curve is remarkably similar to the 2020 one, perhaps indicating some sort of oscillatory mechanism between the two ``QPE phases'' (one at higher eccentricity when weak QPEs are present, the other at lower eccentricity when weak QPEs are absent) observed in RX~J1301.9+2747.
We note however how a transition between the two phases appears to happen on extremely short timescales, between 2022A and 2022B. 
Hence, while the impacts model appears promising in RX J1301.9+2747 (and perhaps even more so in other more regular QPE sources such as GSN 069 and eRO-QPE2), a complete solution that accounts for the two different phases (with and without weak QPEs) has not been reached yet, and further work is needed (Miniutti et al., in prep.).
 
\subsection{Comparison to other QPE sources\label{discu:comparison}}

X-ray QPEs have been observed in six galaxies so far \citep[GSN 069, RX~J1301.9+2747, eRO-QPE1, eRO-QPE2, eRO-QPE3, and eRO-QPE4;][]{2019Natur.573..381M,2020A&A...636L...2G,2021Natur.592..704A,2024A&A...684A..64A}, plus two good candidates with X-ray flaring properties very similar to QPEs \citep[J0249 and Tormund,][]{2021ApJ...921L..40C, 2023A&A...675A.152Q}. 

All these eight galaxies  have an  X-ray quiescent (non-QPE) emission well-represented by the high-energy tail of an accretion disk with inner temperature $\sim 40-100$~eV. 
The quiescent emission of RX~J1301.9+2747 appears more complex than most of the other QPE sources. 
In fact it requires, in addition to the emission of an accretion disk with inner temperature $\sim 50-60$ eV, a harder spectral component from epoch 2019 onward with properties similar to those of the AGN soft X-ray excess.
It is well-reproduced by the Comptonization of the accretion disk seed photons into a plasma with optical depth $\tau=10$ and temperature $\sim 400$ eV, typical of the warm coronae observed in AGN \citep[e.g.,][]{2020A&A...634A..85P, 2024arXiv240614378P}.
A similar spectral component has been observed in eRO-QPE4 during both quiescence and the QPE emission \citep{2024A&A...684A..64A} and is also well-fitted by a Comptonization model in addition to disk emission.

About half of the QPE sources and candidates showed a long-term decay in their X-ray flux consistent with a TDE. 
The quiescent emission of GSN 069 showed a slow long-term decay \citep{2018ApJ...857L..16S} since the slew detection in 2010 \citep{2013MNRAS.433.1764M} for about 3000 days until a sudden rebrightening \citep{2023A&A...670A..93M} and subsequent decline of emission were observed \citep{2023A&A...674L...1M}.
Also the quiescent emission of eRO-QPE3 showed a decline over the $\sim 2.5$ years probed by eROSITA observations, then completely disappeared in subsequent \textit{XMM-Newton }follow-up \citep{2024A&A...684A..64A}. 
J0249 and Tormund are associated with X-ray- and optically-selected TDEs respectively, and show a decay in their quiescent X-ray flux over respectively about 15 years and 6 months \citep{2021ApJ...921L..40C, 2023A&A...675A.152Q}.
There is no direct evidence for a TDE in RX~J1301.9+2747. However, it is a young post-starburst E+A galaxy \citep{1999AJ....117..140C}, a kind of galactic nucleus where a high rate of TDEs is observed \citep{2014ApJ...793...38A, 2021ApJ...908L..20H} and expected \citep{2022MNRAS.511.2885B}. It also likely hosts a nuclear star cluster (NSC) \citep{2017ApJ...837....3S}, and NSCs are also expected to enhance the TDE rate in galactic nuclei \citep{2023arXiv231000038W}. 
The fact that there is no clear evidence of a TDE-like flare in the optical history of RX~J1301.9+2747 is not surprising in presence of a disk. In this case the disruption of stars might not produce the typical TDE flare observed in gas-poor environments (i.e., in quiescent galaxies) but more complex light curves \citep{2024MNRAS.527.8103R}. In particular, if the  disk is dense enough, no luminous TDE flare is expected at all, due to the rapid mixing of the stellar debris with the disk material. In this case the light curve variability would reflect variations in mass accretion rate in the disk and possible changes of states/rearrangements of the inner disk, i.e., long-lived bumpy light curves. This would be compatible with the observed long-term quiescent luminosity evolution of RX~J1301.9+2747.

The X-ray quiescent luminosity variability of RX~J1301.9+2747 is only on the order of a factor of 2.5 over four decades, from the $L_{0.3-2\, \rm{keV}}\sim 2.5\times 10^{41}$ erg s$^{-1}$ of the first \textit{ROSAT} observation to the $L_{0.3-2\, \rm{keV}} \sim 10^{41}$ erg s$^{-1}$ of the last \textit{XMM-Newton} one. In comparison, the GSN 069 quiescent luminosity spans $L_{0.3-2\, \rm{keV}} \sim 10^{42-43}$ erg s$^{-1}$. Interestingly, the observed luminosity variations are non-monotonic. 
Both \textit{ROSAT} observations correspond to a larger flux compared to both previous and following epochs of observations. However, they had low exposure time and moderate X-ray statistics, and the observed short-term variability is only suggestive.
The non-monotonic variations during the last three \textit{XMM-Newton} observations are instead significant, with the first observation of June 2022 (2022A) at a lower X-ray flux than both the previous (2020) and the following (2022B) observation.
The factor of $ \sim 15\%$ increase in flux (and disk luminosity) between 2022A and 2022B happens in  less than one day. 

The quiescent emission in GSN 069 also shows the presence of a quasi-periodic oscillation (QPO) with quasi-period similar to the recurrence time between consecutive QPEs, and shifted by about 10 ks with respect to the QPE peaks in most of the observations when QPEs are present \citep{2023A&A...670A..93M}. The quiescent emission of RX~J1301.9+2747 showed hints of the presence of a QPO during a few segments of the 2020 observation \citep{2020A&A...644L...9S}, but this has not been observed again in the 2022 data, leaving the possibility open of the presence, if any, of a weak and transient QPO in the quiescent emission of RX~J1301.9+2747. However, the QPO observed in GSN 069 has a similar quasi-period as QPEs recurrence times, while the QPO in RX~J1301.9+2747 is detected on much shorter timescales and appears decoupled from the QPE recurrence times, hence they are likely not produced by the same physical phenomenon.

The QPEs in RX~J1301.9+2747 have been observed persistently in all the \textit{XMM-Newton} observations between 2000 and 2022. 
The QPEs of GSN 069 have been very regular in amplitude and recurrence time spacing until the quiescence rebrigthening, when they showed an irregular behavior and the presence of weak QPEs, and then disappeared for a few months \citep{2023A&A...670A..93M}. They then reappeared during the quiescent emission decay \citep{2023A&A...674L...1M}. The quiescence light curve evolution suggests that GSN 069 underwent a partial TDE, and that there is a quiescent luminosity threshold for the appearance of QPEs \citep{2023A&A...674L...1M}.   
The QPEs of eRO-QPE2 displayed a remarkable regularity in amplitude and alternation of long/short recurrence time, similar to GSN 069, in their discovery observations \citep{2021Natur.592..704A}. Subsequent observations have shown an evolution of the QPE properties \citep{2024arXiv240117275A}, which are however still much more regular in amplitude and recurrence times compared to the case of RX J1301.9+2747.
This is different to what is observed in eRO-QPE1, that instead shows QPEs of different amplitudes and spacing in time \citep{2024ApJ...965...12C}, as well as the presence of both isolated QPEs and overlapping ones \citep{2022A&A...662A..49A}. The QPE activity in eRO-QPE1 appears to have significantly weakened over the course of about four years, as shown by  \citet{2024ApJ...965...12C} \citep[see also][]{2024ApJ...963L..47P}.
The quiescent emission of eRO-QPE3 fell under the detection limits about one year after its first detection, however QPEs have been observed also at later times. These have small amplitudes and irregular time separations  \citep{2024A&A...684A..64A}.
The case of eRO-QPE4 appears more similar to RX~J1301.9+2747 in terms of properties of QPEs: it shows in fact both weak and strong QPEs with varying time separations, although of about twice the duration and one order of magnitude higher luminosity of those of RX~J1301.9+2747 \citep[see Fig. 13 of][]{2024A&A...684A..64A}.
For Tormund and J0249 the QPE association is less secure. 
In the first case a portion of a flare with spectral properties very similar to QPEs has been observed about six months after the optical TDE \citep{2023A&A...675A.152Q}. In the second case, one and a half QPE-like flares were observed in archival \textit{XMM-Newton} observations, but no flaring emission was detected in follow-up observations with the same satellite. Interestingly, the quiescent emission of J0249 appears to have developed a harder X-ray spectral component in the 15 years elapsed between the two long \textit{XMM-Newton} observations, on the last of which QPEs were not present \citep{2021ApJ...921L..40C}. 
It might be possible that the presence of a strong X-ray hot corona somehow inhibits the presence of QPEs. For example, in the context of stellar EMRI impact scenarios \citep[e.g.,][]{2023ApJ...957...34L,2023MNRAS.526...69T}, the impacting star might not survive the strong ablation expected if exposed to sustained hard X-rays.

\subsection{QPEs and other repeating X-ray transient phenomena\label{discu:rnt}}

Repeating nuclear transients are not limited to QPEs: in the past few years several extragalactic sources have been observed displaying repeating patterns of flaring emission in the X-ray band. 
These include ESO 243-49 HLX-1 \citep{2009Natur.460...73F}, ASASSN-14ko \citep{2021ApJ...910..125P}, eRASSt J0456-20 \citep{2023A&A...669A..75L}, and Swift J0230+28 \citep{2023NatAs...7.1368E,2024NatAs.tmp...15G}.
However, either the timescales between flares, the X-ray spectral properties, or the detailed spectral-timing evolution of the X-ray flares observed in the afore-mentioned sources differ substantially from the case of QPE sources.

The timescales between flares are much longer in HLX-1 \citep[about one year between bursts of X-ray emission, e.g.,][]{2011ApJ...743....6S,2015ApJ...811...23Y}, eRASSt J0456-20 \citep[about 200 days,][]{2023A&A...669A..75L}, and ASASSN-14ko \citep[about 100 days, e.g.,][]{2021ApJ...910..125P} than in QPE sources (between 2 and 20 hours). The source Swift J0230+28 has the shortest time separation among these nuclear transients \citep[about 22 days,][]{2023NatAs...7.1368E,2024NatAs.tmp...15G}, still being much longer than the typical time separation between QPEs. 
The X-ray spectral properties of HLX-1, ASASSN-14ko, and eRASSt J0456-20 are substantially different to those of QPE sources. In fact they have a well-developed hard X-ray power law, which is negligible or absent in QPE sources.  Swift J0230+28 is again the source most similar to QPEs, having a purely thermal X-ray spectrum with no signal above $\sim 2$ keV. However, the detailed spectral-timing evolution of the X-ray flares observed in Swift J0230+28 is different to the one of QPEs. In Swift J0230+28 in fact the temperature of the X-ray spectra smoothly increases during the flare evolution, until reaching a maximum at the end of the flare. This is opposite to what is observed in all the QPE sources, that show instead a decreasing temperature along the flare evolution.
The detailed flare shape is also substantially different in QPE sources and in these repeating nuclear transients. In particular, eRASSt J0456-20 has a much slower rise than decay, and a plateau in between them; Swift J0230+28 has a slightly slower rise than decay; and HLX-1 has fast rise, exponential decay flares with much faster rise than QPEs. ASASSN-14ko shows most of its quasi-periodic variability in the optical/UV band, with flares with varying shape in a few years, from slower rise/fast decay to more symmetric \citep{2023ApJ...956L..46H}. 
Repeating nuclear transients are generally interpreted as partial TDEs. The difference in physical properties of the X-ray flares compared to QPEs might reflect a different physical configuration for the partial TDE, or a different physical mechanism.

\section{Conclusions\label{sec:conclu}}
  
We presented the analysis of five \textit{XMM-Newton} and three VLA observations of the QPE-emitting galaxy RX~J1301.9+2747, complemented by archival \textit{EXOSAT}, \textit{ROSAT}, and \textit{Chandra} X-ray observations.
Based on the results of this work, any model that aims at reproducing the observed properties of RX~J1301.9+2747 should take into account the following:
\begin{description}
\item[$\bullet$] X-ray QPEs have been observed persistently in all the five \textit{XMM-Newton} observations between 2000 and 2022. The EPIC cameras have detected 34 QPEs. Of these, 8 QPEs are weak and 26 are strong, with a difference in count rate of a factor of $\sim 2$ (Fig.~\ref{FIG1}).\\
\item[$\bullet$] The time separation between the QPE peaks (recurrence time) has a large spread around an average value of 3h~45m (13.5 ks); the minimum is 1h~17m (4.6 ks), and the maximum is 6h~05m (21.9 ks).
The X-ray QPEs of RX~J1301.9+2747 are irregularly spaced in time, but not fully chaotic. 
Strong QPEs happen after a longer recurrence time compared to weak QPEs, while no weak QPE is observed whenever the recurrence time is larger than 1 h~40m (6 ks).
Weak QPEs are not always present, as there are long phases ($>\,$20~h) when only strong QPEs are observed,  in which case QPEs appear to be more regularly spaced. The recurrence time consistently alternates between long and short; in phases where weak QPEs are not present, this alternation is less dramatic in amplitude (Fig.~\ref{FIG3}).
The higher the QPE amplitude, the longer appears to be its duration.\\
\item[$\bullet$] The properties of QPEs as a function of energy (arrival time, duration, intensity) are similar to the properties of other QPE sources and QPE candidates: at higher energies the QPE peaks arrive earlier (with a difference of up to 10 minutes between the $0.2-0.4$ keV and the $1-2$ keV bands in the case of RX J1301.9+2747), last less (with a difference in duration of a factor $2\times$ between low and high energies), and have a much lower count rate (Fig.~\ref{FIG4}). This energy dependence results in a 'hysteresis-like' behavior: the QPE spectral evolution consists of a hard rise followed by a soft decay (because QPEs last less at higher energies), and the peak temperature is reached well before the peak luminosity (because QPEs at higher energies arrive earlier and are less energetic than QPEs at lower energies; see  Fig.~\ref{FIG5} and \ref{FIGB2}).\\ 
\item[$\bullet$] The X-ray quiescent emission of RX~J1301.9+2747 is well-detected at all epochs of observation, varying of a factor $\sim 2\times$ in luminosity during 40 years. The quiescent spectra of RX J1301.9+2747 are super-soft (Fig.~\ref{FIGC1}) and well-represented by the emission of a disk with inner temperature $\sim 50-60$~eV and $L_{BOL}\sim 1-2\times 10^{42}$~erg~s$^{-1}$ ($L_{0.3-2\, \rm{keV}}\sim {0.8-2}\times 10^{41}$ erg s$^{-1}$). 
The inner temperature of the disk slightly decays throughout the 20 years covered by \textit{XMM-Newton} observations. However, its luminosity fluctuates non-monotonically (Fig.~\ref{FIG7}).
After 2019, there is the emergence of a harder spectral component in quiescence, which carries about $1/4$ of the $0.3-2$ keV disk luminosity (Fig.~\ref{FIG6} and \ref{FIGC2}). This component can be modeled with a thermal model with a temperature $kT\sim 180$ eV, with a power law with photon index $\Gamma\gtrsim 3.5$, or with thermal Comptonization of the disk photons on a plasma with optical depth $\tau=10$ and temperature $kT\sim 400$ eV.
The $2-10$ keV bolometric correction $\kappa_{2-10\, \rm{keV}}> 2700$ is very large as the hard X-ray power law is, if present, very weak with $L_{2-10\, \rm{keV}}< 4\times 10^{38}$ erg s$^{-1}$.\\
\item[$\bullet$] The QPE spectra are super-soft with no significant detection at $E>2$ keV. They are well-represented by a blackbody or a Comptonization model; in comparison, a bremmstrahlung model is slightly less-favored, and a power law model is strongly disfavored. If compared to a blackbody model, the QPE peak bolometric luminosity is $\sim 2\times 10^{42}$ erg~s$^{-1}$ and the QPE temperature at the peak luminosity is $130$ eV. Weak QPEs have generally a slightly lower duration ($\sim 35$~m vs $\sim 45$~m), a slightly lower temperature ($\sim 120$~eV), and a lower bolometric luminosity ($\sim 10^{42}$~erg~s$^{-1}$) at their peak compared to strong QPEs. 
The equivalent blackbody radius $R_{bb}$ increases from $\sim 10^{10}$ cm during rise-1 to $\sim 4\times 10^{10}$ cm during decay-2 (Fig.~\ref{FIG11}, lower panel). Weak QPEs have systematically smaller $R_{bb}$ throughout their evolution.
QPEs happening after a longer recurrence time appear to last more and there is a weak positive correlation between the energy emitted by each QPE and the time preceding it (Fig.~\ref{FIG12}).\\
\item[$\bullet$] Compared to the QPEs of the best-studied source GSN 069, the QPEs of RX J1301.9+2747 last for a shorter period of time, are less luminous, and are hotter (Fig.~\ref{FIG14}). In particular, the maximum difference in QPE temperature is during the rise, while the QPE temperatures of the two sources converge during the decay (Fig.~\ref{FIG15}). In RX J1301.9+2747 the decrease in temperature appears to be monotonic from the QPE rise-1 to the QPE decay-2, from $\sim 160$ eV to $\sim 80$ eV (Fig.~\ref{FIG11}, top panel).\\
\item[$\bullet$] There is no correlated radio/X-ray variability during the QPE emission in 2020 or 2022B. Whilst there appears to be a small amount of radio variability associated with the QPE in 2022A, its significance is $< 2\sigma$.
Between epochs of observations of RX J1301.9+2747 the radio source does show variability on the order of 5--10$\%$ (Fig.~\ref{FIG17}). If attributed to ISS, the observed radio variability implies a very compact size of the source emission $< 0.008$~pc. This is comparable to the optical  and to the smallest radio-emitting regions in TDEs ($\sim 0.003$ pc).
The radio emission of RX J1301.9+2747 does not look like that of TDEs, which (when detected) display consistent long-term variability attributable to outflows, and may show emission at timescales $>1000$ days post-disruption \citep[e.g.][]{2020SSRv..216...81A,Goodwin2022,2023arXiv230813595C}.
\end{description}

Given the consistent alternance of short/long recurrence times between QPEs observed throughout our campaign, EMRI scenarios appear the most favoured in explaining the complex properties of RX J1301.9+2747. Different than the cases of much more regular QPE sources such as GSN 069 or eRO-QPE2, the presence of large differences in consecutive recurrence times between the QPEs of RX~J1301.9+2747 is however extremely difficult to explain in any scenario requiring a single interaction per orbit. 
EMRI models assuming impacts on the accretion disk appear promising, but still cannot fully account for the complexities of the timing behaviour in RX~J1301.9+2747. 

Future observations of both the quiescent emission and the QPEs of RX~J1301.9+2747 are necessary to try to solve this cosmic puzzle. 
High-cadence monitoring of the quiescent emission on long timescales might help identifying dynamical effects taking place in the nucleus of this galaxy, while long, uninterrupted light curves are necessary to try to break the enigma of the repetition pattern of its X-ray QPEs.

\begin{acknowledgements}
We thank the referee for their constructive criticism that allowed us to improve our results and their presentation.
MG is supported by the ``Programa de Atracci\'on de Talento'' of the Comunidad de Madrid, grant numbers 2018-T1/TIC-11733 and 2022-5A/TIC-24235.
GM is supported by Grant PID2020-115325GB-C31 funded by MCIN/AEI/10.13039/501100011033.
RA and PK acknowledge support from NASA through the NASA Hubble Fellowship grant HST-HF2-51534.001-A awarded by the Space Telescope Science Institute, which is operated by the Association of Universities for Research in Astronomy, Incorporated, under NASA contract NAS5-26555.
MB acknowledges support provided by MUR under grant ``PNRR - Missione 4 Istruzione e Ricerca - Componente 2 Dalla Ricerca all'Impresa - Investimento 1.2 Finanziamento di progetti presentati da giovani ricercatori ID:SOE\_0163'' and by University of Milano-Bicocca under grant ``2022-NAZ-0482/B''.
XS is supported by NSFC through grant No. 12192221. GP acknowledges financial support from the European Research Council (ERC) under the European Union’s Horizon 2020 research and innovation program Hot Milk (grant agreement No. 865637), from Bando per il Finanziamento della Ricerca Fondamentale 2022 dell’Istituto Nazionale di Astrofisica (INAF): GO Large program and from the Framework per l’Attrazione e il Rafforzamento delle Eccellenze (FARE) per la ricerca in Italia (R20L5S39T9).
We thank Chris Done for extending the \texttt{optxagnf} model to include negative black hole spin values.
The National Radio Astronomy Observatory is a facility of the National Science Foundation operated under cooperative agreement by Associated Universities,  Inc.
This work was supported by the Australian government through the Australian Research Council’s Discovery Projects funding scheme (DP200102471).  This research has made use of: data and/or software provided by the High Energy Astrophysics Science Archive Research Center (HEASARC), which is a service of the Astrophysics Science Division at NASA/GSFC; NASA's Astrophysics Data System Bibliographic Services (ADS); the NASA/IPAC Extragalactic Database (NED) which is operated by the Jet Propulsion Laboratory, California Institute of Technology, under contract with the National Aeronautics and Space Administration. 
Software used: \texttt{SAS} \citep{2004ASPC..314..759G}, \texttt{numpy} \citep{2020Natur.585..357H}, \texttt{Astropy} \citep{2013A&A...558A..33A,2018AJ....156..123A,2022ApJ...935..167A}, \texttt{matplotlib} \citep{Hunter:2007}, \texttt{Ultranest} \citep{2021JOSS....6.3001B}, \texttt{Xspec} \citep{1996ASPC..101...17A}, \texttt{pyXspec} \citep{2021ascl.soft01014G}, \texttt{BXA} \citep{2014A&A...564A.125B}.
Based on observations obtained with \textit{XMM-Newton}, an ESA science mission with instruments and contributions directly funded by ESA Member States and NASA.
\end{acknowledgements}

\bibliographystyle{aa}
\bibliography{biblio}

\begin{appendix}

\section{Observations log\label{appendix:obslog}}
\begin{table*}
\caption{Observation log for the \textit{XMM-Newton} EPIC exposures on RX~J1301.9+2747 between 2000 and 2022. \label{table:obslog}}
\centering
\begin{tabular}{cccc}
\hline \hline
Instrument & Start/End time &  total exposure & quiescent exposure \\
 &      (yyyy-mm-dd hh:mm:ss) & (s)  &  (s)\\
\hline
EPIC-pn & 2000-12-10 21:24:27/2000-12-11 04:00:20  & 29805& 17490 \\
EPIC-MOS1  &  2000-12-10 20:18:02/2000-12-11 03:58:46  & 27400 & 20820 \\
EPIC-MOS2  &  2000-12-10 21:24:27/2000-12-11 04:00:20  & 29805 & 20810 \\
\hline
EPIC-pn & 2019-05-30 20:42:24/2019-05-31 10:09:04  & 45158 & 20390\\
EPIC-MOS1  & 2019-05-30 20:42:24/2019-05-31 10:09:04  & 47015 & 32010\\
EPIC-MOS2  & 2019-05-30 20:42:24/2019-05-31 10:09:04 & 47000 & 29490 \\
\hline
EPIC-pn & 2020-07-11 18:14:03/2020-07-13 06:49:36  & 131651 &  48700\\
EPIC-MOS1  & 2020-07-11 17:48:21/2020-07-13 06:53:48  & 133514 &  81510  \\
EPIC-MOS2  & 2020-07-11 17:48:44/2020-07-13 06:53:49 & 133493 & 74660\\
\hline
EPIC-pn & 2022-06-17 19:47:19/2022-06-19 06:50:39  & 117927 &  37690  \\
EPIC-MOS1  & 2022-06-17 19:47:19/2022-06-19 06:50:39  & 102987 &  52030   \\
EPIC-MOS2 & 2022-06-17 19:47:19/2022-06-19 06:50:39  & 103141 & 50090  \\
\hline
EPIC-pn & 2022-06-19 19:40:51/2022-06-21 06:42:31  & 116421 & 55580   \\
EPIC-MOS1  &  2022-06-19 19:40:51/2022-06-21 06:42:31  & 114633 & 75660     \\
EPIC-MOS2  &  2022-06-19 19:40:51/2022-06-21 06:42:31  & 113290 & 72600 \\ 
\hline
\end{tabular}
\tablefoot{All exposures were taken in full frame mode, using the thin optical filter, and pointed at the source coordinates, except for the 2000 observation that was pointed at the Coma Cluster and used the medium optical filter. Dates of observations are in Coordinated Universal Time (UTC). The quiescent exposure time excludes both QPEs and time intervals of large background flaring.}
\end{table*}

\begin{table}[]
    \centering
    \begin{small}
    \begin{tabular}{ccc}
    \hline
   Telescope & Start time & Total exposure \\
   & (yyyy-mm-dd hh:mm:ss) & (s) \\
    \hline
        \textit{EXOSAT} & 1983-07-16 09:02:12 & $(...)$\\
        \textit{ROSAT} & 1991-06-16 22:46:37 & 21893\\
        \textit{ROSAT} & 1991-06-17 22:45:02 & 22183 \\
        \textit{Chandra CXC} & 2009-06-27 03:11:40 & 5050\\
    \hline
    \end{tabular}
    \caption{Log of the archival X-ray observations of RX~J1301.9+2747 used in this work.}
    \label{tab:otherobs}
    \end{small}
\end{table}

\begin{table}[]
    \centering
    \begin{tabular}{ccc}
    \hline
    Date & Frequency & Flux Density  \\
    (yyyy-mm-dd) & GHz & ($\mu$Jy) \\
    \hline
        2020-07-11 & 6 & 313$\pm$2 \\
        2020-07-12 & 6 & 344$\pm$3\\
        2022-06-17 & 6 & 334$\pm$3\\
        2022-06-19 & 6 & 340$\pm$2\\
    \hline
    \end{tabular}
    \caption{Radio flux density measurements of RX~J1301.9+2747 taken with the VLA. Errors reported are the statistical error of the Gaussian point source fit combined with the image rms, but we note that the absolute flux density scale accuracy of the VLA is limited to 5$\%$.}
    \label{tab:radio_fluxes}
\end{table}

\FloatBarrier
\section{Light curve fit to individual QPEs\label{appendix:lightcurves}}

We fit the EPIC light curves of RX~J1301.9+2747 deriving Bayesian posterior probability distributions with \texttt{UltraNest}.
The Bayes theorem states that:
$$ P(\theta|D)=\frac{\mathcal{L}(\theta)P(\theta)}{\mathcal{Z}}$$
where $\mathcal{L}(\theta)=P(D|\theta)$ is the likelihood that specifies the frequency of producing the dataset $D$ given the parameters $\theta$, $P(\theta)$ is the assumed prior, and $P(\theta|D)$ is the posterior PDF. The Bayesian evidence, or marginal likelihood $\mathcal{Z} = \int\mathcal{L}(\theta)P(\theta) d\theta$, is the posterior normalization \citep[e.g.,][]{2023arXiv230905705B}.

We compared the data to the model \texttt{[constant + Gaussian]}, comprised of a constant to represent the quiescent emission, plus Gaussian emission lines to represent the QPEs. We tried to assess if there is variability of the quiescent emission, but the results were inconclusive except for the first segment of the 2020 observation, where a clear decrease in the quiescent flux emission is observed, although of modest amplitude. We therefore assumed the quiescent count rate to be constant, although it may vary between different QPEs (i.e., the fit is performed locally around each QPE peak, with time intervals shown in the x-axis of each panel of Fig.~\ref{FIGB1}). The emission QPE profile was also fit to a profile akin to that used by \citet{2022A&A...662A..49A} to model the QPEs of eRO-QPE1, and no significant differences in terms of QPE peak position and QPE duration were found. We therefore used the simpler Gaussian to model the QPEs of RX~J1301.9+2747.
We used a Gaussian log-likelihood function, i.e., we assumed that the data were drawn from a Gaussian distribution, with mean equal to the model prediction and standard deviation equal to the measurement error.
The prior probability distributions for the quiescence count rate CR$_{quie}$, the QPE count rate CR$_{qpe}$ and width $\sigma_{qpe}$ were assumed to be flat and uninformed, with uniform distributions between $[10^{-7}-10^{-1}]$ ct s$^{-1}$, $[0.01-2]$ ct s$^{-1}$, and $[100-1000]$ s, respectively. The QPE peak position $t_{peak}$ was also assumed to have a flat prior probability distribution within a time interval visually centered around the peak and typically lasting between 2000 and 4000 seconds. The analysis was performed on the pn data and on the merged MOS1 and MOS2 data simultaneously, assuming that $t_{peak}$ is the same for the three instruments.
Results of the fit are shown in Fig.~\ref{FIGB1}, in light blue for the pn data and in dark blue for the MOS data, and reported in Table~\ref{table:QPEprop}. 

\begin{figure*}[ht!]
\centering
\includegraphics[ width=4.5cm]{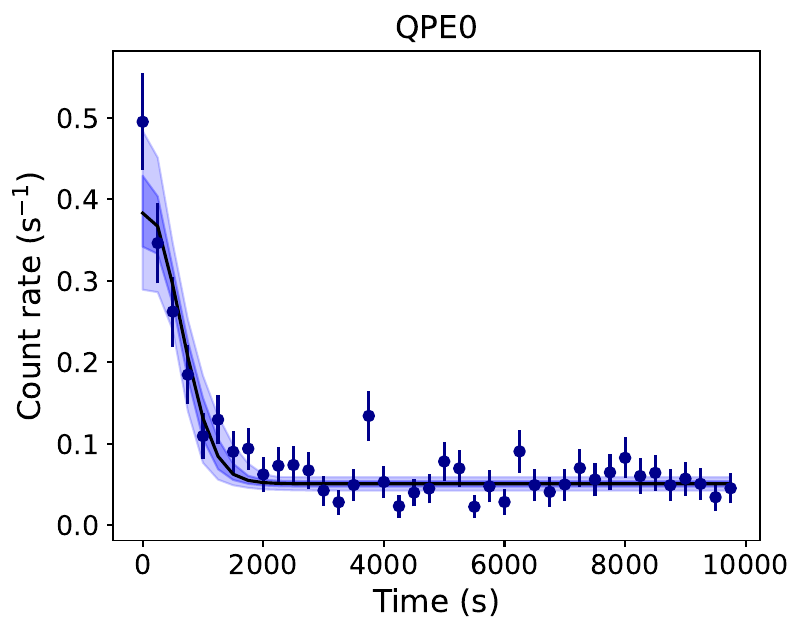}
\includegraphics[ width=4.5cm]{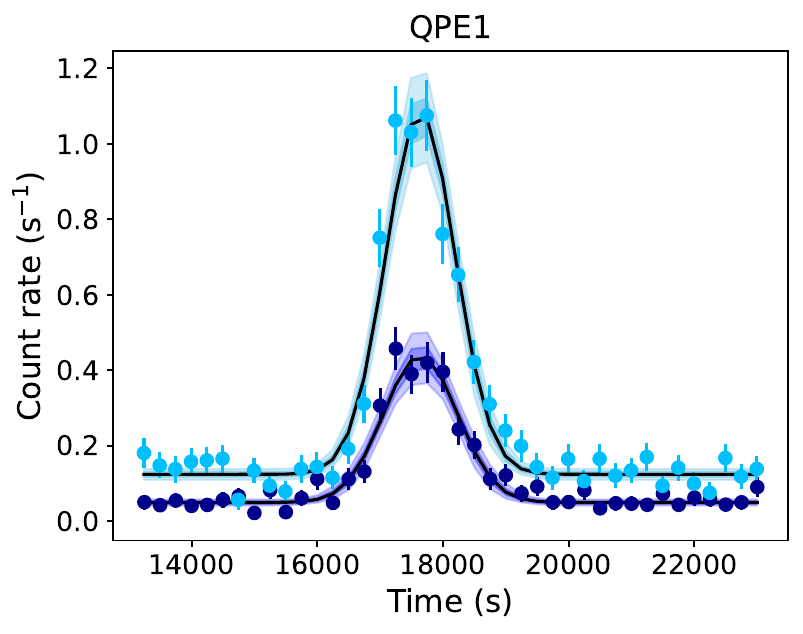}
\includegraphics[ width=4.5cm]{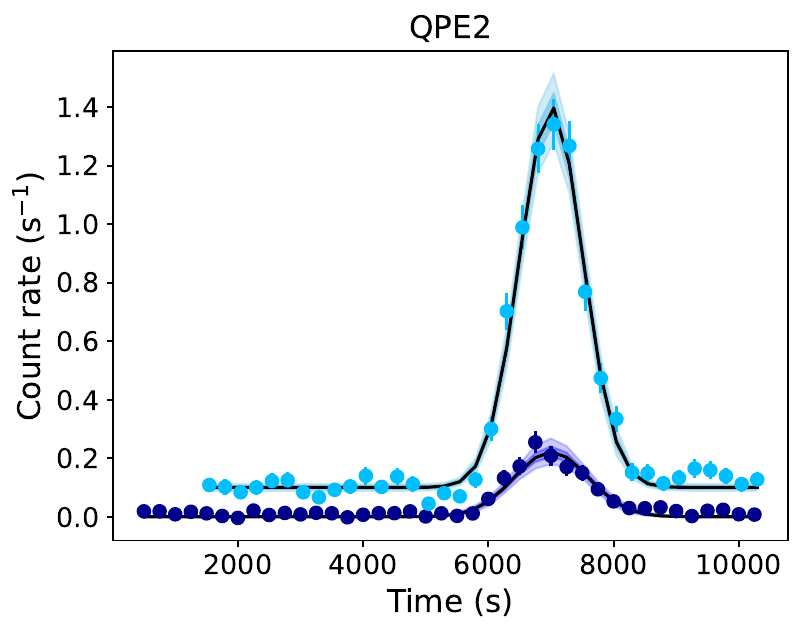}
\includegraphics[ width=4.5cm]{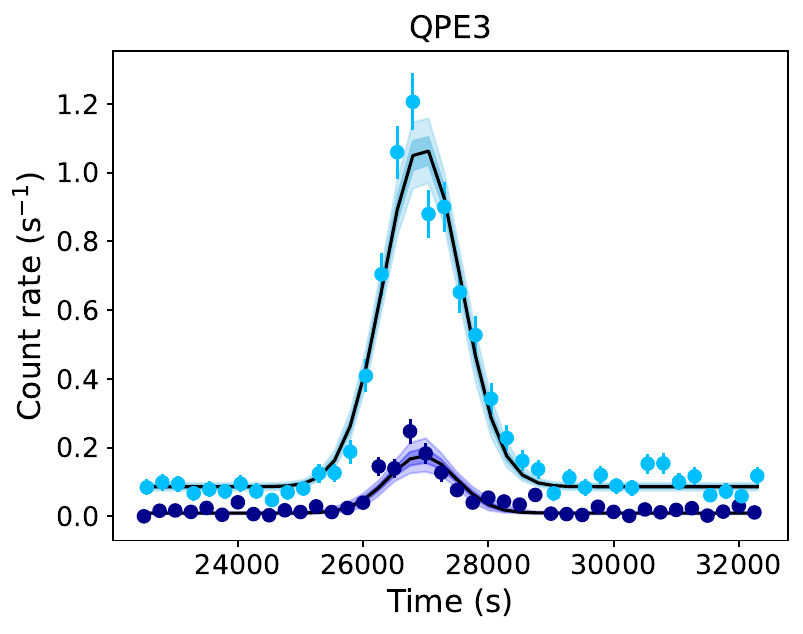}
\includegraphics[ width=4.5cm]{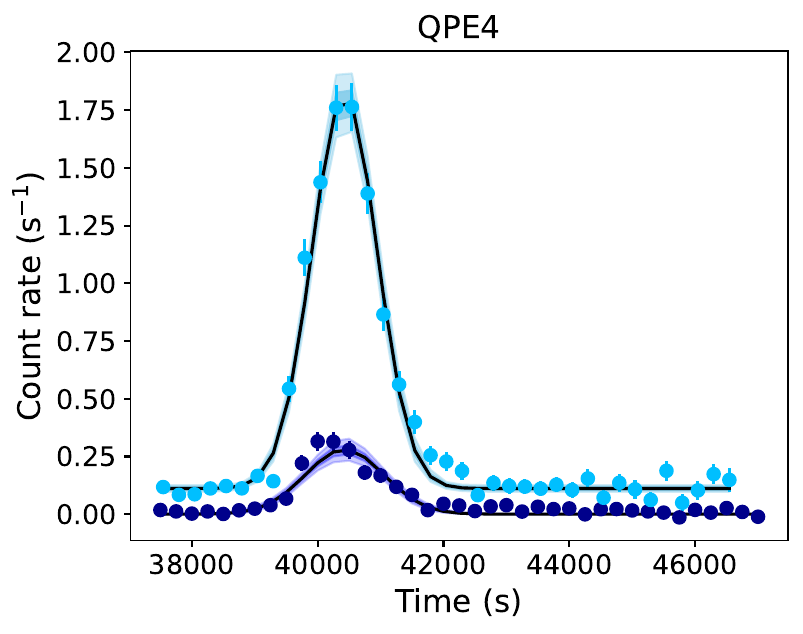}
\includegraphics[ width=4.5cm]{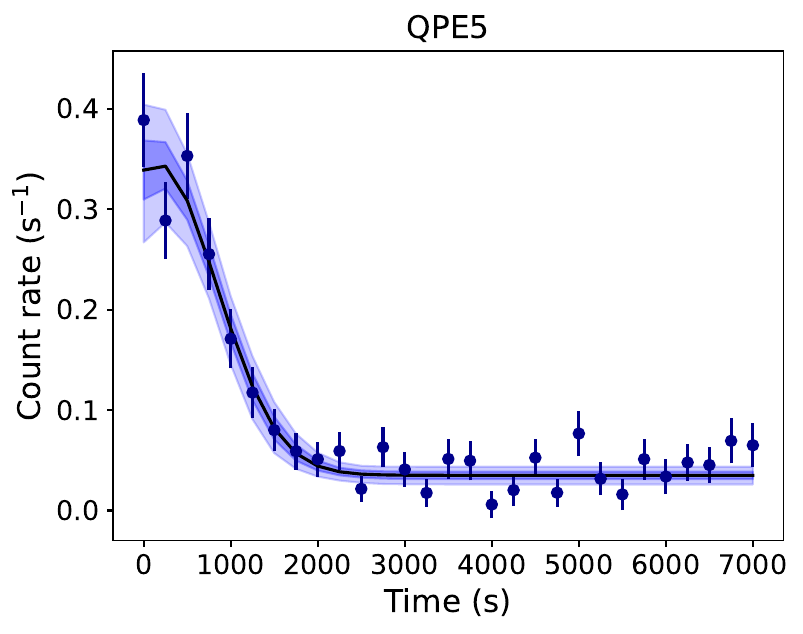}
\includegraphics[ width=4.5cm]{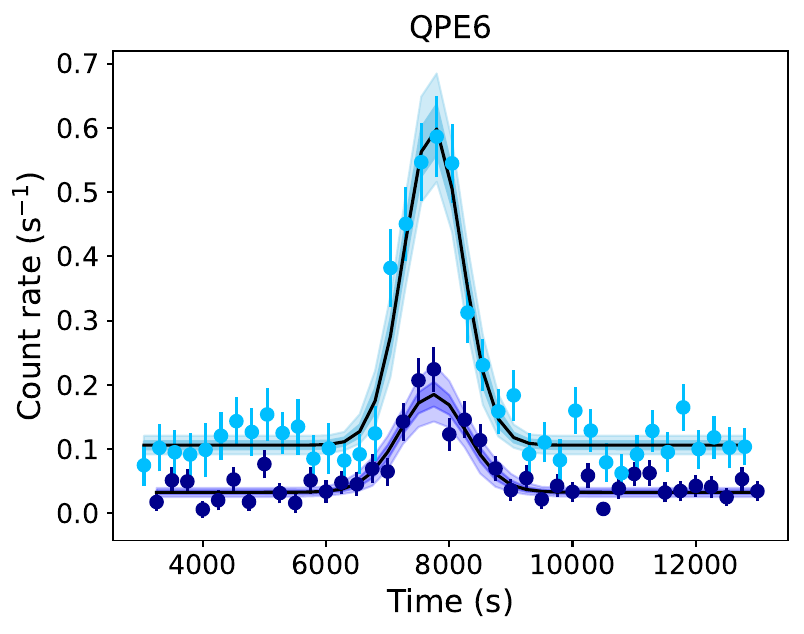}
\includegraphics[ width=4.5cm]{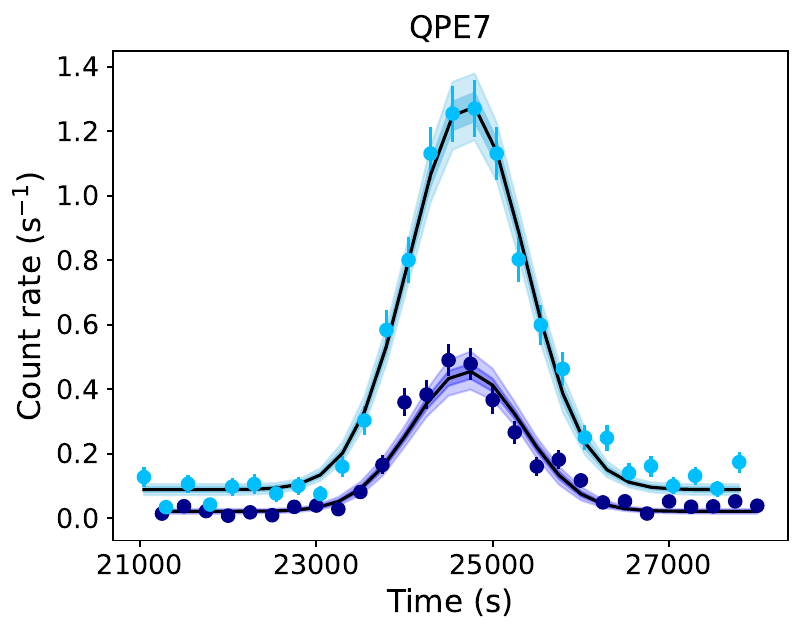}
\includegraphics[ width=4.5cm]{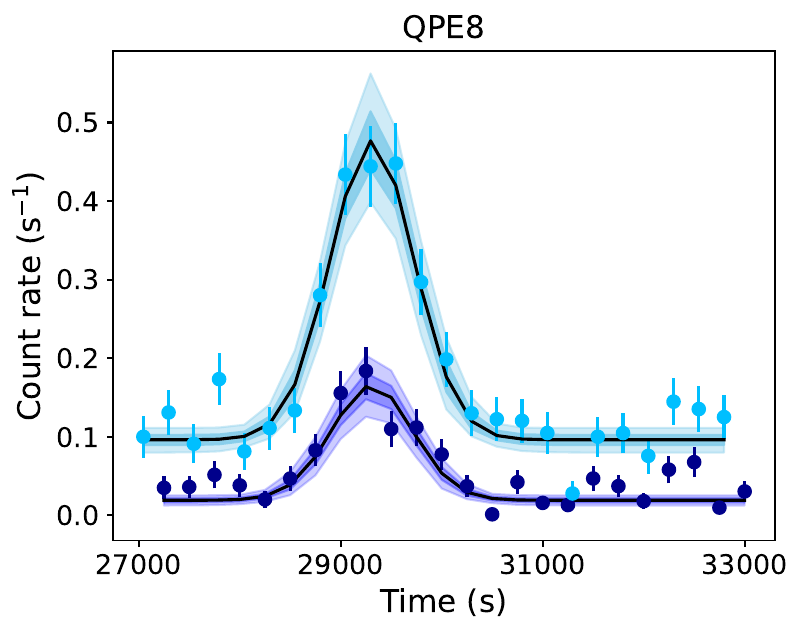}
\includegraphics[ width=4.5cm]{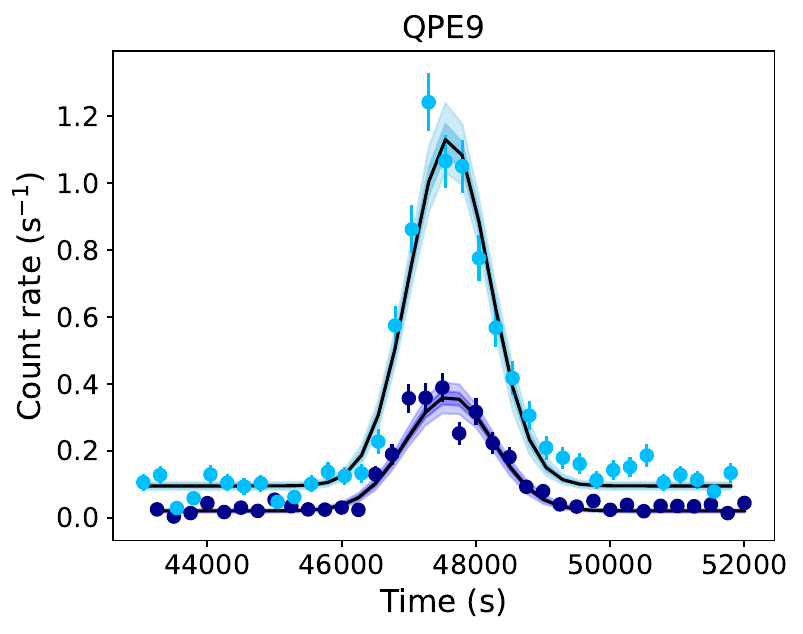}
\includegraphics[ width=4.5cm]{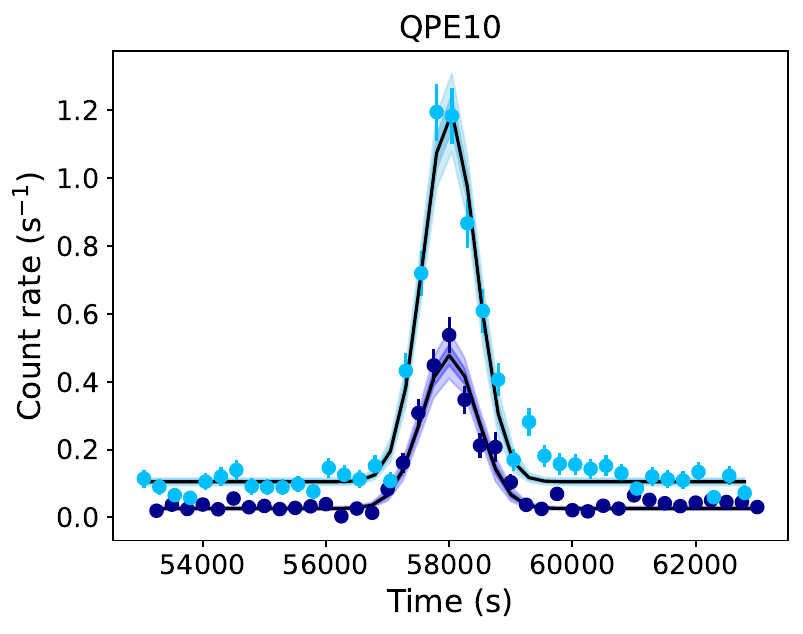}
\includegraphics[ width=4.5cm]{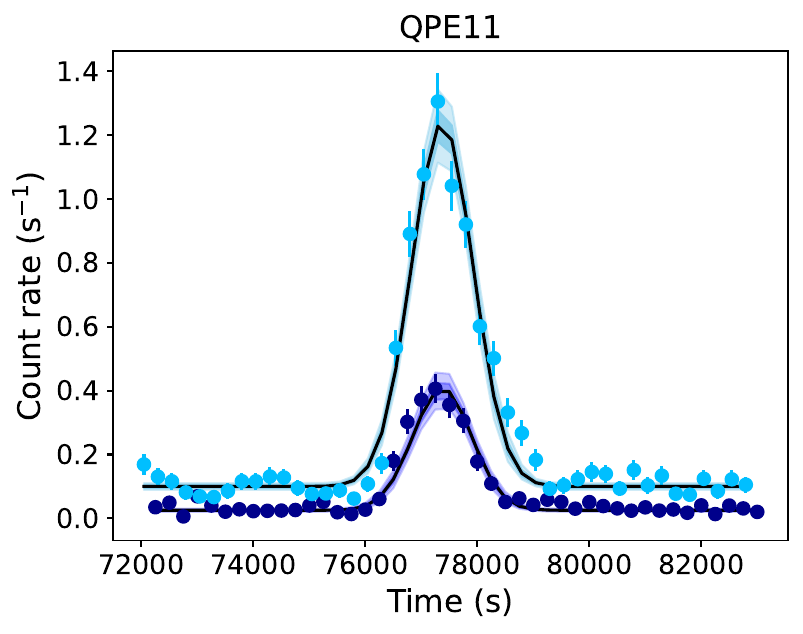}
\includegraphics[ width=4.5cm]{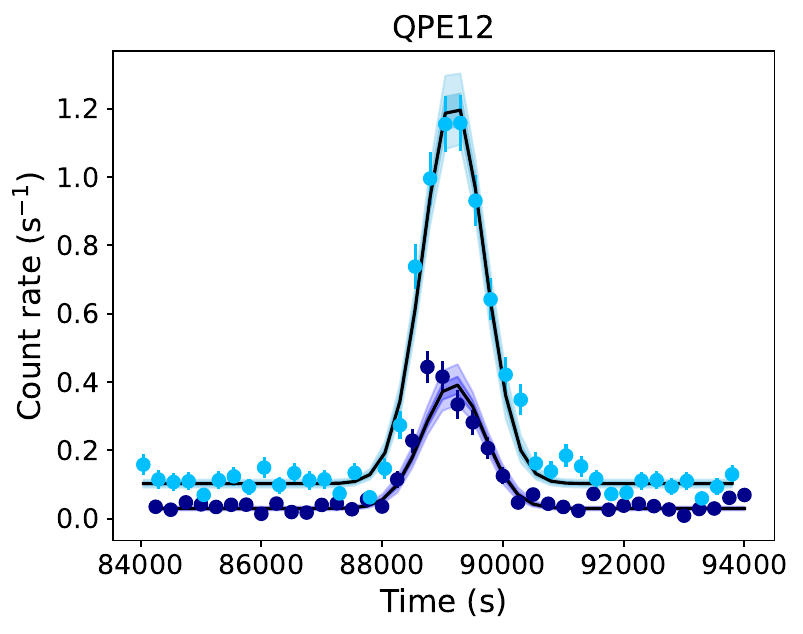}
\includegraphics[ width=4.5cm]{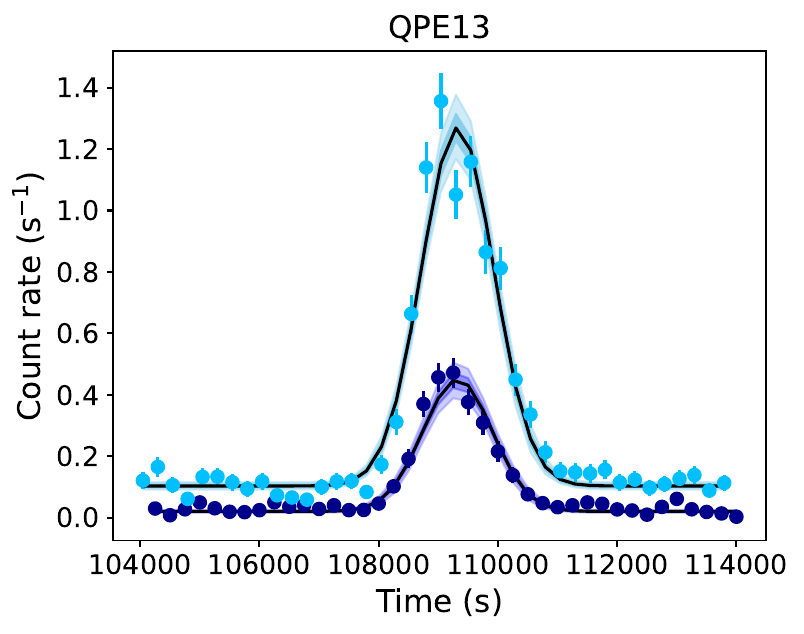}
\includegraphics[ width=4.5cm]{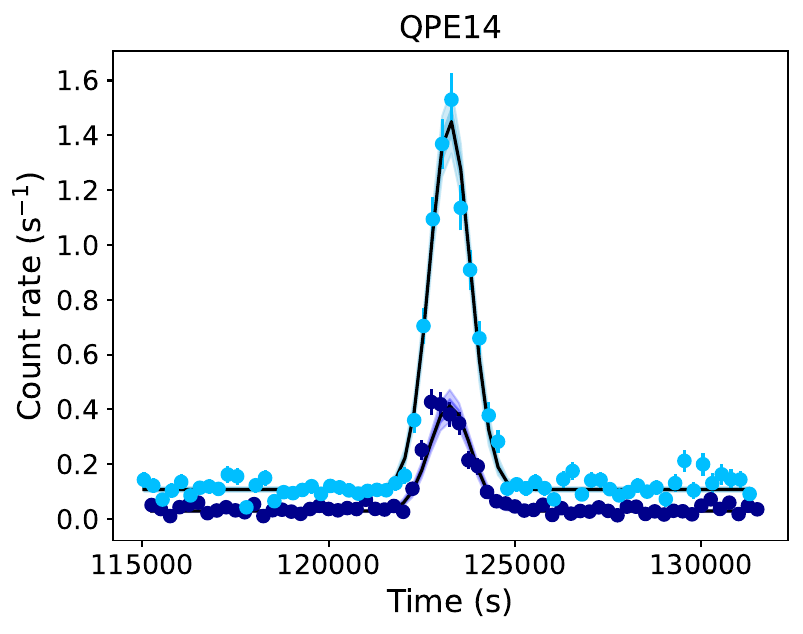}
\includegraphics[ width=4.5cm]{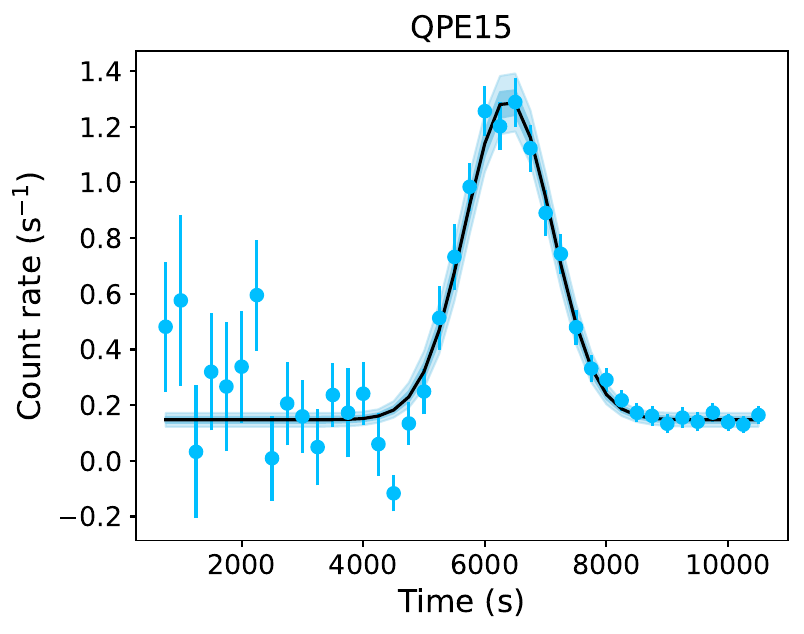}
\includegraphics[ width=4.5cm]{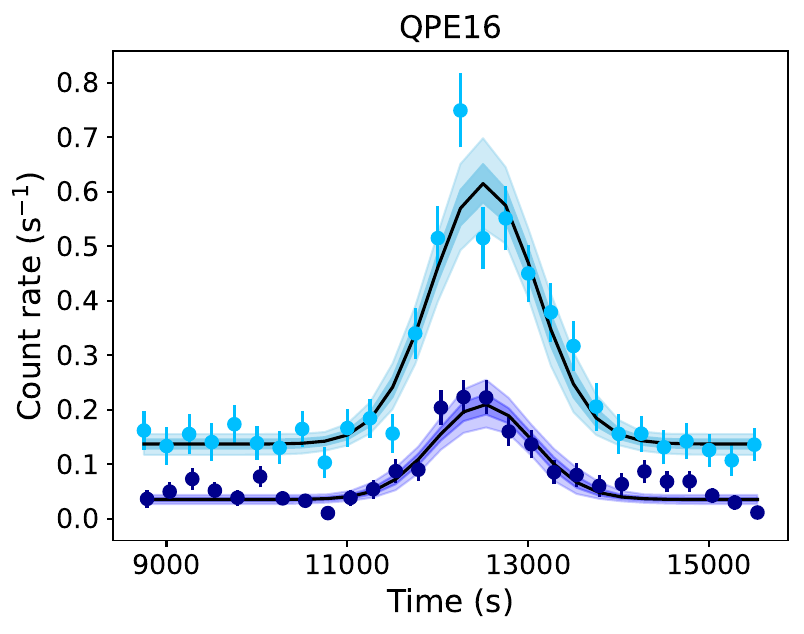}
\includegraphics[ width=4.5cm]{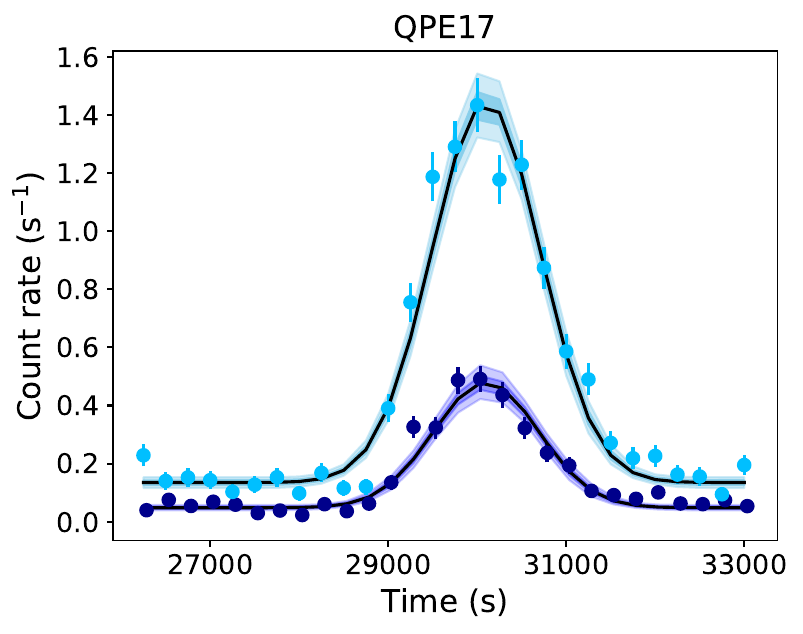}
\includegraphics[ width=4.5cm]{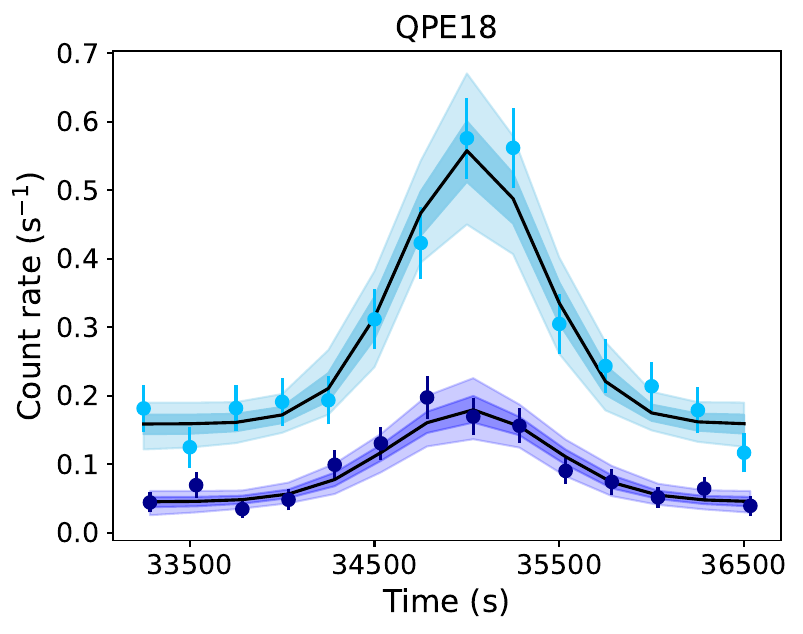}
\includegraphics[ width=4.5cm]{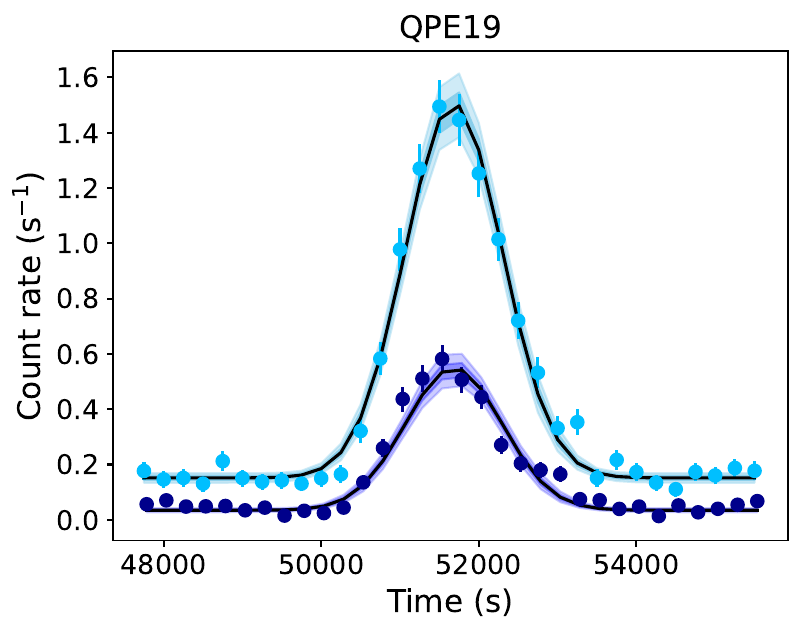}
\includegraphics[ width=4.5cm]{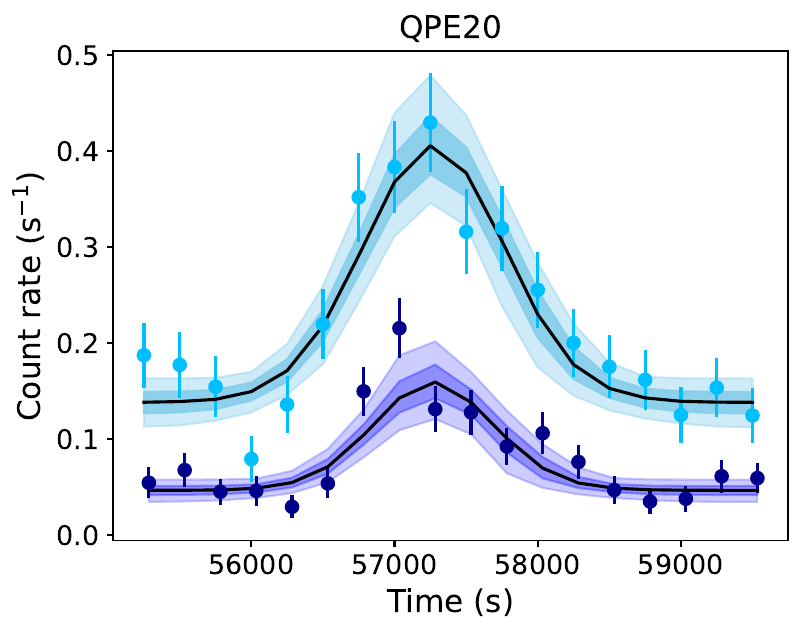}
\includegraphics[ width=4.5cm]{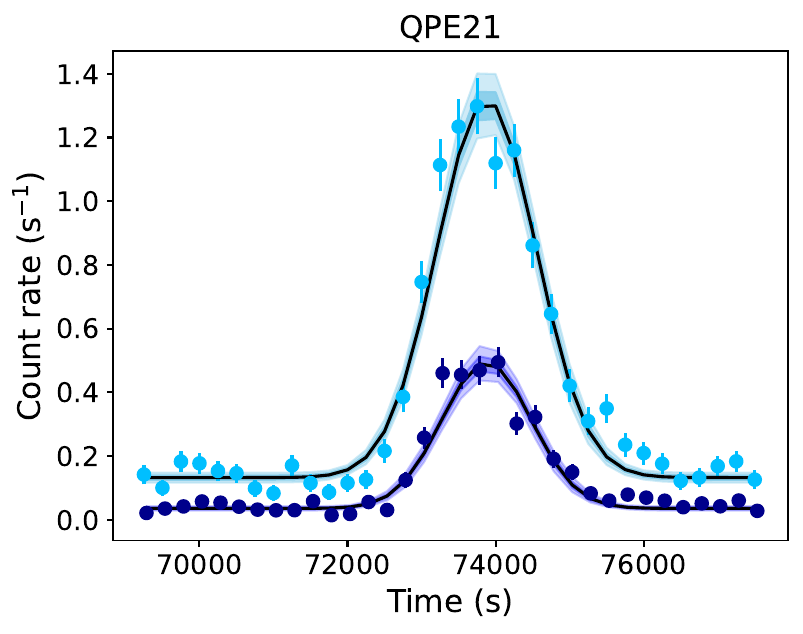}
\includegraphics[ width=4.5cm]{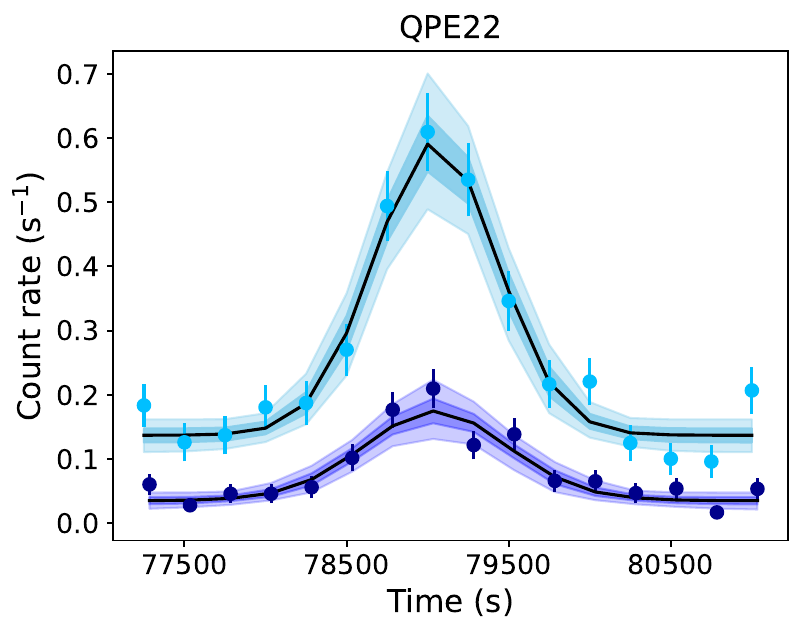}
\includegraphics[ width=4.5cm]{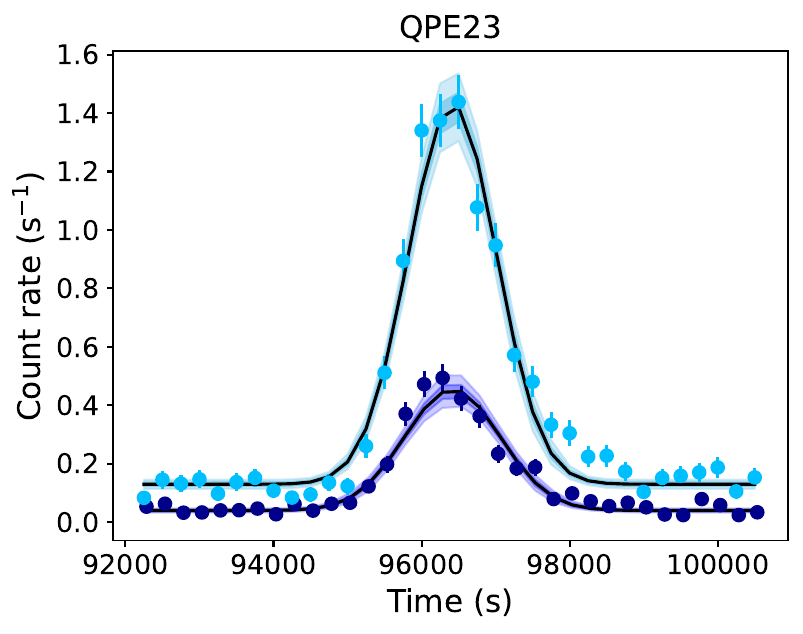}
\caption{Zoom on the light curves of RX~J1301.9+2747 centered on the QPEs detected by the EPIC cameras (pn in light blue, MOS in dark blue) between 2000 and 2022. The black line represent the median of the posterior PDF of the model \texttt{[constant + Gaussian]} used to fit the light curves, while the shaded areas represent the 1$\sigma$ and 2$\sigma$ equivalent percentiles. Time on the x-axis is computed from the beginning of each observation in Mission Reference Time: 92866770 s for QPE0$-$QPE1; 675637593 s for QPE2$-$QPE4; 710876963 s for QPE5$-$QPE14; 771890338 s for QPE15$-$QPE24; and 772064151 s for QPE25$-$QPE33.} \label{FIGB1} 
\end{figure*}
\begin{figure*}[ht!]\ContinuedFloat
\centering
\includegraphics[ width=4.5cm]{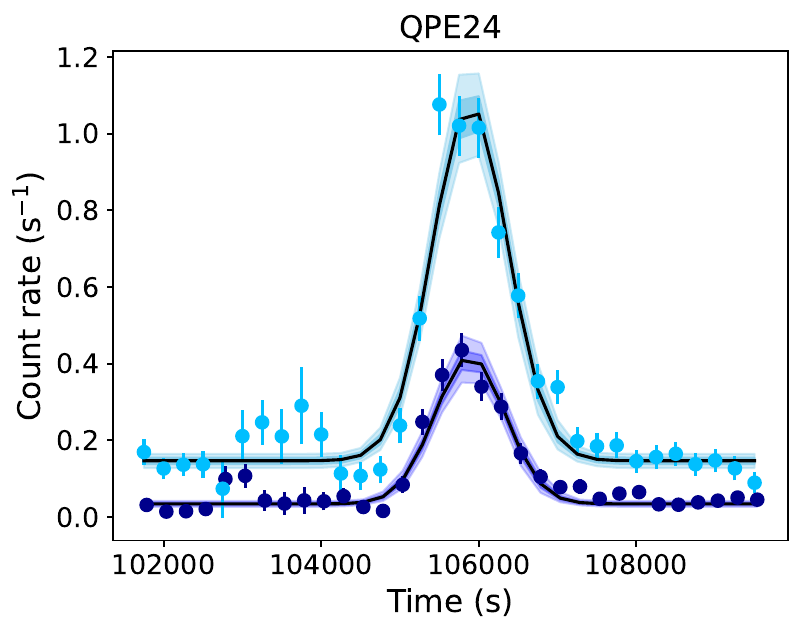}
\includegraphics[ width=4.5cm]{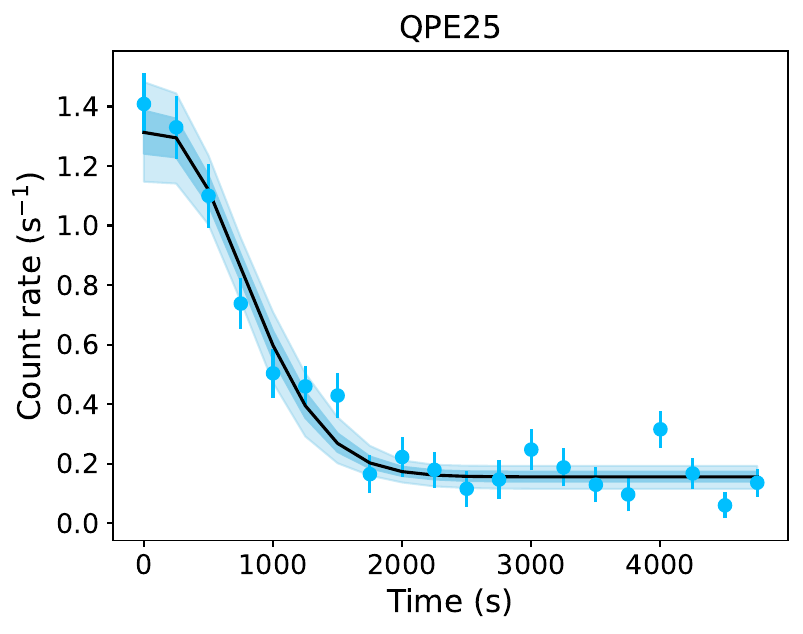}
\includegraphics[ width=4.5cm]{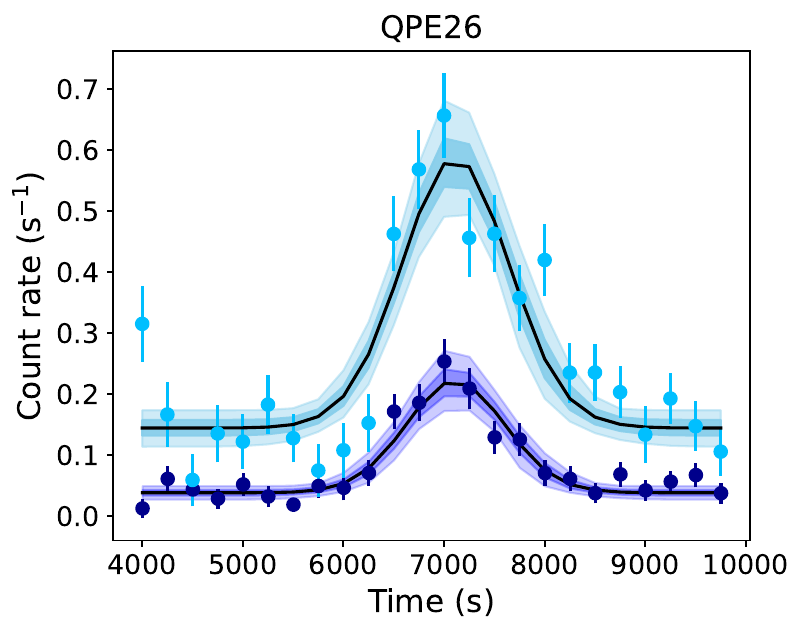}
\includegraphics[ width=4.5cm]{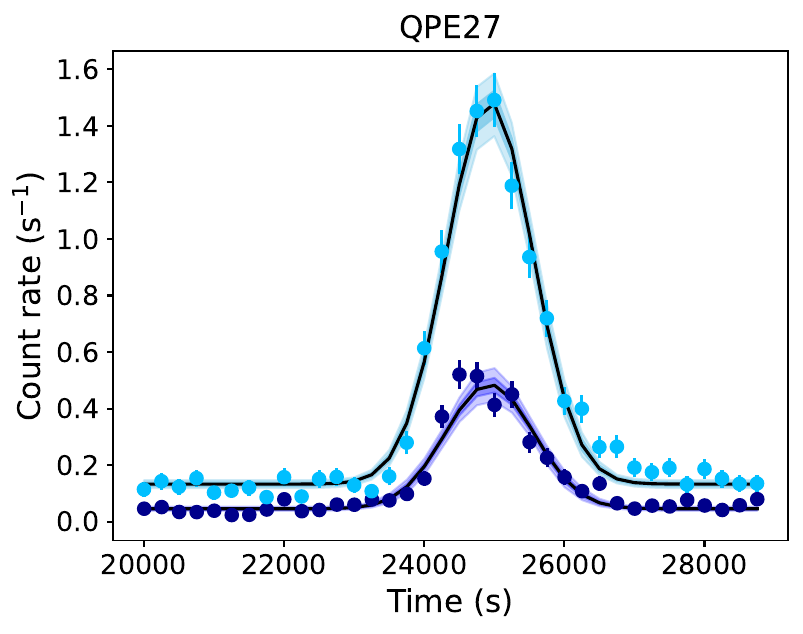}
\includegraphics[ width=4.5cm]{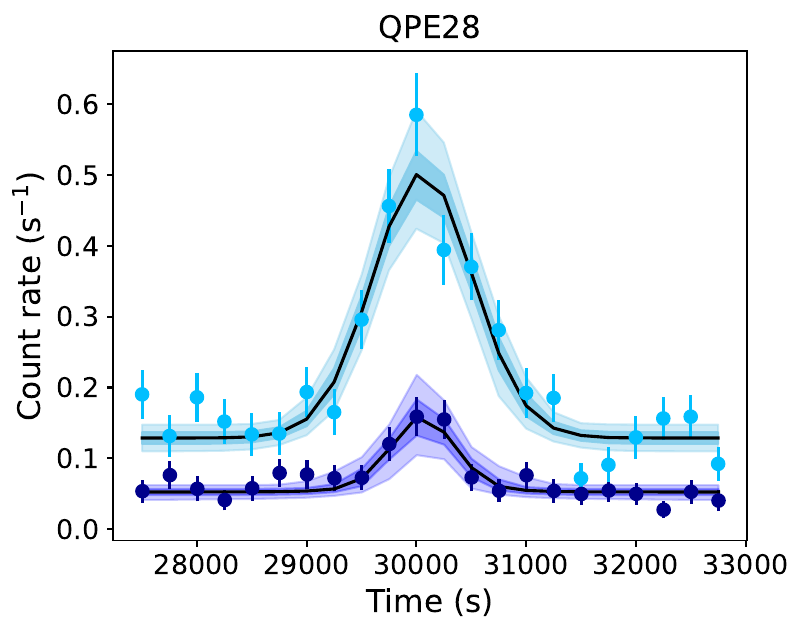}
\includegraphics[ width=4.5cm]{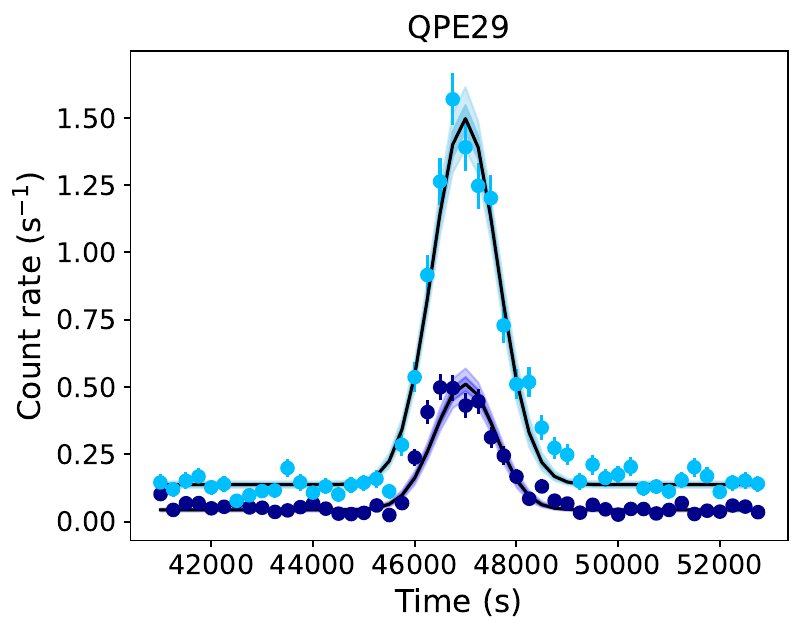}
\includegraphics[ width=4.5cm]{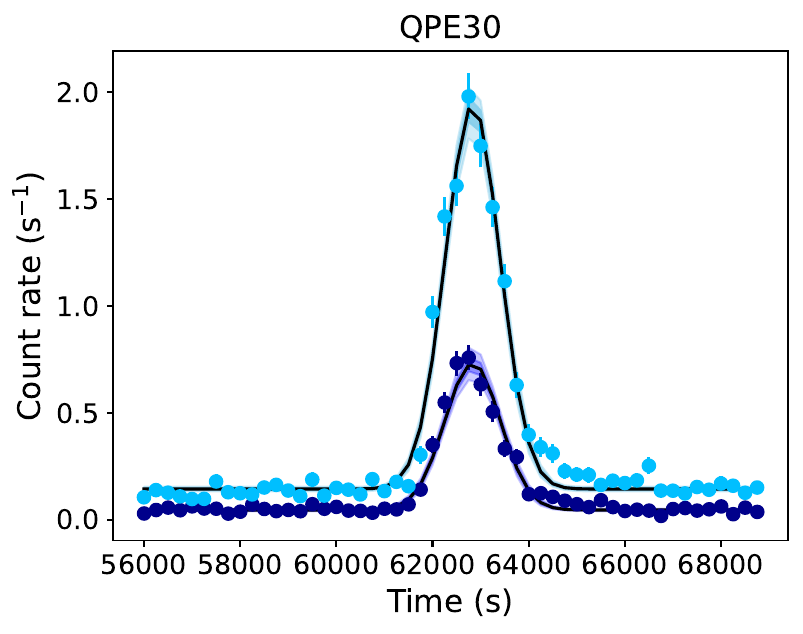}
\includegraphics[ width=4.5cm]{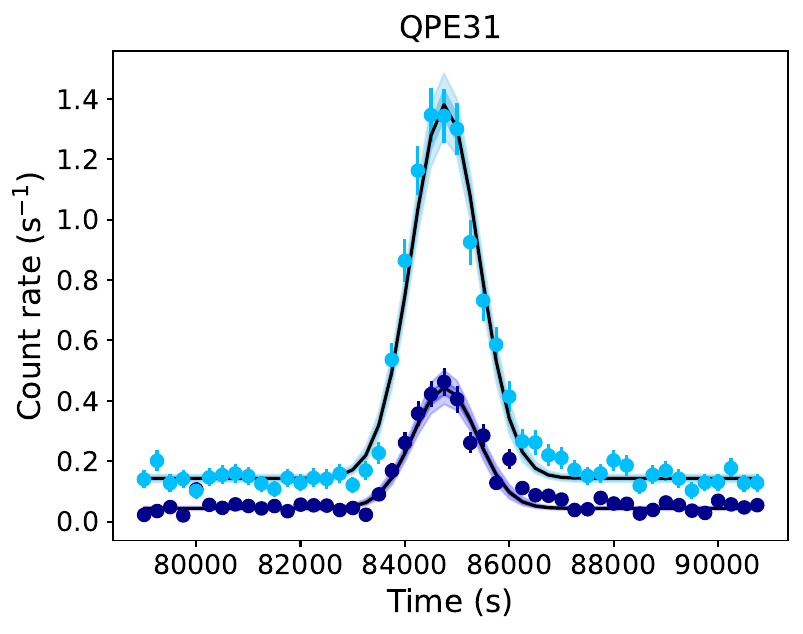}
\includegraphics[ width=4.5cm]{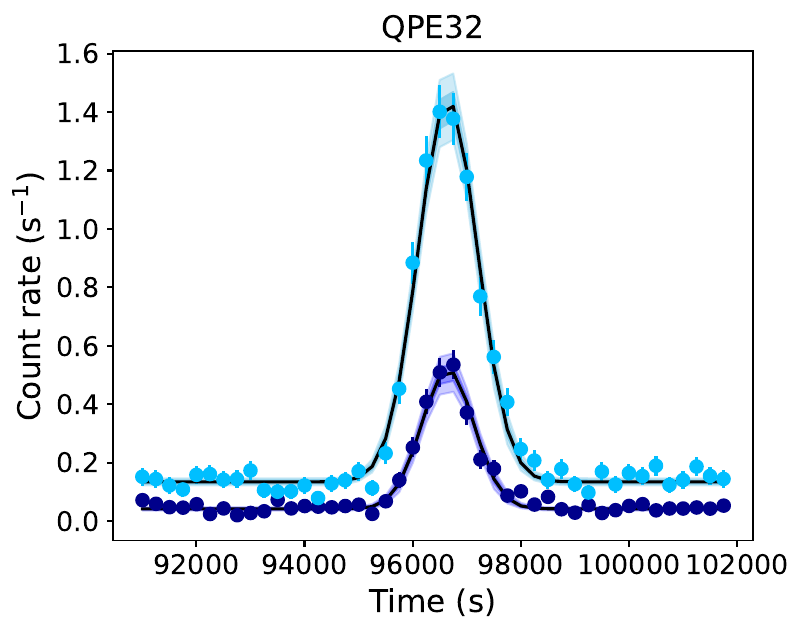}
\includegraphics[ width=4.5cm]{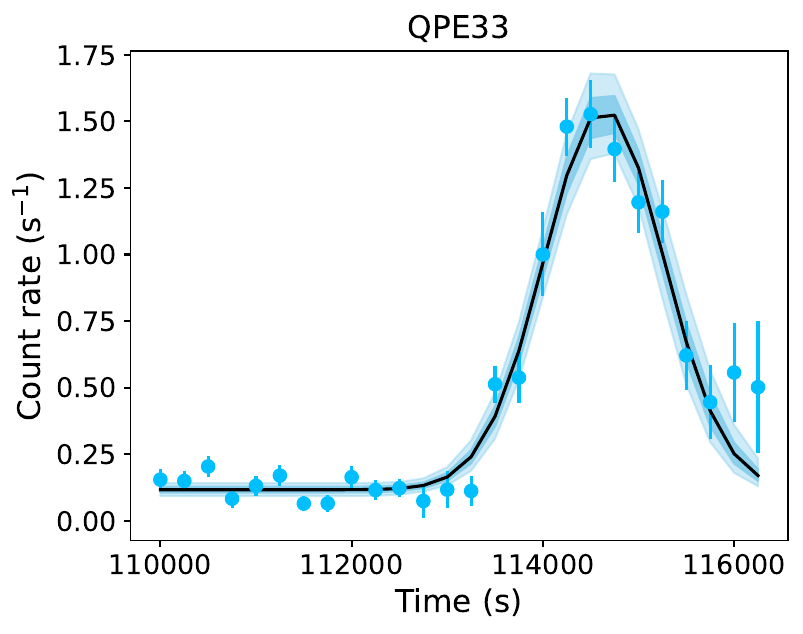}
\caption{-- continued.}
\end{figure*}

\begin{table*}
\caption{QPE properties\label{table:QPEprop}}
\centering
\begin{tabular}{cccccccc}
\hline \hline
(1) & (2) & (3) & (4) & (5) & (6) & (7) & (8) \\
QPE  &  $t_{peak}$ & CR$_{QPE}$ & $\sigma$ & Duration & $kT^{peak}$ & $L_{0.3-2\, \rm{keV}}^{peak}$ & $L_{\rm{BOL}}^{peak}$\\
 & (s) & (s$^{-1}$) & (s) & (m) & (eV) & ($10^{42}$ erg s$^{-1}$) & ($10^{42}$ erg s$^{-1}$) \\
\hline
0$^{(*)}$  & $92866819\pm{50}$ & $0.34\pm{0.04}$ & 
$557\pm{86}$ &  $43.7\pm{0.5}$ & $(...)$ & $(...)$ & $(...)$\\ 
1  & $92884412\pm{24}$ & $0.96\pm{0.05}$ & $551\pm{25}$ & $43.3\pm{0.4}$ & $133\pm{6}$ & $1.46\pm{0.12}$ & $1.90\pm{0.18}$\\ 
\hline
2  & $675644600\pm{17}$ & $1.30\pm{0.05}$ & $504\pm{14}$ & $39.5\pm{0.2}$  & $141\pm{6}$ & $1.76\pm{0.12}$  & $2.21\pm{0.17}$\\ 
3  & $675664534\pm{23}$ & $0.99\pm{0.04}$ & $617\pm{21}$ & $49.0\pm{0.4}$ & $126\pm{5}$ & $1.40\pm{0.09}$ & $1.87\pm{0.15}$\\ 
4  & $675678021\pm{15}$ & $1.71\pm{0.06}$  & $516\pm{14}$ & $40.5\pm{0.2}$ & $161\pm{5}$ & $2.36\pm{0.22}$ & $2.79\pm{0.29}$\\ 
\hline
5$^{(*)}$  & $710877090\pm{91}$ &  $0.31\pm{0.03}$ & $700\pm{75}$ &  $54.8\pm{1.3}$ & $(...)$ & $(...)$ & $(...)$\\
6  & $710884700\pm{33}$ &  $0.50\pm{0.04}$ &  $473\pm{34}$ & $37.1\pm{0.7}$ & $(...)$ & $(...)$ & $(...)$ \\
7  & $710901674\pm{19}$ &  $1.20\pm{0.04}$ &  $650\pm{20}$ & $51.1\pm{0.4}$ & $126\pm{5}$ & $2.24\pm{0.20}$ & $3.02\pm{0.33}$\\
8  & $710906274\pm{30}$ &  $0.38\pm{0.04}$ & $415\pm{36}$ &  $32.6\pm{0.6}$ & $145\pm{5}$ & $0.49\pm{0.07}$ & $0.61\pm{0.10}$\\
9  & $710924568\pm{20}$ &  $1.04\pm{0.05}$ & $593\pm{21}$ &  $46.6\pm{0.4}$ & $124\pm{9}$ & $1.62\pm{0.11}$ & $2.21\pm{0.18}$\\
10 & $710934969\pm{20}$ &  $1.09\pm{0.05}$ & $430\pm{17}$ &  $33.7\pm{0.3}$ & $139\pm{9}$ & $1.71\pm{0.11}$ & $2.16\pm{0.16}$\\
11 & $710954337\pm{19}$ &  $1.11\pm{0.05}$ & $551\pm{14}$ &  $43.3\pm{0.4}$ & $128\pm{6}$ & $1.59\pm{0.11}$ & $2.12\pm{0.17}$\\
12 & $710966143\pm{18}$ &  $1.11\pm{0.05}$ & $504\pm{17}$ &  $39.6\pm{0.4}$ & $146\pm{6}$ & $1.45\pm{0.10}$ & $1.79\pm{0.13}$\\
13 & $710986288\pm{18}$ &  $1.17\pm{0.04}$ & $607\pm{18}$ &  $47.6\pm{0.4}$ & $134\pm{15}$ & $1.60\pm{0.10}$ & $2.07\pm{0.14}$\\
14 & $711000219\pm{17}$ &  $1.35\pm{0.05}$ & $545\pm{15}$ &  $42.8\pm{0.3}$ & $128\pm{15}$ & $2.07\pm{0.13}$ & $2.75\pm{0.20}$\\
\hline
15 & $771896728\pm{31}$ &  $1.15\pm{0.05}$ & $712\pm{29}$ &  $55.9\pm{0.5}$ &  $(...)$ & $(...)$ & $(...)$\\
16 & $771902847\pm{37}$ &  $0.48\pm{0.03}$ & $577\pm{43}$ &  $45.3\pm{0.7}$ & $(...)$  &$(...)$   & $(...)$ \\
17 & $771920440\pm{19}$ &  $1.32\pm{0.05}$ & $610\pm{18}$ &  $47.9\pm{0.3}$ & $124\pm{5}$ & $1.69\pm{0.11}$ & $2.29\pm{0.17}$\\
18 & $771925355\pm{32}$ &  $0.40\pm{0.05}$ & $380\pm{53}$ &  $29.8\pm{0.8}$ & $114\pm{5}$  & $0.65\pm{0.09}$ & $0.94\pm{0.16}$\\
19 & $771942019\pm{18}$ &  $1.36\pm{0.05}$ & $617\pm{19}$ &  $48.4\pm{0.3}$ & $124\pm{6}$ & $1.69\pm{0.10}$ & $2.30\pm{0.15}$\\
20 & $771947612\pm{55}$ &  $0.27\pm{0.03}$ & $497\pm{59}$ &  $39.0\pm{1.4}$ & $104\pm{6}$  & $0.36\pm{0.05}$ & $0.56\pm{0.13}$\\
21 & $771964216\pm{19}$ &  $1.19\pm{0.04}$ & $672\pm{20}$ &  $52.8\pm{0.3}$ & $125\pm{5}$ & $1.60\pm{0.09}$ & $2.16\pm{0.15}$\\
22 & $771969387\pm{28}$ &  $0.46\pm{0.04}$ & $377\pm{37}$ &  $29.6\pm{0.7}$ & $133\pm{5}$ & $0.54\pm{0.07}$ & $0.71\pm{0.10}$ \\
23 & $771986755\pm{19}$ &  $1.30\pm{0.05}$ & $595\pm{19}$ &  $46.7\pm{0.3}$ & $128\pm{6}$ & $1.70\pm{0.11}$ & $2.26\pm{0.17}$\\
24 & $771996228\pm{20}$ &  $0.93\pm{0.05}$ & $479\pm{23}$ &  $37.6\pm{0.3}$ & $135\pm{6}$ & $1.32\pm{0.09}$ & $1.70\pm{0.14}$\\
\hline
25$^{(**)}$  & $772064237\pm{39}$ &  $1.17\pm{0.07}$ & $650\pm{52}$ &  $51.0\pm{0.9}$ & $(...)$ & $(...)$ & $(...)$\\
26  & $772071265\pm{42}$ &  $0.44\pm{0.04}$ & $535\pm{56}$ &  $42.0\pm{0.8}$ & $(...)$ & $(...)$ & $(...)$ \\
27  & $772089084\pm{18}$ &  $1.36\pm{0.05}$ & $617\pm{18}$ &  $48.4\pm{0.3}$ & $131\pm{5}$ & $1.66\pm{0.14}$ & $2.18\pm{0.21}$\\
28  & $772094209\pm{34}$ &  $0.38\pm{0.04}$ & $458\pm{43}$ &  $35.9\pm{0.7}$ & $96\pm{5}$ & $0.59\pm{0.08}$ & $1.01\pm{0.22}$\\
29  & $772111144\pm{19}$ &  $1.36\pm{0.05}$ & $637\pm{19}$ &  $50.0\pm{0.3}$ & $129\pm{5}$ & $1.58\pm{0.09}$ & $2.10\pm{0.14}$\\
30  & $772126986\pm{14}$ &  $1.80\pm{0.05}$ & $567\pm{13}$ &  $44.5\pm{0.2}$ & $144\pm{5}$ & $2.23\pm{0.12}$ & $2.78\pm{0.16}$\\
31  & $772148922\pm{20}$ &  $1.24\pm{0.05}$ & $643\pm{20}$ &  $50.5\pm{0.3}$ & $123\pm{5}$ & $1.62\pm{0.10}$ & $2.23\pm{0.16}$\\
32  & $772160801\pm{16}$ &  $1.31\pm{0.05}$ & $551\pm{18}$ &  $43.3\pm{0.3}$ & $137\pm{5}$ & $1.72\pm{0.11}$ & $2.20\pm{0.15}$\\
33  & $772178789\pm{37}$ &  $1.43\pm{0.07}$ & $627\pm{33}$ &  $49.2\pm{0.4}$ & $(...)$ & $(...)$ & $(...)$\\
\hline 
\end{tabular}
\tablefoot{QPE timing and spectral properties at the peak. The timing properties were derived using Bayesian inference on the \texttt{[constant + Gaussian]} model given the $0.2-2$ keV EPIC-pn and EPIC-MOS data. The QPE spectral properties at the peak were derived using Bayesian inference on the redshifted blackbody model \texttt{[zashift*bbodyrad]}, given the $0.3-2$ keV pn and MOS data, assuming that QPEs are additive component superimposed over a constant quiescent emission. Column (1): consecutive QPE number. Col (2): time of the QPE peak in Mission Reference Time. Col (3): QPE count rate defined as the peak normalization of the best-fitting Gaussian line. Col. (4): QPE duration, defined as twice the FWHM of the best-fitting Gaussian line. Col. (5): QPE temperature at the peak. Col. (6): QPE luminosity at the peak in the $0.3-2$ keV band. Col. (7): QPE bolometric luminosity at the peak.  
QPEs marked with (*)/(**) have been half-detected at the beginning of the observation by the MOS/pn cameras only, and therefore report quantities derived used such data. The horizontal lines divide the five different epochs of \textit{XMM-Newton }observations of RX~J1301.9+2747: 2000, 2019, 2020, 2022A, and 2022B. 
}
\end{table*}

\subsection{Hardness ratios}
We show in Fig.~\ref{FIGB2} the hardness ratio plot for each QPE observed by \textit{XMM-Newton} in the galaxy RX~J1301.9+2747 from 2000 to 2022. The hardness ratio is computed as HR = CR($0.6-2$)/CR($0.2-2$), where CR($0.6-2$) is the count rate in the $0.6-2$ keV band and CR($0.2-2$) is the total count rate in the $0.2-2$ keV band.
\begin{figure*}[ht!]
\centering
\includegraphics[ width=4.5cm]{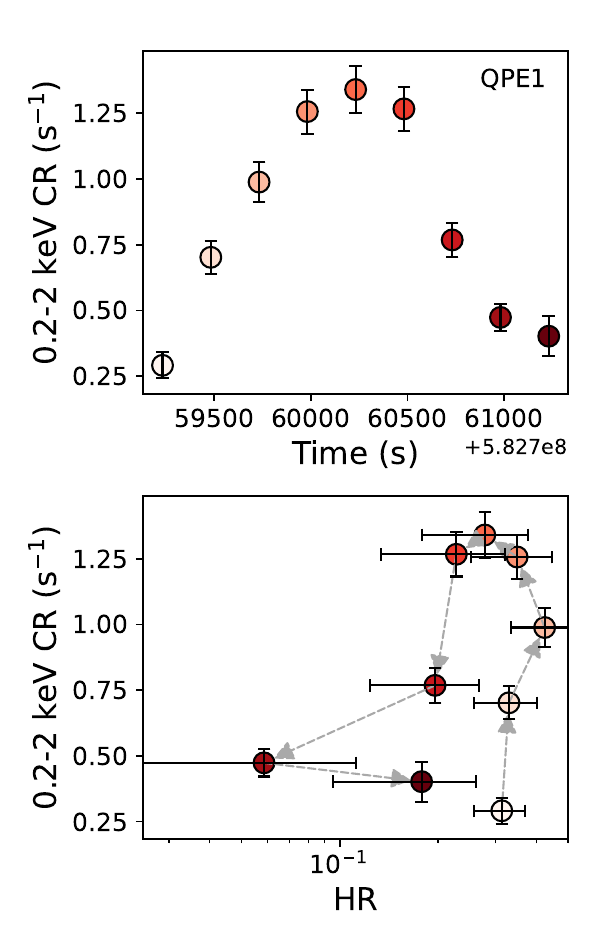}
\includegraphics[ width=4.5cm]{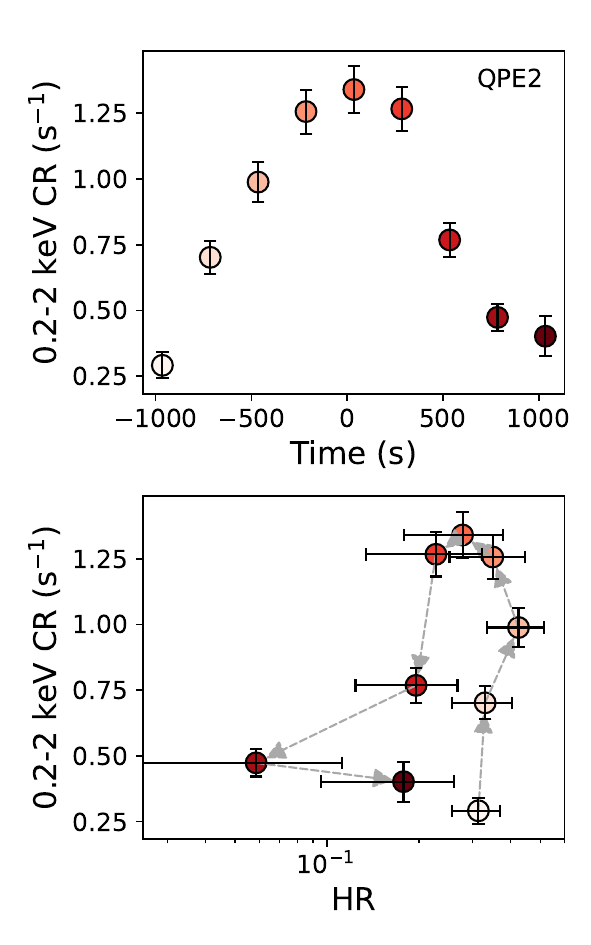}
\includegraphics[ width=4.5cm]{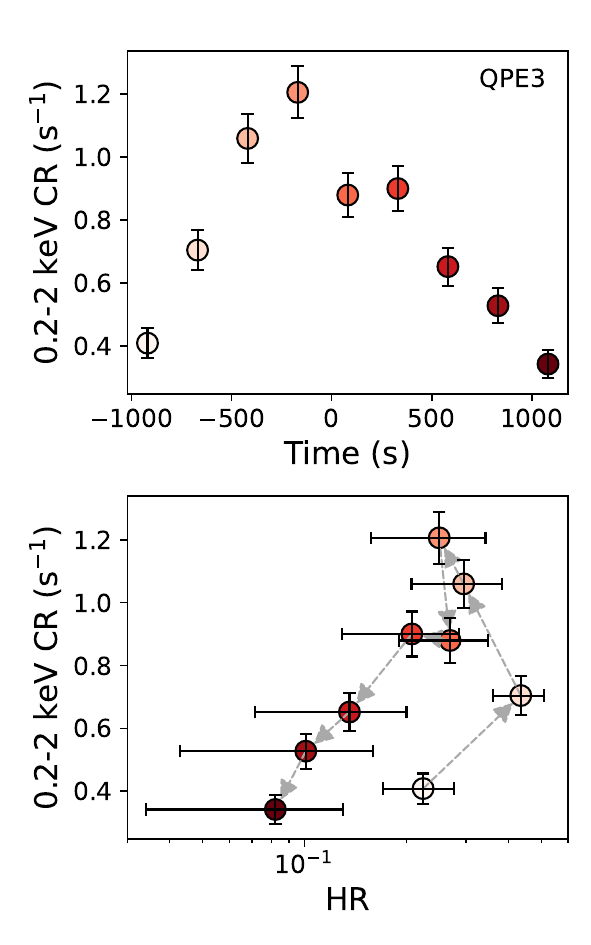}
\includegraphics[ width=4.5cm]{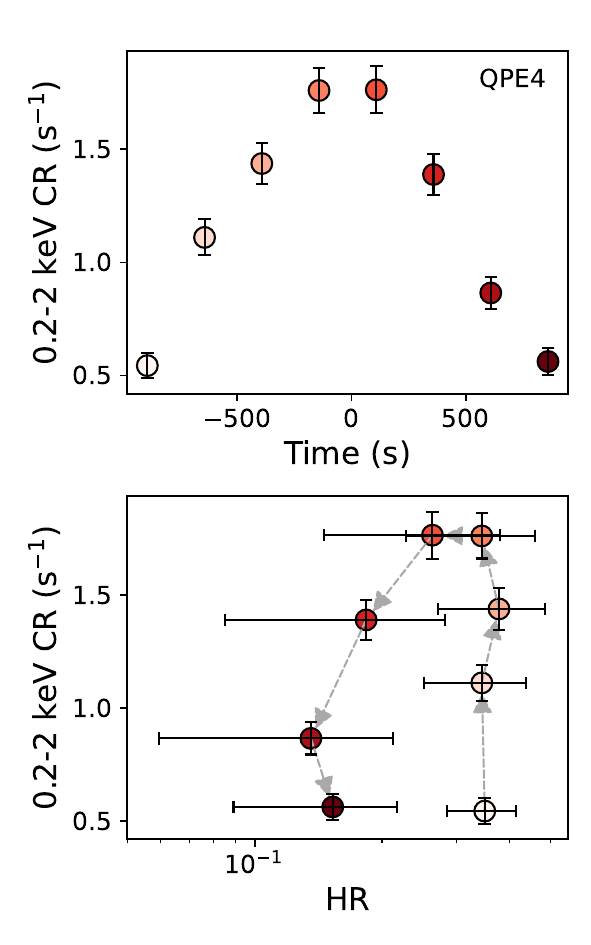}
\includegraphics[ width=4.5cm]{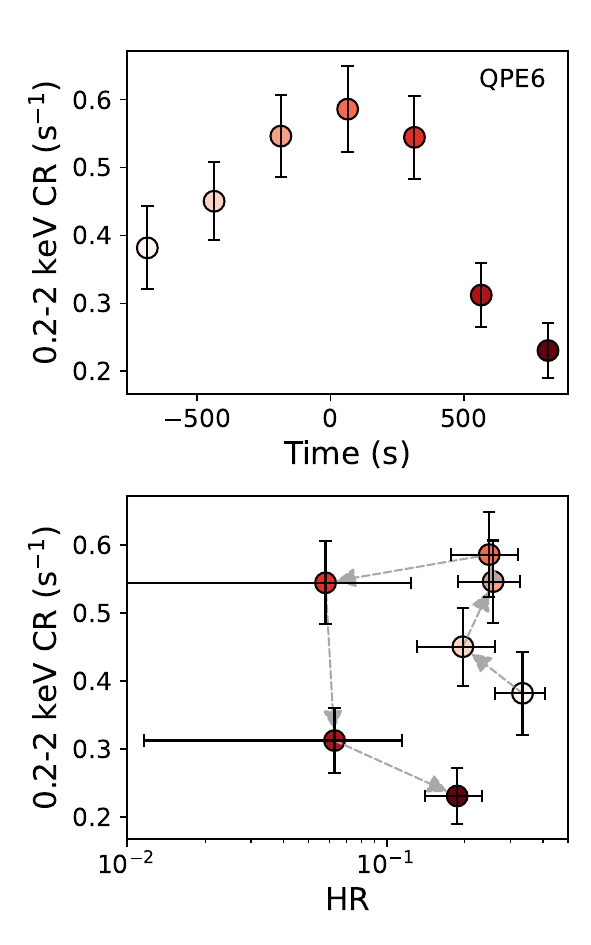}
\includegraphics[ width=4.5cm]{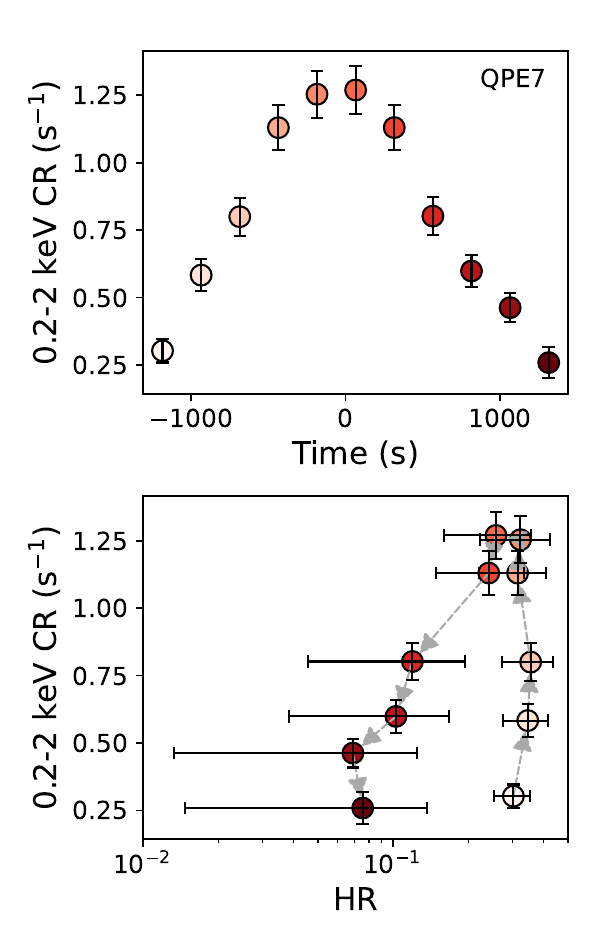}
\includegraphics[ width=4.5cm]{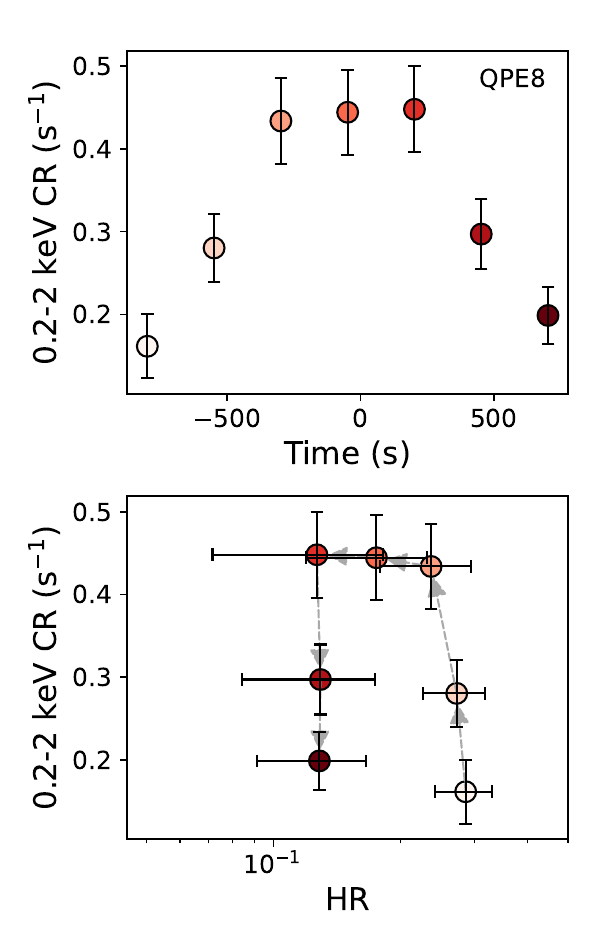}
\includegraphics[ width=4.5cm]{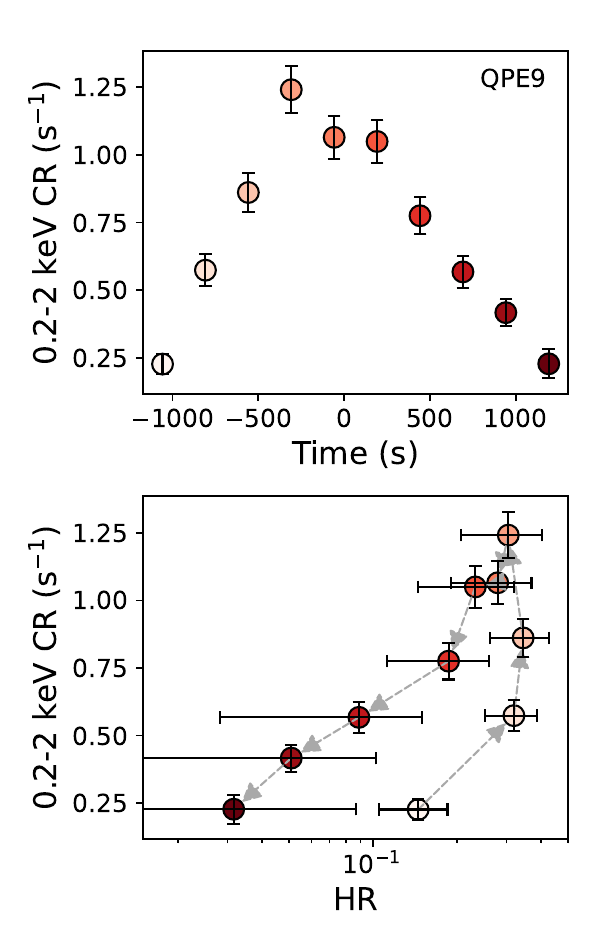}
\includegraphics[ width=4.5cm]{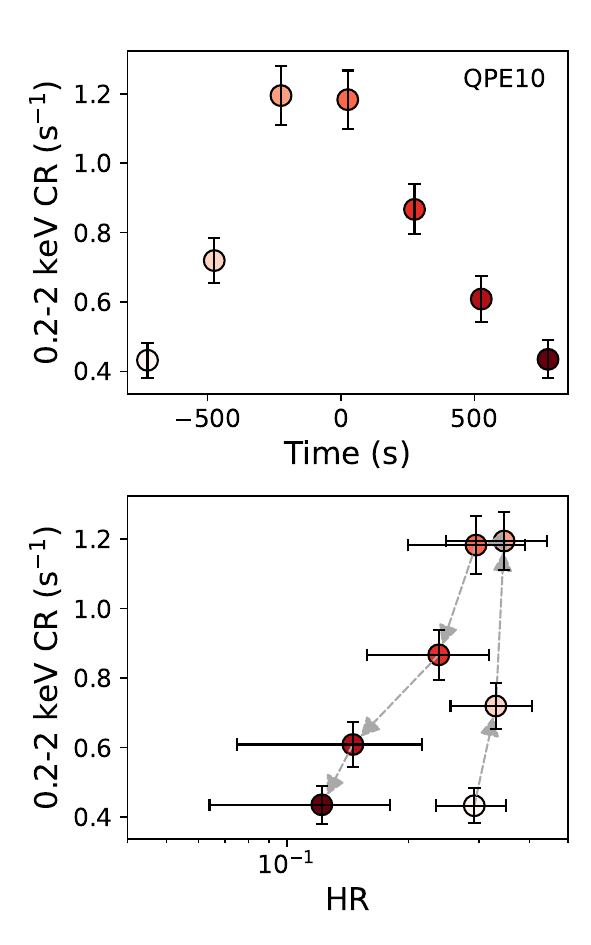}
\includegraphics[ width=4.5cm]{zzHR_QPE11zzz.pdf}
\includegraphics[ width=4.5cm]{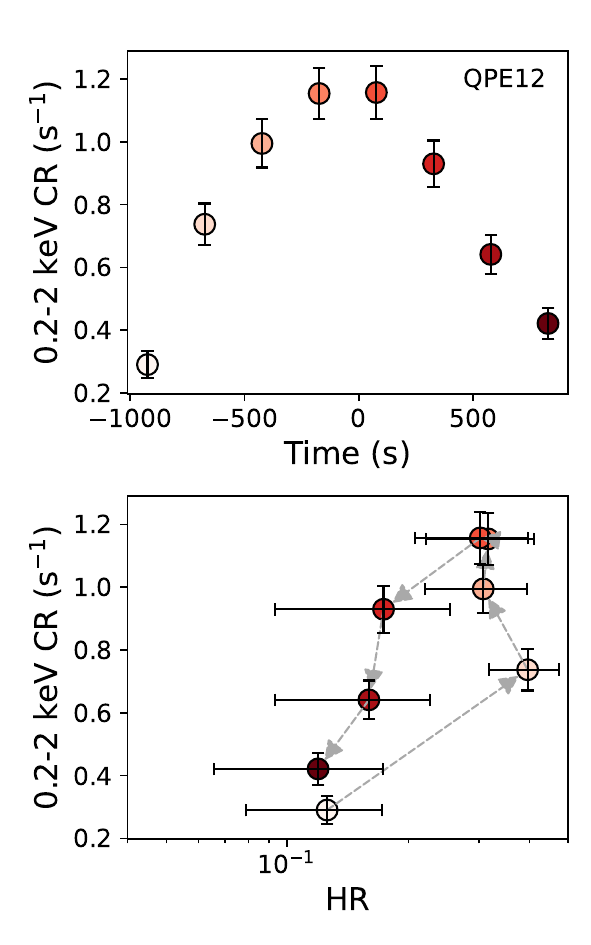}
\includegraphics[ width=4.5cm]{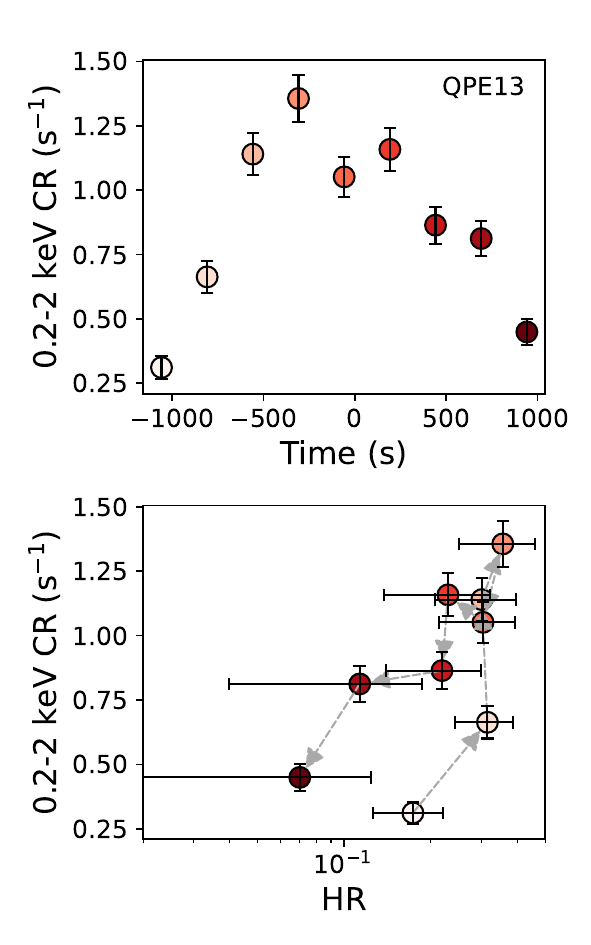}
\caption{For each QPE we plot in the top panel the $0.2-2$ keV background-subtracted QPE light curve, centered around the QPE peak and binned to 250 s; and in the lower panel the hardness ratio HR = CR($0.6-2$)/CR($0.2-2$), where CR($0.6-2$) is the count rate in the $0.6-2$ keV band and CR($0.2-2$) is the total count rate in the $0.2-2$ keV band. Each point is color-coded following the time evolution of the QPE in the panel above. The HR shows a counter-clockwise evolution, as highlighted by the dashed arrow. All these quantities have been computed using the pn data only. \label{FIGB2} }
\end{figure*}
\begin{figure*}[ht!]\ContinuedFloat
\centering
\includegraphics[ width=4.5cm]{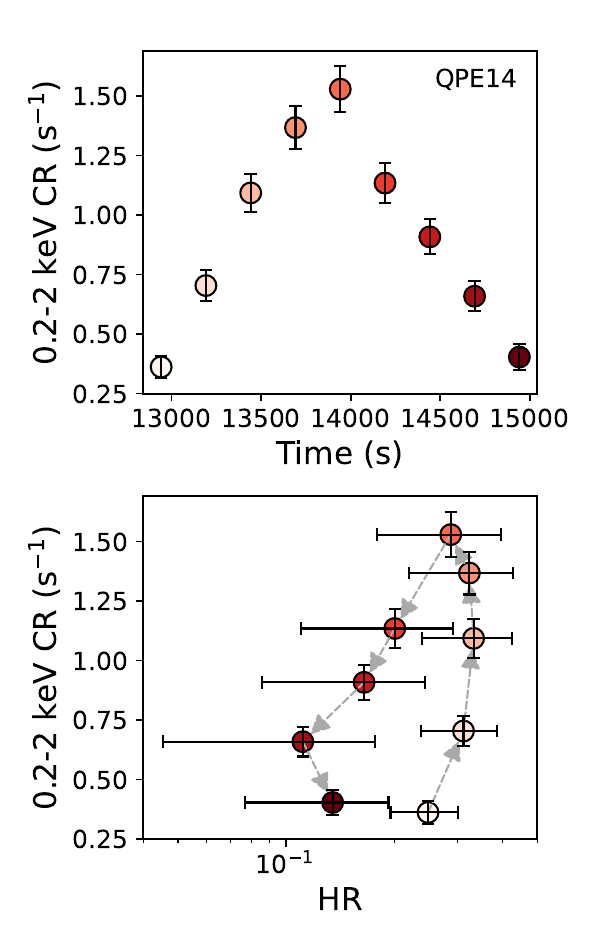}
\includegraphics[ width=4.5cm]{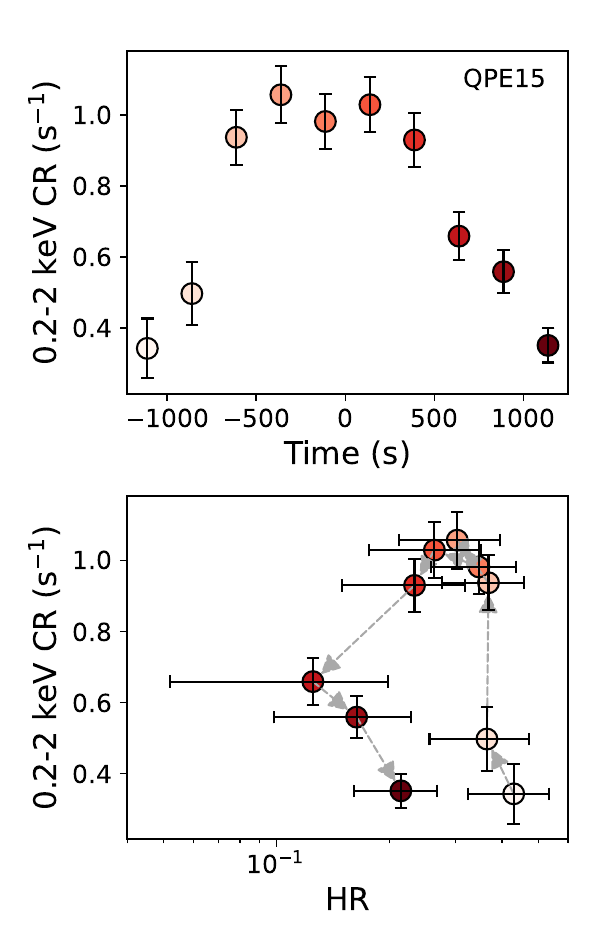}
\includegraphics[ width=4.5cm]{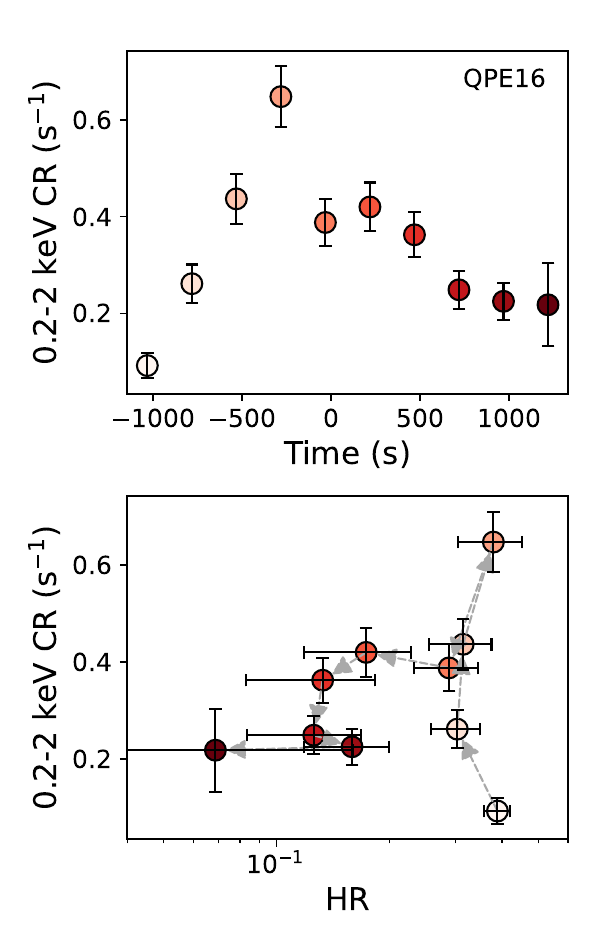}
\includegraphics[ width=4.5cm]{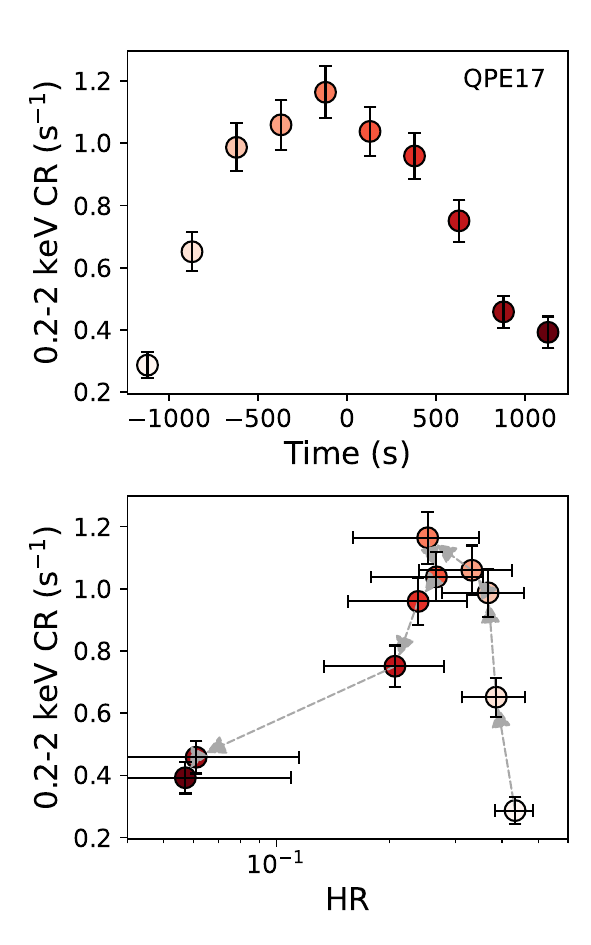}
\includegraphics[ width=4.5cm]{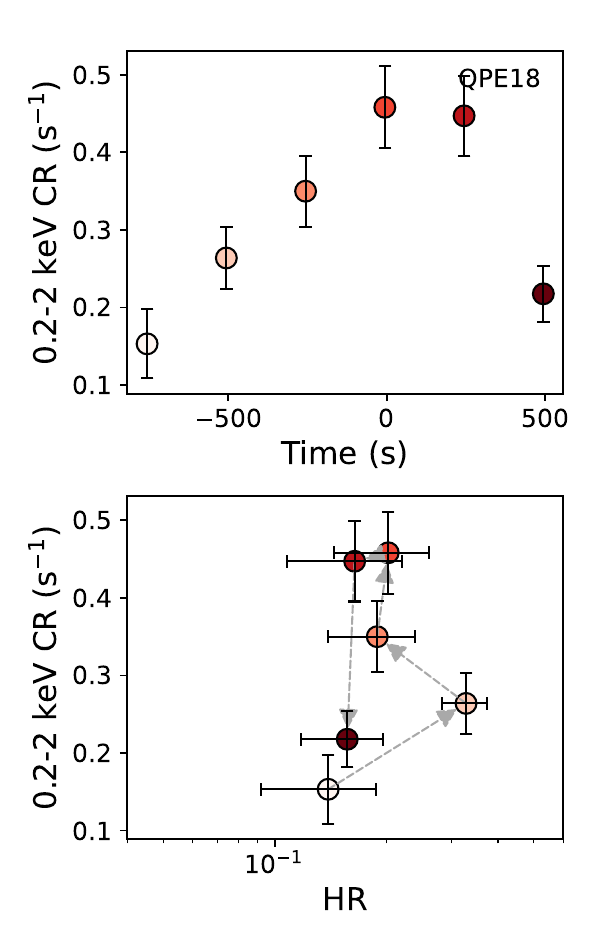}
\includegraphics[ width=4.5cm]{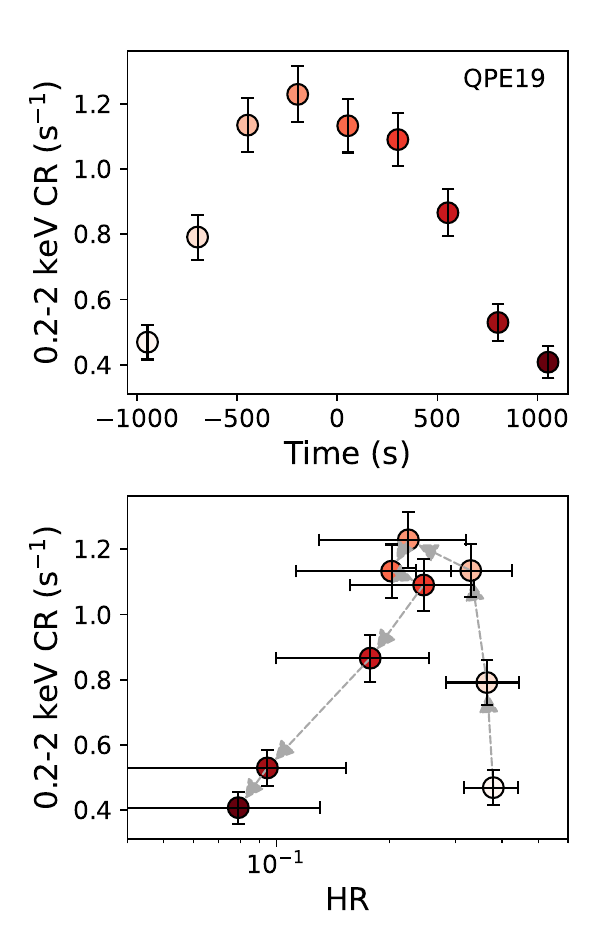}
\includegraphics[ width=4.5cm]{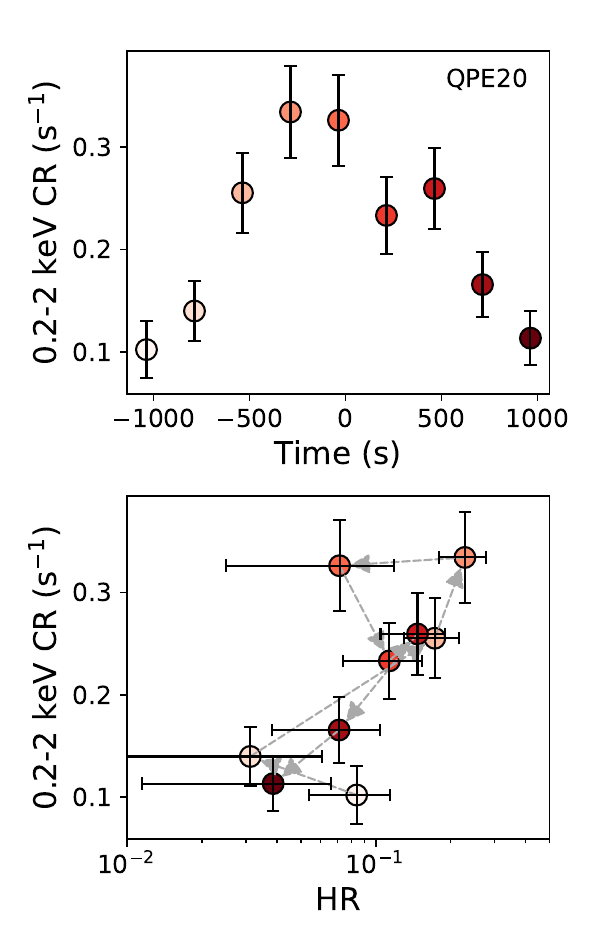}
\includegraphics[ width=4.5cm]{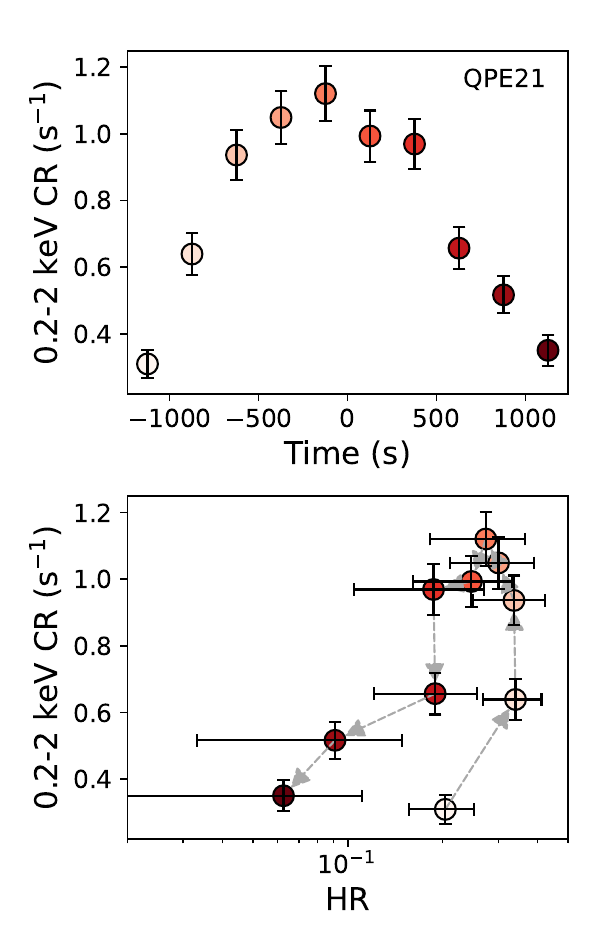}
\includegraphics[ width=4.5cm]{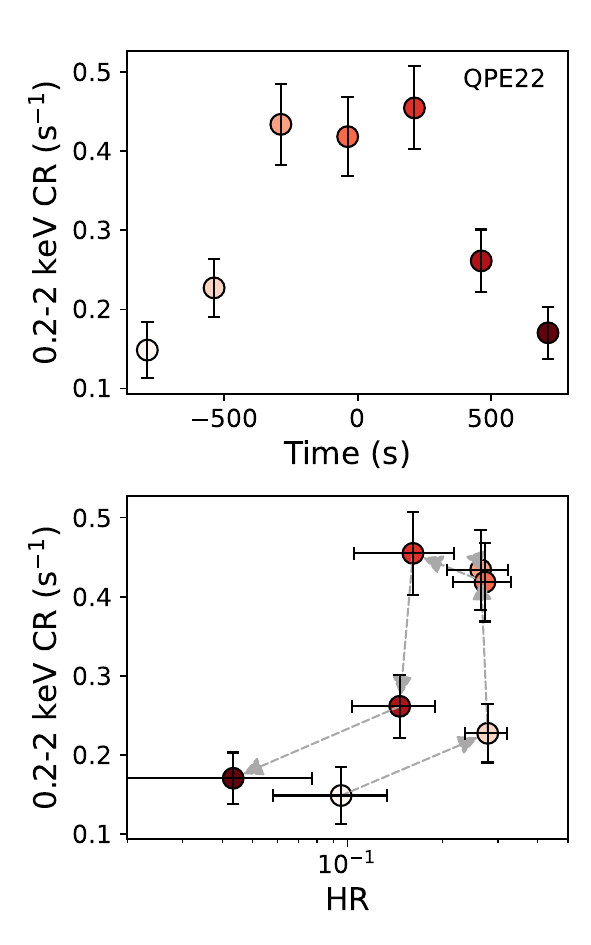}
\includegraphics[ width=4.5cm]{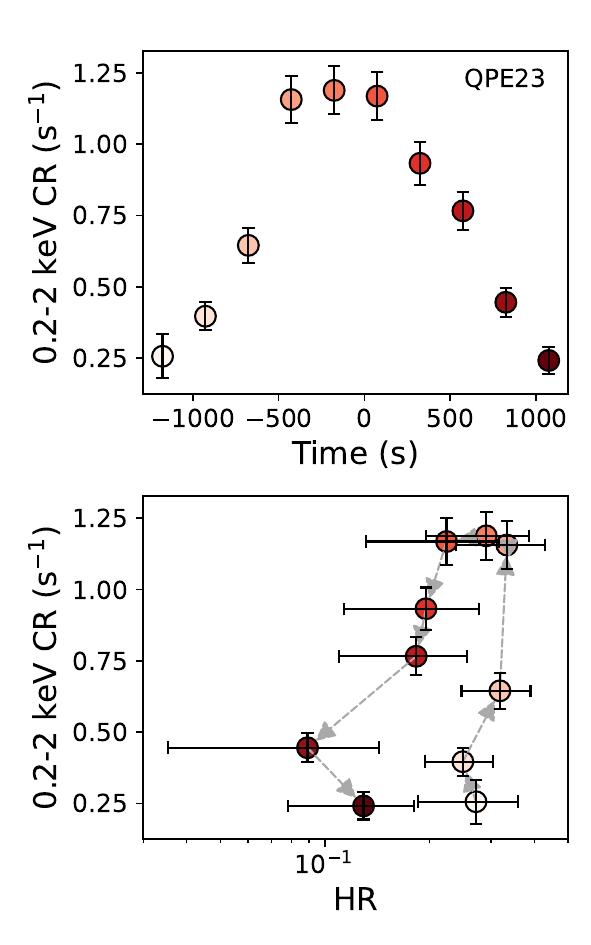}
\includegraphics[ width=4.5cm]{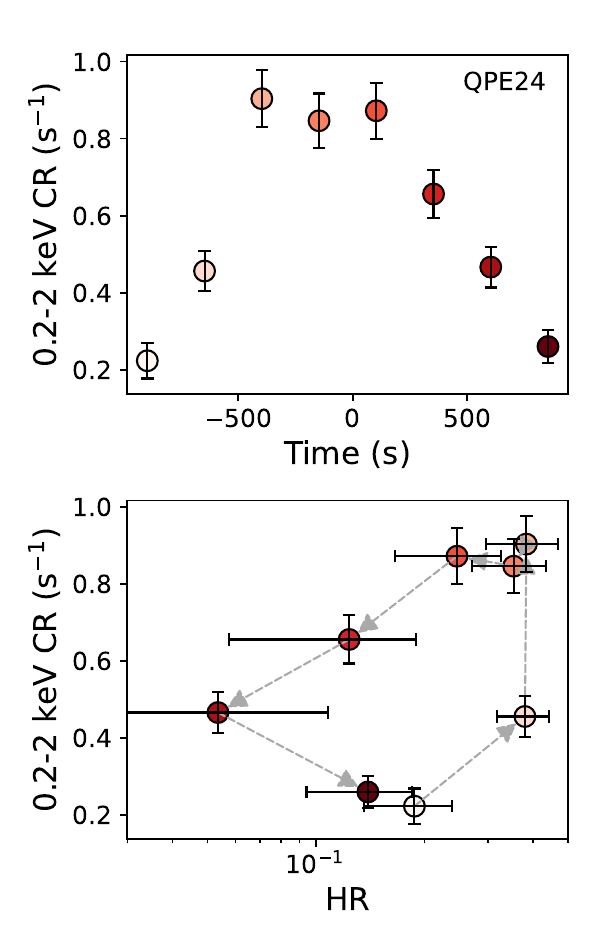}
\includegraphics[ width=4.5cm]{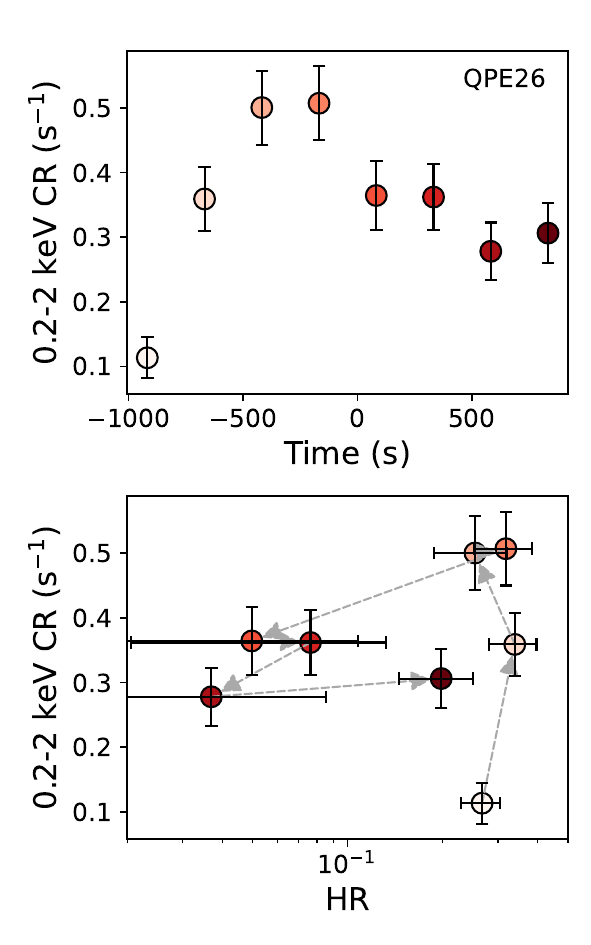}
\caption{-- continued.}
\end{figure*}
\begin{figure*}[ht!]\ContinuedFloat
\centering
\includegraphics[ width=4.5cm]{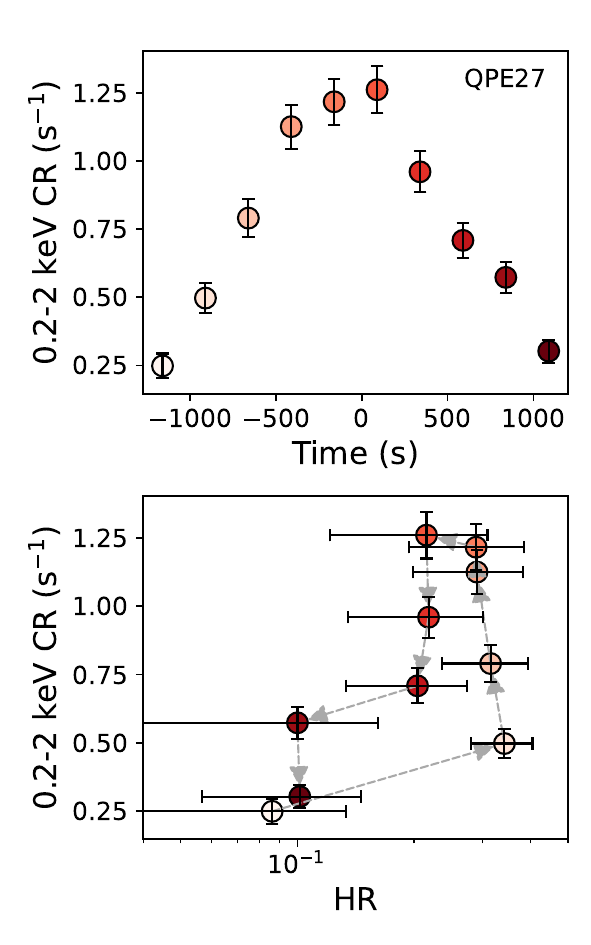}
\includegraphics[ width=4.5cm]{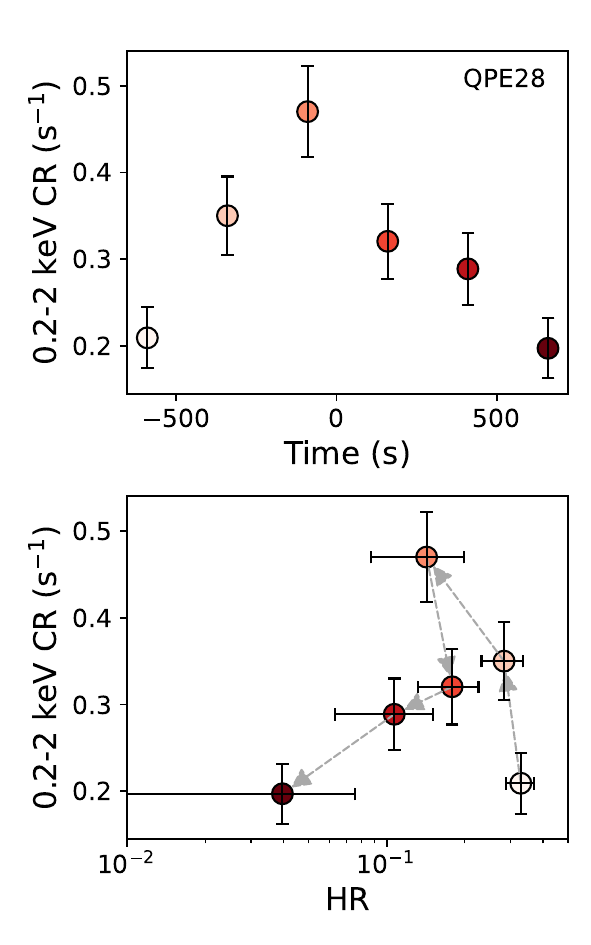}
\includegraphics[ width=4.5cm]{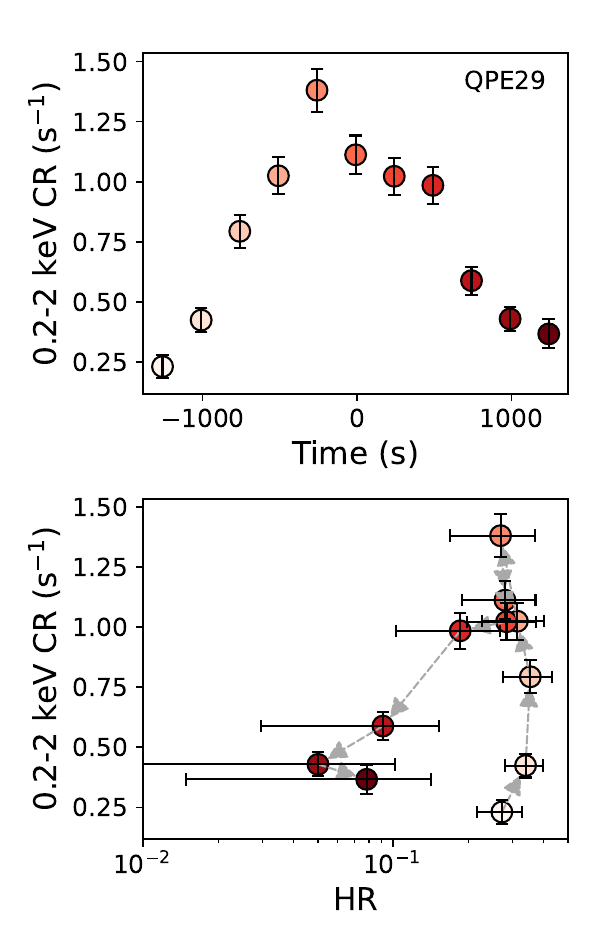}
\includegraphics[ width=4.5cm]{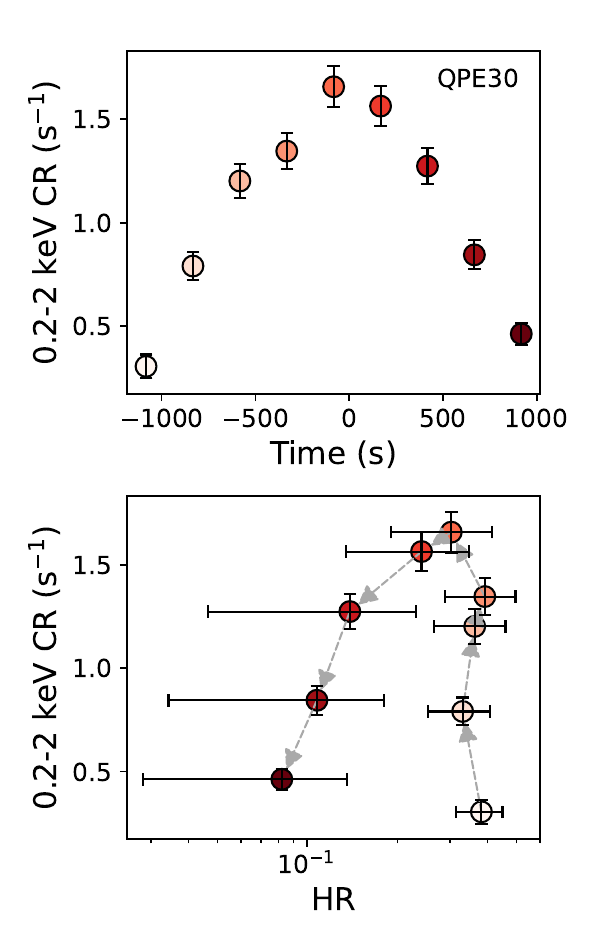}
\includegraphics[ width=4.5cm]{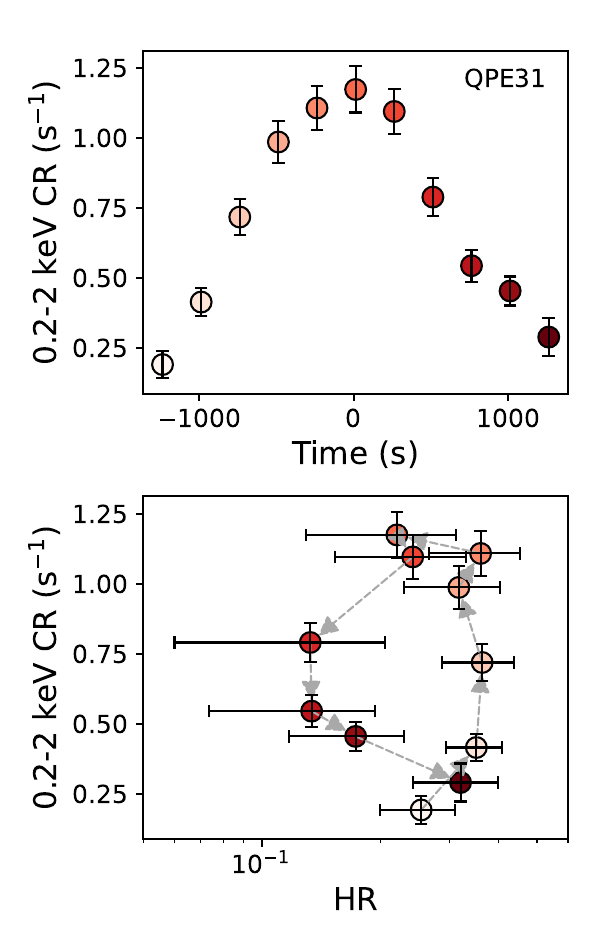}
\includegraphics[ width=4.5cm]{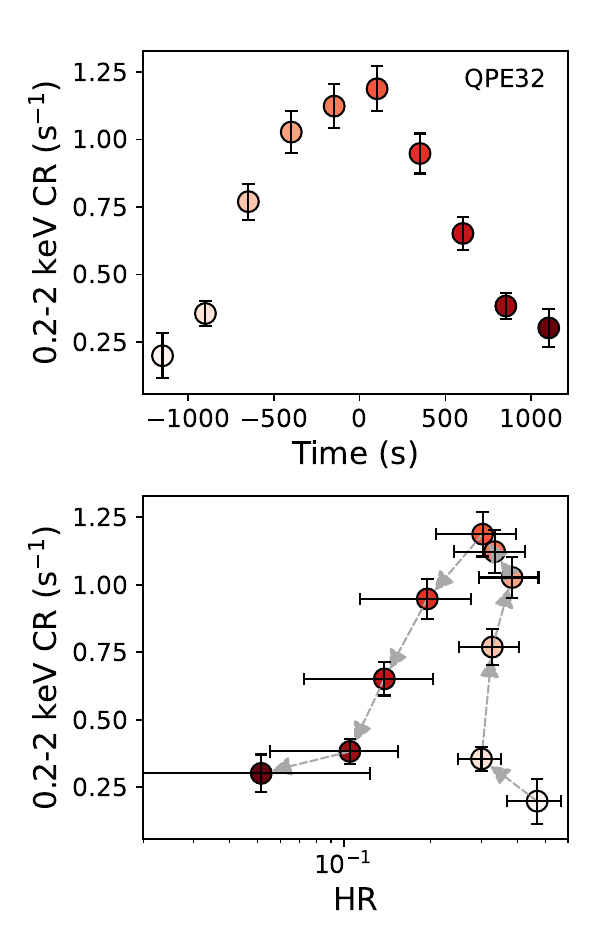}
\includegraphics[ width=4.5cm]{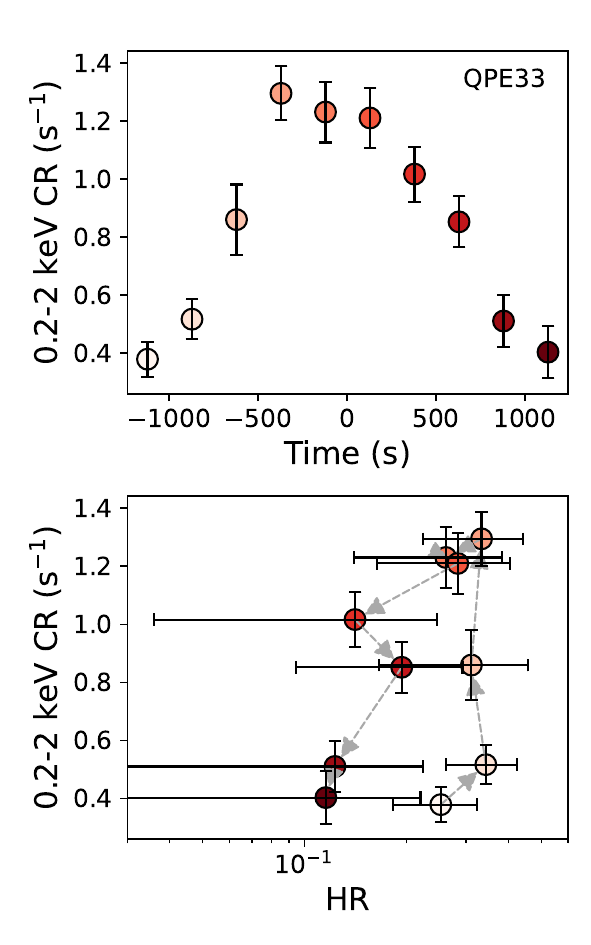}
\caption{-- continued.}
\end{figure*}

\section{Fit to the quiescent spectra\label{appendix:quiescence}}

 \begin{figure*}[htb]
\centering
\includegraphics[width=18cm]{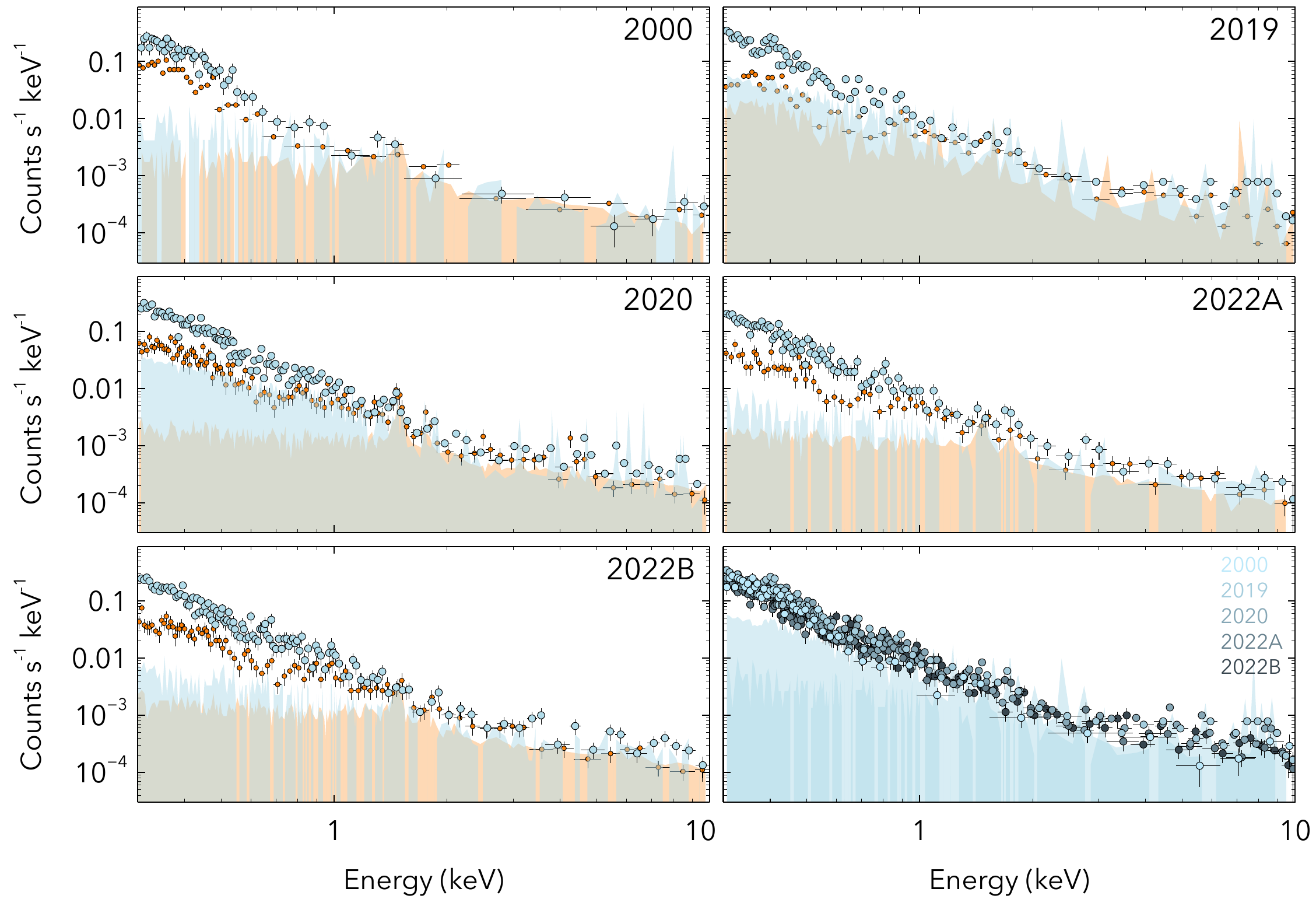}
\caption{\textit{XMM-Newton} EPIC quiescent spectra (i.e., excluding time intervals when QPEs are present) of RX J3101.9+2747, rebinned for visual purposes to $3\sigma$ significance. For each epoch of observation the pn(MOS) spectra are plotted with blue(orange) circles, while the background spectra with shaded areas of the same color. The last panel reports the five epochs together, omitting the MOS data and plotting only the highest-level (2019) and lowest-level (2022A) pn backgrounds for clarity.
\label{FIGC1}}
\end{figure*}

The time intervals including QPEs identified with the light curve analysis were excluded in order to extract the spectra during quiescence.
We used the \texttt{BXA} package, assuming a Poissonian likelihood function and using the W statistic\footnote{The W statistic, or wstat, is the implementation of the Cash statistic \citep{1979ApJ...228..939C} when the background is not modeled.} on the unbinned $0.3-10$ keV spectra.
The background spectra were grouped to a minimum of 3 counts per energy bin in order to avoid statistical biases\footnote{See the discussion in \url{https://giacomov.github.io/Bias-in-profile-poisson-likelihood/}}.
The ten EPIC quiescent spectra, one pn and one merged MOS1+MOS2 spectrum for each of the five epochs of \textit{XMM-Newton} observations of RX~J1301.9+2747, are shown in Fig.~\ref{FIGC1}. Each epoch of observation (2000, 2019, 2020, 2022A, and 2022B) was fit independent of the other. The pn and MOS spectra of each epoch were fit together to the same model, with a multiplicative constant factor \texttt{const} to take into account instrumental cross-calibration uncertainties. The priors on \texttt{const} were always assumed to be uniform between $[0.8,1.2]$. The value of the column density was allowed to reach $10^{21}$ cm$^{-2}$ during the fit, assuming uniform priors. 

\begin{figure*}[htb]
\centering
\includegraphics[width=6cm]{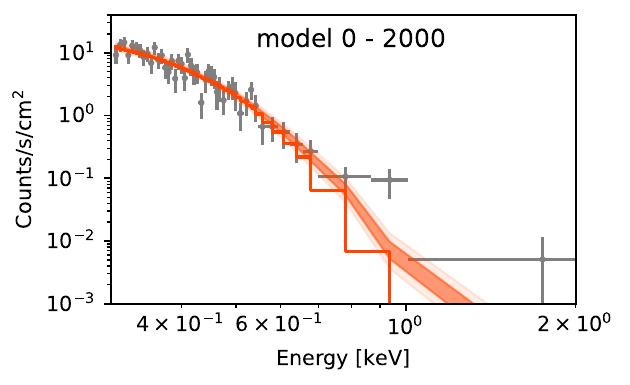}
\includegraphics[width=6cm]{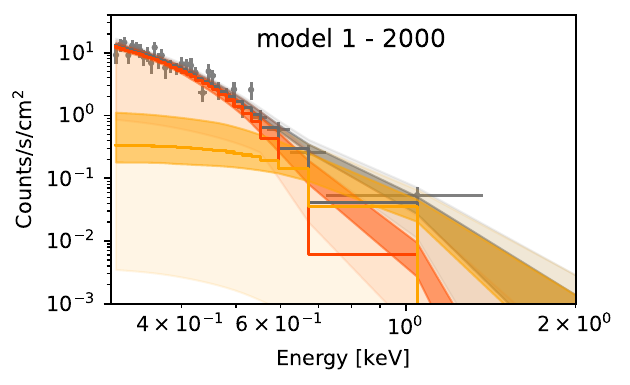}
\includegraphics[width=6cm]{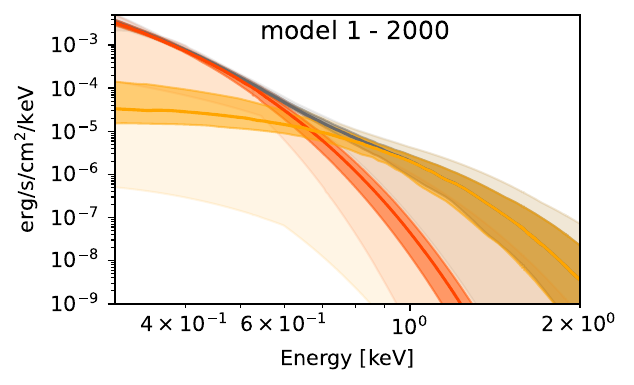}
\includegraphics[width=6cm]{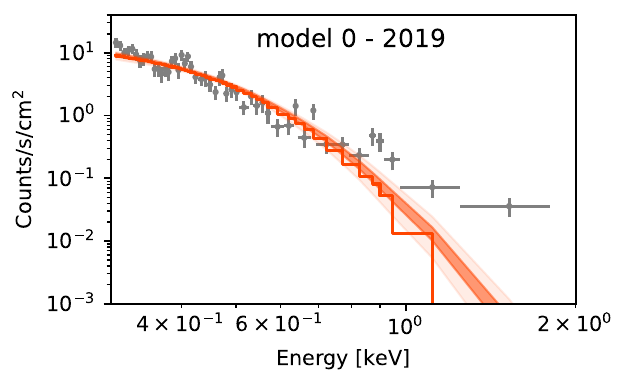}
\includegraphics[width=6cm]{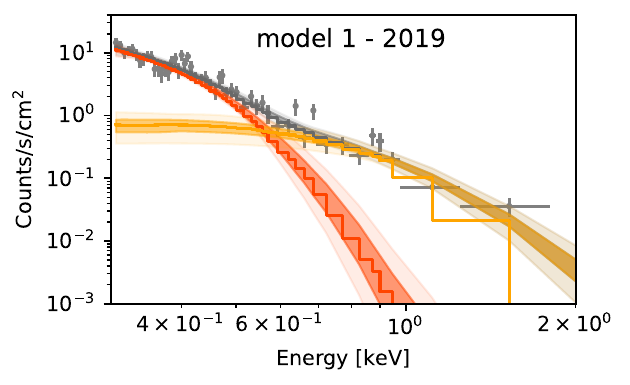}
\includegraphics[width=6cm]{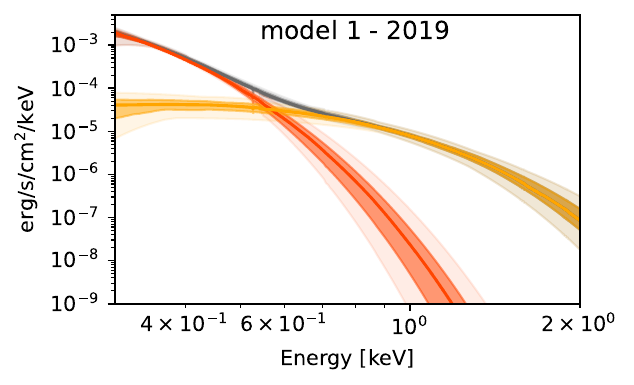}
\includegraphics[width=6cm]{convolved_posteriorz2020_model0.pdf}
\includegraphics[width=6cm]{convolved_posterior_2020_model1.pdf}
\includegraphics[width=6cm]{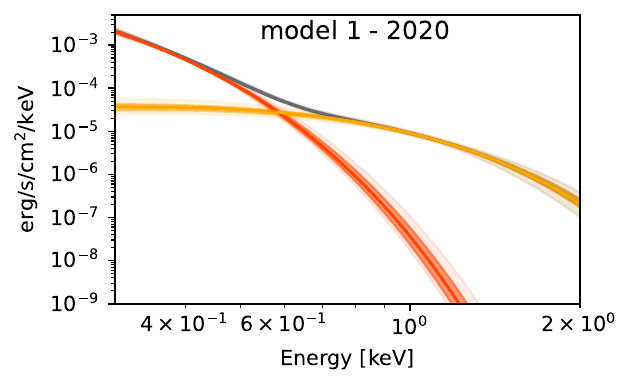}
\includegraphics[width=6cm]{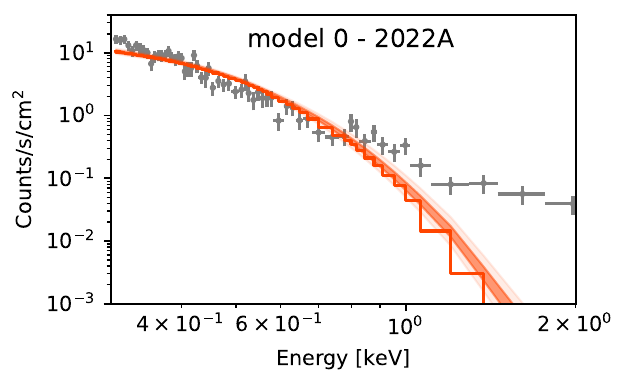}
\includegraphics[width=6cm]{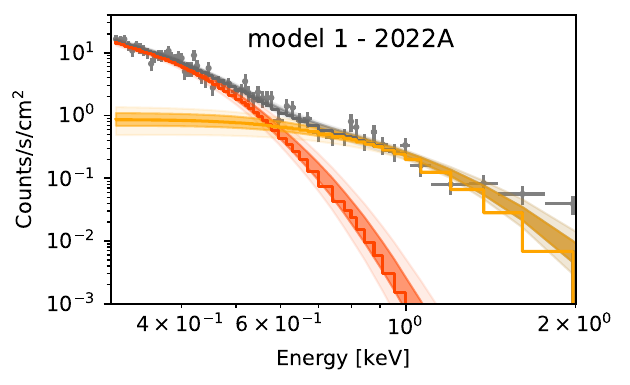}
\includegraphics[width=6cm]{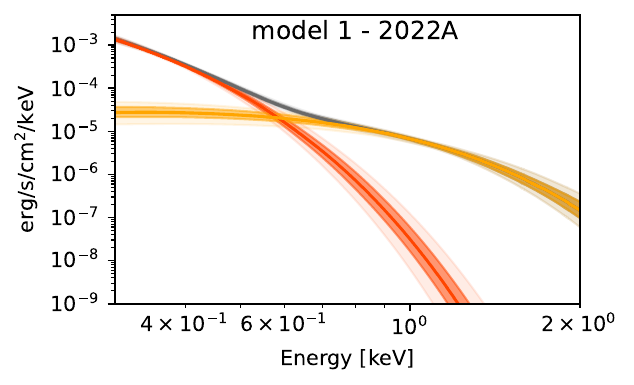}
\includegraphics[width=6cm]{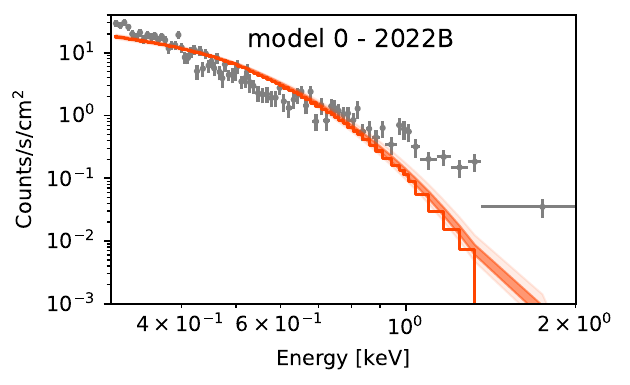}
\includegraphics[width=6cm]{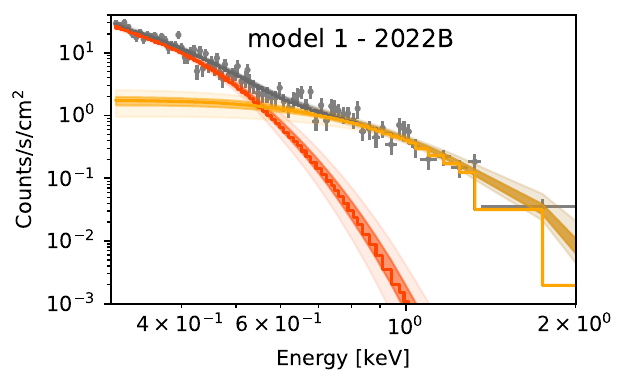}
\includegraphics[width=6cm]{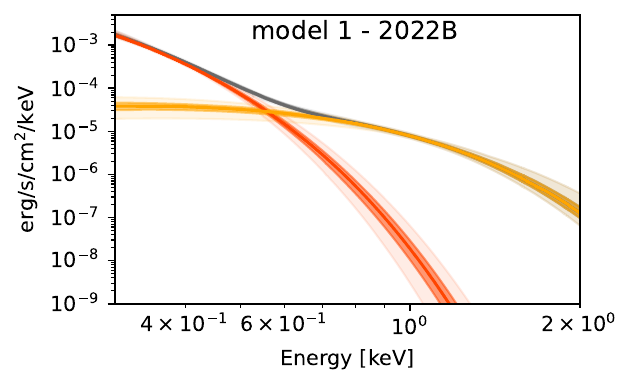}
\caption{The posterior probability distribution of \texttt{[TBabs*zashift*diskbb]} (model 0, left column) and \texttt{[TBabs*zashift*(diskbb + bbodyrad)]} (model 1, central column) applied to the quiescent data, for the five epochs of \textit{XMM-Newton} observation of RX~J1301.9+2747. Data are shown for the EPIC-pn camera only and rebinned to 3$\sigma$ significance. The posterior PDF unconvolved from the instrumental response is plotted for each epoch for the best-fitting model 1 in the right column. The \texttt{[diskbb]} posterior PDF is plotted in dark orange, the \texttt{[bbodyrad]} one in dark yellow, the total model 1 in grey.
\label{FIGC2}}
\end{figure*}

First, we compare the quiescent spectra to two soft thermal models, \texttt{[zashift*bbodyrad]} and \texttt{[zashift*diskbb]}. The \texttt{[zashift]} component takes into account the cosmological redshift of the emitted photons and was always fixed to $z=0.024$.
We assumed flat, uninformed priors on the model parameters, with normalizations spanning several orders of magnitude between $[10-10^{6}]$ and the temperatures ranging between $[10-300]$ eV.
We found that the \texttt{[zashift*diskbb]} model is preferred over the \texttt{[zashift*bbodyrad]} one, given the Bayesian evidences found (with a difference of $\log\mathcal{Z}> 2$ in every epoch). 
We call the model \texttt{[zashift*diskbb]} 'model 0'.
Results of the fit to the five epochs of observation of the quiescent spectra of RX~J1301.9+2747 to model 0 are reported in the first column of Fig.~\ref{FIGC2}, where we plot only the pn data, rebinned visually to 3$\sigma$ significance, in the $0.3-1.5$ keV band (i.e., up to where the spectra are not background-dominated, see Fig.~\ref{FIGC1}). Although the spectra  are very soft and most of the flux contribution is carried by the disk component \texttt{diskbb}, this component alone is not able to well reproduce the data at energies $E > 0.7$ keV, especially from 2019 onward.

We then compared the quiescent spectra of the five epochs 
to two equally complex models: \texttt{[zashift*(diskbb + bbody)]} (model 1) and \texttt{[zashift*(diskbb + powerlaw)]} (model 2).
In this model scheme, \texttt{diskbb} represents the thermal emission from an accretion disk with inner temperature $kT$; \texttt{bbody} represents a soft X-ray excess-like warm thermal component, with temperature $kT$; and the  \texttt{powerlaw} model represents the the hard X-ray coronal emission with photon index $\Gamma$. 
The priors were all uninformed, with these ranges: $[20-200]$ eV for the \texttt{diskbb} inner temperature and $[10^{-4}-10^8]$ for its normalization; $[50-500]$ eV for the \texttt{bbody} temperature and $[10^{-6}-10^3]$ for its normalization; $[1.5-4]$ for the \texttt{powerlaw} photon index, and $[10^{-13}-10^{-5}]$ for its normalization.

Results are reported in Table~\ref{table:quiescence}, where we report 
the median value of the posterior PDF of the model parameters along with their 1 standard deviation upper and lower credible values, the observed flux, the unabsorbed luminosity, and the logarithms of the Bayesian evidence $\mathcal{Z}$ and of the Akaike information criterion \citep[AIC$\:=C - 2\times m$, where $C$ is the likelihood and $m$ is the number of degrees of freedom,][]{1974ITAC...19..716A} normalized to their highest values, for each epoch of observation.
The model that gives the most probable representation of the data will have the highest $\mathcal{Z}$ and AIC.

Given the number of degrees of freedom and the Bayesian evidence $\mathcal{Z}$, models 1 and 2 are always strongly preferred to model 0 except for epoch 2000, when model 0 gives $\Delta\log\mathcal{Z} = 0.9$ compared to model 1, and $\Delta\log\mathcal{Z} = - 0.5$ compared to model 2.
Model 1 is the most probable representation of the data at all epochs except for 2022A and 2022B, when model 2 is the most probable one.
The differences in $\log\mathcal{Z}$ between model 1 and model 2 are however very small,  $\Delta\log\mathcal{Z}< 1.5$, except for 2019 (when $\Delta\log\mathcal{Z}=6.1$). 
There are two physical reasons why model 1 can be preferred over model 2.
First, the photon index $\Gamma$ of the \texttt{powerlaw} component of model 2 is found to be very steep compared to the typical AGN values, and even steeper than the most extreme observed values. Such a steep power law is likely not produced by Comptonization in a standard X-ray hot corona. On the contrary, the temperature $kT$ of the \texttt{bbody} component found in model 1 is typical of the soft X-ray excess observed in AGN. 
Second, as the power law model diverges at low energies, the bolometric luminosity estimate in model 2 is up to one order of magnitude larger than the estimate with model 0, while it is larger by a factor of $\sim 4$ in model 1 compared to model 0. Given that the spectra are evidently dominated by the \texttt{diskbb} component, we conservativately favor the model that produces a bolometric luminosity with a value more similar to this component alone. We thus adopt model 1 as our best-fitting model.

The posterior PDFs of model 0 and model 1 folded with the instrumental response are plotted in the left and central columns of Fig.~\ref{FIGC2} compared to the observed data. The right column of the same figure reports the unconvolved posterior PDFs for the best-fitting model 1, where the accretion disk \texttt{diskbb} is plotted in dark orange and the soft excess \texttt{bbody} in light orange. We limit these plots to the $0.3-2$ keV band as above these energies the signal is strongly dominated by the background, however the data were considered up to $10$~keV.
That the disk alone is not sufficient can be seen in the left column of Fig.~\ref{FIGC2} and in Table~\ref{table:quiescence}, where Model 0 provides the lowest $0.3-2$ keV flux compared to all the other models, signaling that some harder-than-the-disk flux is not modeled by \texttt{diskbb} alone. When using model 0, the inner temperature of the 
accretion disk seems to increase from 2000 onward (top left panel in Fig.~\ref{FIG8}). 
Once the harder component is introduced, the disk inner temperature stabilizes around $kT\sim 60$ eV in each epoch and the $0.3-2$ keV flux is recovered well (it is constant between models 1, 2, 3).
This temperature is independent of the harder component adopted, \texttt{bbody} or \texttt{powerlaw}.
When using model 2, during epoch 2000 the power law  is insensitive to the fit (see the third row of Fig.~\ref{FIG8}). This means that the \texttt{powerlaw} component is weakly required by the data during this epoch compared to the other epochs of observation.
As the \texttt{powerlaw} model mostly models soft X-ray photons (given its very steep photon index), this result should be seen also when using model 1. Indeed, also when using model 1 the warm thermal component \texttt{bbody} is well constrained in each epoch of observation, except for epoch 2000 (second row of Fig.~\ref{FIG8}). 
This confirms that the power law component mostly models soft X-ray photons.

Finally, we compared the data of the quiescent emission of RX~J1301.9+2747 to a disk plus Comptonization model, \texttt{[diskbb + compTT]} (model 4). We used the model developed by \citet{1994ApJ...434..570T} and assumed a disk geometry for the Comptonizing region, with a seed photon temperature equal to the inner temperature of the accretion disk.
As there is complete degeneracy between the Comptonizing plasma temperature and optical depth, we fixed the latter to a series of values $\tau=1, 5,10$, and left the former as free parameter.
We assumed uninformed priors for all parameters except for epoch 2000, when we used Gaussian priors for the temperature of the accretion disk ($60\pm{10}$ eV) and of the Comptonizing plasma ($400\pm{100}$ eV).
The limits of the parameter values were $[30-100]$ eV and $[10^{5}-10^5]$ for the disk temperature and normalization, $[0.1-10]$ keV for the Comptonizing region temperature, and $[10^{-9}-10^{-1}]$ for its normalization.
We found that a good representation of the soft X-ray excess-like spectral component is given by Comptonization of the disk seed photons on a plasma with average temperature of $ 400\pm{25}$ eV for an optical depth $\tau=10$. The plasma average temperature would instead be $1.1\pm{0.1}$ keV  for $\tau=5$, and $7.9\pm{0.7}$ keV  for $\tau=1$.
These values are typical of the warm coronae observed in AGN \citep[e.g.,][]{2018A&A...611A..59P}.
The SED of \texttt{compTT} is broader than the one of \texttt{bbody}, thus when using the model \texttt{[diskbb + compTT]} the inferred temperature of the accretion disk is lower than when using \texttt{[diskbb + bbody]} (model 1). This effect is larger for lower optical depths of the Comptonizing region, going for example from $51\pm{2}$~eV in the case of $\tau=10$ to $44\pm{2}$~eV in the case of $\tau=1$.

\begin{table*}
\caption{Fit results to the \textit{XMM-Newton} quiescent spectra of RX~J1301.9+2747 \label{table:quiescence}}
\centering
\begin{scriptsize}
\begin{tabular}{cccccccc|cc|ccccc}
\hline \hline
model &  $kT_{\rm{disk}}$ & $\log N_{\rm{disk}}$ & $kT$ & $\log N$ & $\Gamma$ & $\log N_{\rm{pow}}$ & $N_H$  & $\log\mathcal{Z}$ & $AIC$ &  $f_{0.3-2}$ & $f_{2-10}$ & $L_{0.3-2\, \rm{keV}}$ & $L_{2-10\, \rm{keV}}$ & $L_{BOL}$\\
(1) & (2) & (3) & (4) & (5) & (6) & (7) & (8) & (9) & (10) & (11) & (12) & (13) & (14) & (15)\\
\hline
\multicolumn{15}{c}{Epoch: 2000}\\
 0 & $66^{+3}_{-3}$ & $3.7^{+0.1}_{-0.2}$ & - & - & - & - & $<0.2$& $-3.4$ &  $-1.7$ & $1.34^{+0.07}_{-0.07}$ & - & $0.19^{+0.01}_{-0.01}$&  - & $1.7^{+0.3}_{-0.3}$\\[0.12cm]
 1 & $59^{+6}_{-6}$ & $4.1^{+0.3}_{-0.4}$  & $160^{+40}_{-60}$ & $-0.06_{-1.22}^{+1.18}$ & - & - & $<0.5$ & $-2.5$ &   0 & $1.40^{+0.08}_{-0.08}$ & $<0.01$ & $0.20^{+0.02}_{-0.02}$ &  $<0.02$ & $2.45^{+1.00}_{-1.00}$\\[0.12cm]
 2 & $65^{+3}_{-3}$ & $3.7^{+0.2}_{-0.2}$  & - & - & $2.9^{+0.8}_{-0.9}$  & $-9.0_{-2.7}^{+2.7}$ & $<0.2$ & $-3.9$ &   $-4.2$ & $1.36^{+0.07}_{-0.07}$ & $<0.11$ & $0.19^{+0.02}_{-0.01}$ &  $<0.14$ & $1.8^{+0.3}_{-0.4}$\\[0.12cm]
 3 & $59^{+4}_{-4}$ & $4.0^{+0.3}_{-0.2}$  & $150^{+30}_{-30}$ & $0.09^{+0.79}_{-0.49}$ & $1.9^{+0.2}_{-0.2}$ & $ -10.0^{-+2.1}_{-2.0}$ & $<0.6$ & $0$ &   $-4.0$ & $1.40^{+0.08}_{-0.08}$ & $<0.04$ & $0.20^{+0.02}_{-0.01}$ &  $<0.06$ & $2.4^{+0.7}_{-0.6}$\\[0.12cm]
  4 & $60^{+4}_{-4}$ & $3.9^{+0.4}_{-0.3}$  & $330^{+40}_{-50}$ & $-4.1^{+1.0}_{-1.2}$ & $-$ & $ -$ & $<0.3$ & $-0.4$ &   $-1.0$ & $1.38^{+0.08}_{-0.06}$ & $<0.06$ & $0.20^{+0.01}_{-0.02}$ &  $<0.08$ & $1.9^{+0.7}_{-0.7}$\\[0.12cm]
 \hline
\multicolumn{15}{c}{Epoch: 2019}\\
0 & $94^{+4}_{-3}$ & $2.3^{+0.2}_{-0.1}$ & - & - & - & - & $<0.4$ & $-50.9$ &   $-52.3$ & $0.85^{+0.03}_{-0.03}$ & - & $0.117^{+0.004}_{-0.005}$ &  - & $0.45^{+0.04}_{-0.04}$\\[0.12cm]
1 & $57^{+5}_{-4}$ & $3.9^{+0.5}_{-0.4}$ & $175^{+15}_{-15}$ & $0.4^{+0.2}_{-0.3}$ & - & - & $<3.2$ & $-2.62$ &   $0$ & $1.02^{+0.04}_{-0.05}$ & $0.07_{-0.05}^{+0.05}$ & $0.15^{+0.04}_{-0.05}$ &  $0.11^{+0.08}_{-0.07}$ & $1.6^{+1.4}_{-1.4}$\\[0.12cm]
2 & $51^{+6}_{-6}$ & $4.1^{+0.5}_{-0.4}$  & - & - & $3.8^{+0.2}_{-0.1}$  & $-5.17^{+0.05}_{-0.04}$ & $<2.1$ & $-8.5$ &  $-6.7$ &  $1.01^{+0.05}_{-0.04}$ & $1.6^{+0.3}_{-0.4}$ & $0.15^{+0.03}_{-0.03}$ & $2.1^{+0.4}_{-0.4}$ & $3.6^{+1.5}_{-1.5}$\\[0.12cm]
3 & $58^{+4}_{-4}$ & $3.8^{+0.3}_{-0.3}$  & $180^{+10}_{-20}$ & $0.35^{+0.21}_{-0.20}$  & $1.9^{+0.2}_{-0.2}$  & $-10.0^{+2.1}_{-2.0}$ & $<1.6$ & $0$ &   $-3.9$ & $1.02^{+0.04}_{-0.04}$ & $0.09^{+0.06}_{-0.07}$ & $0.15^{+0.02}_{-0.05}$ & $0.14^{+0.09}_{-0.10}$ & $1.5^{+0.7}_{-0.6}$\\[0.12cm]
4 & $52^{+4}_{-5}$ & $4.0^{+0.5}_{-0.2}$  & $360^{+40}_{-40}$ & $-3.1^{+0.3}_{-0.3}$  & $-$  & $-$ & $<1.4$ & $-7.2$ &   $-3.2$ & $1.01^{+0.04}_{-0.04}$ & $0.4^{+0.2}_{-0.2}$ & $0.15^{+0.02}_{-0.02}$ & $0.6^{+0.3}_{-0.3}$ & $1.9^{+0.9}_{-0.9}$\\[0.12cm]
\hline
\multicolumn{15}{c}{Epoch: 2020}\\
0 & $107^{+2}_{-3}$ & $2.07^{+0.07}_{-0.06}$ & - & - & - & - & $<0.1$ & $-223.6$ &   $-224.6$ & $0.82^{+0.02}_{-0.02}$ & - & $0.112^{+0.003}_{-0.002}$ &  - & $0.35^{+0.02}_{-0.02}$\\[0.12cm]
1 & $61^{+2}_{-3}$ & $3.7^{0.1}_{-0.2}$ & $200^{+10}_{-10}$ & $0.16^{+0.12}_{-0.11}$ & - & - & $<0.2$ & $-3.0$ &   $0$ & $1.02^{+0.03}_{-0.02}$ & $0.18_{-0.05}^{+0.06}$ & $0.145^{+0.006}_{-0.005}$ &  $0.29^{+0.08}_{-0.09}$ & $1.4^{+0.2}_{-0.2}$ \\[0.12cm]
2 & $49^{+4}_{-3}$ & $4.3^{+0.3}_{-0.3}$  & - & - & $3.6^{+0.1}_{-0.1}$  & $-5.06^{+0.03}_{-0.02}$ & $<0.4$ & $-4.2$ &  $-2.1$ & $1.03^{+0.03}_{-0.03}$ & $2.4^{+0.5}_{-0.4}$ & $0.15^{+0.01}_{-0.01}$ & $3.3^{+1.1}_{-1.2}$ & $3.3^{+1.1}_{-1.2}$\\[0.12cm]
3 & $60^{+2}_{-2}$ & $3.7^{+0.1}_{-0.1}$  & $200{+10}_{-10}$ & $0.2^{+0.1}_{-0.1}$  & $1.9^{+0.2}_{-0.2}$  & $-9.5^{+2.5}_{-3.0}$ & $<0.2$ & $0$ &  $-3.9$ & $1.02^{+0.03}_{-0.02}$ & $0.2^{+0.1}_{-0.1}$ & $0.145^{+0.005}_{-0.005}$ & $0.3^{+0.2}_{-0.2}$ & $1.4^{+0.2}_{-0.2}$\\[0.12cm]
4 & $52^{+3}_{-3}$ & $4.1^{+0.2}_{-0.3}$  & $430^{+30}_{-30}$ & $-3.2^{+0.1}_{-0.2}$  & $-$  & $-$ & $<0.5$ & $-4.3$ &  $0$ & $1.03^{+0.03}_{-0.03}$ & $0.9^{+0.2}_{0.2}$ & $0.15^{+0.01}_{-0.01}$ & $1.3^{+0.03}_{-0.03}$ & $1.9^{+0.5}_{-0.4}$\\[0.12cm]
\hline
\multicolumn{15}{c}{Epoch: 2022A}\\
0 & $107^{+4}_{-3}$ & $1.9^{+0.1}_{-0.1}$ & - & - & - & - & $<0.1$ & $-90.3$ &  $-91.9$ & $0.58^{+0.02}_{-0.03}$ & - & $0.080^{+0.004}_{-0.002}$ &  - & $0.24^{+0.02}_{-0.01}$\\[0.12cm]
1 & $61^{+4}_{-3}$ & $3.5^{+0.2}_{-0.2}$ & $195^{+15}_{-15}$ & $0.06^{+0.18}_{-0.19}$ & - & - & $<0.2$ & $-2.7$  & $-1.0$ & $0.71^{+0.03}_{-0.03}$ & $0.12_{+0.06}^{-0.07}$ & $0.100^{+0.005}_{-0.004}$ &  $0.18^{+0.10}_{-0.09}$ & $0.9^{+0.2}_{-0.2}$ \\[0.12cm]
2 & $48^{+5}_{-5}$ & $4.1^{+0.5}_{-0.3}$  & - & - & $3.7^{+0.2}_{-0.2}$  & $-5.20^{+0.03}_{-0.03}$ & $<0.5$ & $-1.8$  & $-1.4$ & $0.72^{+0.03}_{-0.02}$ & $1.6^{+0.4}_{-0.4}$ & $0.103^{+0.006}_{-0.006}$ & $2.2^{+0.5}_{-0.6}$ & $2.5^{+1.0}_{-1.0}$\\[0.12cm]
3 & $61^{+3}_{-4}$ & $3.5^{+0.2}_{-0.2}$  & $190^{+14}_{-10}$ & $0.09^{+0.17}_{-0.16}$  & $1.9^{+0.2}_{-0.2}$  & $-9.6^{+1.2}_{-2.5}$ & $<0.2$ & $0$ &  $-4.75$ & $0.71^{+0.03}_{-0.04}$ & $0.13^{+0.10}_{-0.11}$ & $0.101^{+0.004}_{-0.005}$ & $0.21^{+0.14}_{-0.16}$ & $0.91^{+0.18}_{-0.18}$\\[0.12cm]
4 & $52^{+4}_{-4}$ & $3.9^{+0.3}_{-0.3}$  & $420^{+60}_{-60}$ & $-3.3^{+0.2}_{-0.2}$  & $-$  & $-$ & $<0.3$ & $-1.8$ &  $0$ & $0.72^{+0.03}_{-0.02}$ & $0.6^{+0.3}_{-0.2}$ & $0.103^{+0.006}_{-0.005}$ & $0.9^{+0.3}_{-0.3}$ & $1.3^{+0.4}_{-0.4}$\\[0.12cm]
\hline
\multicolumn{15}{c}{Epoch: 2022B}\\
0 & $107^{+3}_{-2}$ & $1.94^{+0.06}_{-0.07}$ & - & - & - & - & $<0.1$ & $-177.2$ &  $-179.4$ & $0.67^{+0.02}_{-0.02}$ & - & $0.91^{+0.03}_{-0.03}$ &  - & $0.28^{+0.02}_{-0.01}$\\[0.12cm]
1 & $58^{+2}_{-3}$ & $3.7^{+0.2}_{-0.1}$ & $190^{+10}_{-10}$ & $0.2^{+0.1}_{-0.1}$ & - & - & $<0.2$ & $-2.7$ &  $-1.1$ & $0.84^{+0.02}_{-0.03}$ & $0.10_{+0.04}^{-0.04}$ & $0.119^{+0.005}_{-0.004}$ &  $0.16^{+0.06}_{-0.06}$ & $1.3^{+0.2}_{-0.2}$ \\[0.12cm]
2 & $45^{+3}_{-5}$ & $4.5^{+0.4}_{-0.4}$  & - & - & $3.7^{+0.1}_{-0.2}$  & $-5.15^{+0.03}_{-0.03}$ & $<1.2$ & $-1.7$ &  $-1.0$ & $0.85^{+-0.02}_{-0.02}$ & $1.8^{+0.3}_{-0.4}$ & $0.12^{+0.01}_{-0.01}$ & $2.4^{+0.5}_{-0.4}$ & $3.6^{+1.6}_{-1.6}$\\[0.12cm]
3 & $58^{+2}_{-3}$ & $3.7^{+0.2}_{-0.1}$  & $187^{9}_{10}$ & $0.2^{+0.1}_{-0.1}$  & $1.9^{+0.2}_{-0.2}$  & $-9.9^{+2.3}_{-2.1}$ & $<0.2$ & $0$ &  $-5.3$ & $0.84^{+0.02}_{-0.03}$ & $0.12^{+0.07}_{-0.07}$ & $0.119^{0.004}_{0.004}$ & $0.19^{0.09}_{0.10}$ & $1.3^{+0.2}_{-0.2}$\\[0.12cm]
4 & $49^{+3}_{-4}$ & $4.2^{+0.3}_{-0.2}$  & $410^{+30}_{-30}$ & $-3.2^{+0.2}_{-0.4}$  & $-$  & $-$ & $<0.3$ & $-1.8$ &  $0$ & $0.85^{+0.02}_{-0.03}$ & $0.7^{+0.2}_{0.3}$ & $0.121^{0.005}_{0.005}$ & $0.9^{+0.3}_{-0.2}$ & $1.9^{+0.6}_{-0.6}$\\[0.12cm]
 \hline 
\end{tabular}
\tablefoot{Col. (1) model adopted, model 0: \texttt{[diskbb]}; model 1: \texttt{[diskbb + bbody]}; model 2: \texttt{[diskbb + powerlaw]}; model 3: \texttt{[diskbb + bbody + powerlaw]}; model 4: \texttt{[diskbb + compTT]}. Col. (2)  \texttt{diskbb} inner temperature (eV). Col. (3) logarithm of the \texttt{diskbb} normalization. Col. (4) \texttt{bbody} or \texttt{compTT} temperature (eV).  Col. (5) logarithm of the \texttt{bbody}  or \texttt{compTT}  normalization. Col. (6) \texttt{powerlaw} photon index. Col. (7) logarithm of the \texttt{powerlaw} normalization. Col. (8) 
absorbing column density ($10^{20}$ cm$^{-2}$). Col. (9) logarithm of the Bayesian evidence normalized to the maximum value.  Col. (10) Akaike Information Criterium AIC$\:=C - 2\times m$, where $C$ is the likelihood and $m$ is the number of degrees of freedom, normalized to the maximum value. Col. (11)  observed flux in the $0.3-2$ keV band ($10^{-13}$ erg cm$^{-2}$ s$^{-1}$).  Col. (12) observed flux in the $2-10$ keV band ($10^{-15}$ erg cm$^{-2}$ s$^{-1}$).  Col. (13) intrinsic luminosity computed in the $0.3-2$ keV band ($10^{42}$ erg s$^{-1}$).   Col. (14) intrinsic luminosity computed in the $2-10$ keV band ($10^{39}$ erg s$^{-1}$). Col (15) total bolometric luminosity of the model ($10^{42}$ erg s$^{-1}$). }
\end{scriptsize}
\end{table*}

\section{Fit to the QPE spectra\label{appendix:qpes} }
 
The results of the Gaussian fit to the $0.2-2$ keV light curves of Sect.~\ref{sec:lcurves} were used to select time intervals to extract individual QPE spectral slices. We used the median values of the posterior PDFs for the QPE arrival time $t_{QPE}$ and width $\sigma$, given the pn and MOS data in each epoch of \textit{XMM-Newton }observation.  
We used a $4\sigma$-wide time interval centered around $t_{QPE}$ and divided in five equally-spaced time intervals to extract the QPE spectral slices: two rises (rise-1 and rise-2), one peak, and two decays (decay-1 and decay-2).
Each spectral slice has therefore a different duration, proportional to the total duration of each QPE. One example of the spectral decomposition is shown in Fig.~\ref{FIG10} for QPE2.

For the spectral analysis we used event tables cleaned from strong background flares. These are present typically at the beginning and at the end of \textit{XMM-Newton} revolutions and affect the data of both EPIC cameras (pn and MOS) for 6 out of 34 QPEs, (number 5, 6, 15, 25, 26, and 33), which must be discarded. The pn camera is more sensitive to background flares than the MOS and lost three further QPEs (number 4, 7, 16). Due to the very low photon statistics we discarded QPE16, that is a weak one, while retained the MOS data of the strong QPE4 and QPE7. QPE0 was also excluded from the spectral analysis as it is only half-detected by the MOS. This gives a total of 26 QPEs (21 strong and 5 weak) for which it was possible to extract spectral information.

The QPEs were assumed to be  additive components on top of a constant quiescent emission: we considered the $0.3-2$ keV net QPE spectra by using the average quiescent spectra of each epoch as a background.
The source spectra were analysed unbinned, while the background spectra were binned to a minimum of 3 counts per bin, and the W-statistics was used. The pn and MOS data were fitted jointly for each spectral slice, using a constant factor between the two datasets.

We compared the QPE peak spectra to a blackbody (\texttt{[bbodyrad]}), a bremsstrahlung (\texttt{[bremss]}), and a Comptonization model (\texttt{[compTT]}). We found slightly lower Bayesian evidences for the blackbody model compared to the bremsstrahlung ($\Delta\log\mathcal{Z}\sim 0.5$), with both models strongly favored over the Comptonization one ($\Delta\log\mathcal{Z}\gg 2$). 
The model \texttt{[const*TBabs*zashift*bbodyrad]} was then used to extract information ($L_{BOL}$, $kT$, $R_{bb}$) from the QPE spectral slices, with the caveat that the luminosity estimates are the lowest among the three models tested (i.e., the \texttt{[bbodyrad]} has the narrowest spectral energy distribution among the three).

The presence of absorption during the QPE was tested by allowing the column density of \texttt{[TBabs]} to be free. Uninformed priors were adopted for the cross-instrumental constant (between 0.8 and 1.2), the column density (between $10^{15}$ and $5\times 10^{21}$ cm$^{-2}$), the blackbody temperature (between 30 and 300 eV) and normalization (between $10^{-2}$ and $10^6$). The redshift was always fixed to $z=0.024$.
As the value of $N_H$ is consistent with the Galactic one for 132/135 of the QPE slices, we repeat the fit to the model \texttt{[const*TBabs*zashift*bbodyrad]} fixing the column density value to the Galactic one.

\subsection{Further complexities}
We tested for the presence of ionized absorption along the line of sight both during the QPE spectral evolution and during quiescence. 
In the former case, we compared the data of each QPE spectral slice to the \texttt{[const*TBabs*zedge*zashift*bbodyrad]} model, where \texttt{zedge} is a simple absorption edge at the redshift of the source. 
The absorption edge had uninformed priors for its energy (between $400$ and $900$ eV) and its optical depth (between $10^{-3}$ and 5).
The resulting Bayesian evidences $\mathcal{Z}$ were compared to those obtained with the simpler \texttt{const*TBabs*zashift*bbodyrad} model using the Jeffrey's scale \citep[see][]{2014A&A...564A.125B}, where a difference in $\log\mathcal{Z}$ of 2 is considered decisive, of 1.5 very strong evidence, and above 1 strong evidence that the more complex model provides a better representation of the data than the simpler model.
Out of 135 QPE spectral slices, a difference in  $\log\mathcal{Z}> 1$ (strong evidence for a better data representation after the inclusion of the edge) is found in 10 slices, corresponding to 9 individual QPEs. These are QPEs number 1, 4, 10, 12, 17, 19, 21, 24, and 30.
When replacing the absorption edge with a Gaussian absorption line  the statistical evidence drops in most of the QPE spectral slices, therefore we do not interpret these results as physical, but rather as statistical fluctuations or  spurious effect due to complexities in the continuum emission not taken into account. 
In the latter case we applied the \texttt{zedge} component or a photoionized gas model generated with \texttt{Xstar} to both model 1 and model 4, finding no strong statistical evidence for the presence of absorption in any epoch of \textit{XMM-Newton} observation.

Ionized absorption with an outflowing velocity of $\sim 3000$ km s$^{-1}$ was recently detected in the high-resolution reflection grating spectrometer (RGS) data of GSN 069, both during quiescence and during the QPEs \citep{2024arXiv240617105K}. 
We reduced and analysed the RGS data of RX J1301.9+2747 by stacking all the quiescent good time intervals from 2000 to 2022B. We found that the signal to noise ratio is still completely dominated by the background, thus preventing us from searching for traces of ionized absorption in the high-resolution data. We note how even at the higher X-ray flux level of GSN 069 compared to RX J1301.9+2747, a total of about 2 Ms of exposure time had to be accumulated in order to detect the absorption lines in the high-resolution spectra. In order to  better test for the presence of intrinsic ionized absorption in RX J1301.9+2747, new observations are needed: either very long ones with existing X-ray telescopes, or using future detectors with a very large soft X-ray effective area.
\end{appendix}
\end{document}